\documentclass[prl,aps,superscriptaddress,twocolumn]{revtex4-2}

\usepackage{bm,bbm,color,float,dcolumn,amsmath,amssymb,graphicx,subfigure}

\usepackage[colorlinks=true]{hyperref}
\hypersetup{linkcolor=blue,citecolor=blue,urlcolor=blue}

\begin{document}

\title{Crosscap states and duality of Ising field theory in two dimensions}

\author{Yueshui Zhang}
\affiliation{Faculty of Physics and Arnold Sommerfeld Center for Theoretical Physics, Ludwig-Maximilians-Universit\"at M\"unchen, 80333 Munich, Germany}

\author{Ying-Hai Wu}
\affiliation{School of Physics and Wuhan National High Magnetic Field Center, Huazhong University of Science and Technology, Wuhan 430074, China}

\author{Lei Wang}
\affiliation{Institute of Physics, Chinese Academy of Sciences, Beijing 100190, China}
\affiliation{Songshan Lake Materials Laboratory, Dongguan, Guangdong 523808, China}

\author{Hong-Hao Tu}
\email{h.tu@lmu.de}
\affiliation{Faculty of Physics and Arnold Sommerfeld Center for Theoretical Physics, Ludwig-Maximilians-Universit\"at M\"unchen, 80333 Munich, Germany}

\date{\today}

\begin{abstract}
We propose two distinct crosscap states for the two-dimensional (2D) Ising field theory. These two crosscap states, identifying Ising spins or dual spins (domain walls) at antipodal points, are shown to be related via the Kramers-Wannier duality transformation. We derive their Majorana free field representations and extend bosonization techniques to calculate correlation functions of the 2D Ising conformal field theory (CFT) with different crosscap boundaries. Away from criticality, we develop a conformal perturbation theory to calculate the Klein bottle entropy (norm-square of the crosscap overlap) as a universal scaling function [\href{https://journals.aps.org/prl/abstract/10.1103/PhysRevLett.130.151602}{Phys. Rev. Lett. 130, 151602 (2023)}]. For the Ising field theory, our analytical results support the conjectured monotonicity of the Klein bottle entropy under relevant perturbations. The formalism provides a general framework for studying perturbed 2D CFTs on non-orientable manifolds.
\end{abstract}

\maketitle

Although the two-dimensional (2D) classical Ising model was invented one century ago~\cite{Ising1925,McCoy-Book}, studies on this model continue to deepen our understanding of critical phenomena at the present time. In the absence of external fields, Kramers-Wannier duality~\cite{Kramers1941} and the exact solution~\cite{Onsager1944,Kaufman1949a,Kaufman1949b,YangCN1952b,Schultz1964} reveal a second-order phase transition at a critical temperature. At this critical point, scale invariance is promoted to conformal invariance~\cite{Polyakov1970,Belavin1984,Smirnov2010}, identifying the critical theory as the simplest conformal field theory (CFT) in two dimensions, known as the 2D Ising CFT~\cite{Francesco-Book}. In the vicinity of the critical point, the correlation length is much larger than the lattice spacing, so the long-wavelength physics can be captured using a continuous field theory description. Considering the two relevant perturbations (thermal and magnetic perturbations, driven by temperature and external field, respectively), the scaling limit of the 2D classical Ising model is described by the 2D Ising field theory~\cite{Zamolodchikov1989,Fonseca2002}, which has attracted considerable interest~\cite{Delfino2004,Rutkevich2005,Delfino2006,Konechny2017,Gabai2022,XuXL2024}.

There have been significant advances in the study of 2D CFTs on non-orientable manifolds~\cite{Ishibashi1989,Fioravanti1994,Pradisi1995,Pradisi1996,Fuchs2000,Brunner2004,WeiZX2025,Blumenhagen-Book}, such as the Klein bottle and the real projective plane ($\mathbb{RP}^2$), but the 2D classical Ising model is relatively less explored in this regard~\cite{LuWT2001,Chui2002,cimasoni2024,Shimizu2024}. In general, crosscap boundary states (and the associated crosscap coefficients) are crucial information for understanding the properties of 2D CFTs on non-orientable manifolds. For the 2D Ising CFT, the crosscap coefficients and certain two-point correlators on the $\mathbb{RP}^2$ manifold have been employed to conduct a nontrivial benchmark in the bootstrap program~\cite{Nakayama2016a,Nakayama2016b}.

In this Letter, we demonstrate that the 2D Ising field theory supports two distinct crosscap states, related by Kramers-Wannier duality, which can be explicitly constructed both at and away from criticality. We begin with a lattice formulation (quantum Ising chain) and propose two physically motivated crosscap states on the lattice. These lattice crosscap states identify Ising spins and dual spins (domain walls) at antipodal points, respectively. Remarkably, the overlaps between the lattice crosscap states and the eigenstates of the critical Ising chain are universal (without finite-size corrections), enabling us to derive Majorana free field representations of the crosscap states in the continuum limit. Within the 2D Ising CFT, one of them is the standard Pradisi-Sagnotti-Stanev (PSS) crosscap state~\cite{Fioravanti1994}, while the other corresponds to an equal-weight superposition of the standard PSS crosscap state and a generalized PSS crosscap state labeled by a simple current~\cite{Pradisi1995,Huiszoon1999,Harada2025}. Away from criticality, we develop a conformal perturbation theory to calculate the overlap of crosscap states with perturbed ground states, which we call \emph{crosscap overlap}. This formalism is applicable to any 2D CFT perturbed by relevant operators, thus providing a systematic method to calculate the Klein bottle entropy~\cite{TuHH2017,TangW2017,ChenL2017,WangHX2018,TangW2019,Vanhove2022} (norm-square of the crosscap overlap) as a universal scaling function of dimensionless coupling constants~\cite{ZhangYS2023}. For the $A$-series unitary minimal models, our conformal perturbation theory analysis provides a perturbative proof of the conjectured monotonicity of the Klein bottle entropy~\cite{ZhangYS2023}. For the 2D Ising field theory, our leading-order analytical results further support this monotonicity, strengthening the conjecture.

{\em Ising crosscap states} --- We start with the Hamiltonian formulation of the 2D Ising field theory~\cite{Fonseca2002}
\begin{align}
H = H_{0} - g_1 \int_0^L\mathrm{d}x \, \varepsilon(x) - g_2 \int_0^L\mathrm{d}x \, \sigma(x) \, ,
\label{eq:IsingFT}
\end{align}
where $H_{0}$ is the Hamiltonian of the 2D Ising CFT with central charge $c=1/2$, and $\varepsilon$ and $\sigma$ are primary fields of the Ising CFT with conformal weight $(1/2,1/2)$ and $(1/16,1/16)$, respectively. $g_1$ and $g_2$ are the couplings of the two relevant perturbations. The Hamiltonian~\eqref{eq:IsingFT} is defined on a circle of length $L$ and can be viewed as the generator of the transfer matrix for the 2D classical Ising model in the scaling limit.

To reveal different crosscap states, we first focus on the critical point ($g_1 = g_2 = 0$) and consider the critical Ising chain as its faithful lattice realization:
\begin{align}
    H_{\mathrm{latt}} = -\sum_{j=1}^N \sigma^x_j\sigma^x_{j+1} - \sum_{j=1}^N\sigma^z_j \, ,
\label{eq:Ising-Ham}
\end{align}
where $\sigma^\alpha_j$ ($\alpha =x,z$) are the Pauli operators at site $j$ and $N$ is the total number of sites. We consider even $N$ and adopt periodic boundary condition ($\sigma_{N+1}^\alpha=\sigma_1^\alpha$) throughout this work. The two relevant perturbations in the Ising field theory [Eq.~\eqref{eq:IsingFT}] are realized by adding $H^{\prime}_{\mathrm{latt}} = (1-h_z)\sum_{j=1}^N\sigma_j^z -h_x \sum_{j=1}^N\sigma_j^x$, where the primary field $\varepsilon (x)$ ($\sigma (x)$) is identified as $-\sigma_j^z$ ($\sigma^x_j$) with coupling $g_1\sim 1-h_z$ ($g_2 \sim h_x$).

The Ising chain in Eq.~\eqref{eq:Ising-Ham} has a global $\mathbb{Z}_2$ symmetry, $[H_{\mathrm{latt}},Q]=0$ with $Q=\prod_{j=1}^N \sigma_j^z$. The eigenvalues of $Q$ define $\mathbb{Z}_2$ even ($Q=1$) and $\mathbb{Z}_2$ odd ($Q = -1$) subspaces, which, following the CFT convention~\cite{Francesco-Book}, are called Neveu-Schwarz (NS) and Ramond (R) sectors, respectively. If we restrict ourselves to the NS sector, the critical Ising chain~\eqref{eq:Ising-Ham} is also invariant (self-dual) under the Kramers-Wannier duality transformation. For our purpose, we define the Kramers-Wannier unitary operator as $U_{\mathrm{KW}} = e^{i\frac{\pi}{4}N}\prod_{j=1}^{N-1} (e^{-i\frac{\pi}{4}\sigma_j^z} e^{-i\frac{\pi}{4}\sigma^x_{j}\sigma^x_{j+1}}) e^{-i\frac{\pi}{4}\sigma_N^z}$~\cite{ChenBB2022,Seiberg2024,Append}, which acts on lattice operators as $U_{\mathrm{KW}} \sigma_j^z U^{\dag}_{\mathrm{KW}} = \sigma_j^x \sigma_{j+1}^x $ and $U_{\mathrm{KW}} \sigma_j^x U^{\dag}_{\mathrm{KW}} = \sigma_1^y \prod_{l=2}^{j} \sigma_{l}^z$ in the NS sector.

\begin{figure}[ht]
\centering
\includegraphics[width=0.48\textwidth]{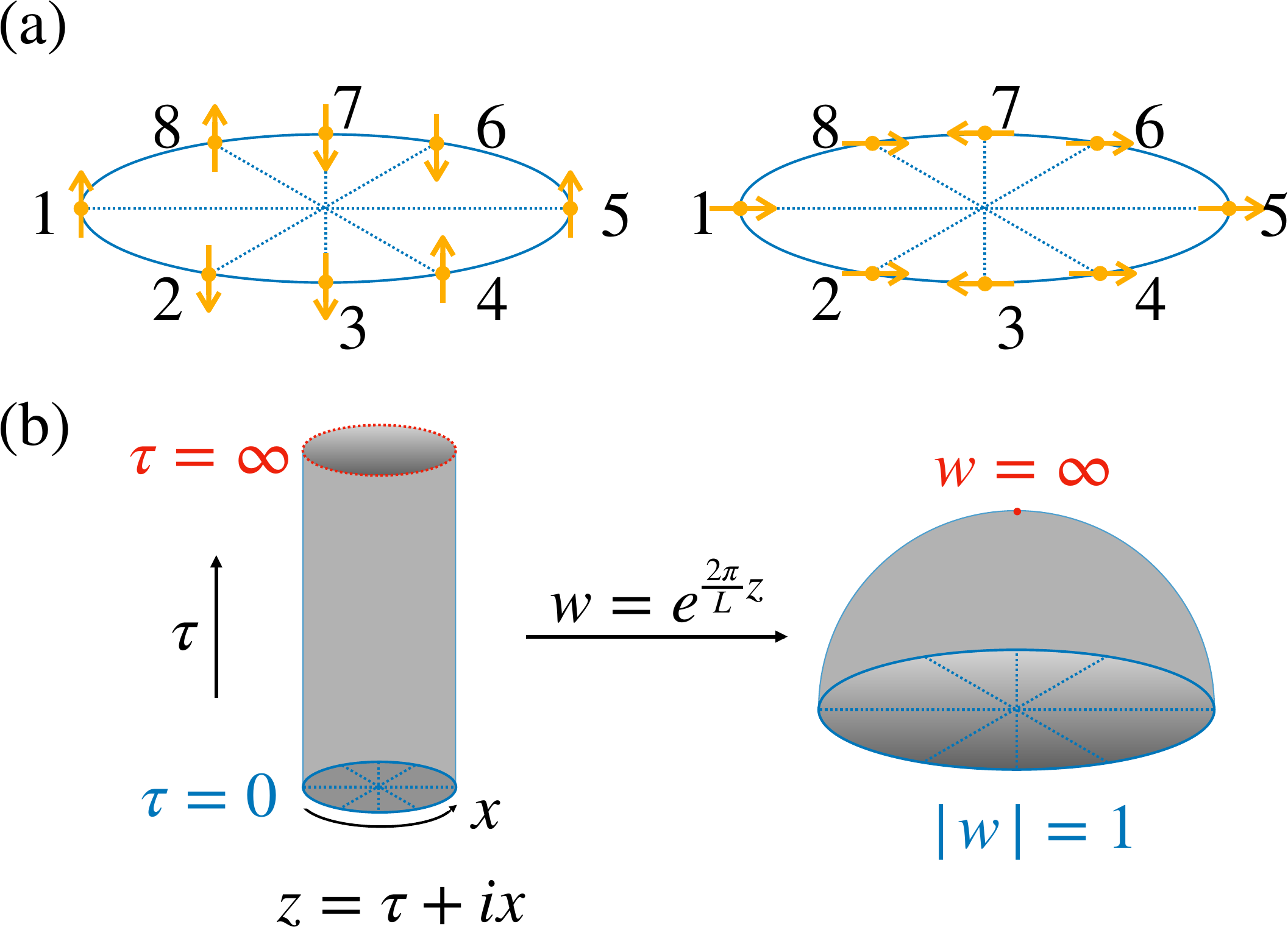}
\caption{Schematics of (a) typical configurations from lattice crosscap states $|\mathcal{C}^+_{\mathrm{latt}}\rangle$ (left panel) and $|\mathcal{C}^-_{\mathrm{latt}}\rangle$ (right panel) in the quantum Ising chain and (b) the conformal transformation from the semi-infinite cylinder with a crosscap boundary to the real projective plane ($\mathbb{RP}^2$).}
\label{fig:Figure1}
\end{figure}

We propose one of the two lattice crosscap states as follows:
\begin{align}
|\mathcal{C}^+_{\mathrm{latt}}\rangle =  \prod_{j=1}^{N/2}(1+\sigma^x_j\sigma^x_{j+N/2}) |\Uparrow\rangle \, ,
\label{eq:cross-latt}
\end{align}
where $|\Uparrow\rangle \equiv |\uparrow_1 \uparrow_2 \cdots \uparrow_N\rangle$ is the fully polarized state in the $\sigma^z$-basis. When placing $|\mathcal{C}^+_{\mathrm{latt}}\rangle$ on the 1D circle, maximally entangled pairs $|\uparrow \uparrow\rangle + |\downarrow \downarrow\rangle$ identify Ising spins between two antipodal sites $j$ and $j+N/2$~\cite{Caetano2022,Ekman2022,Gombor2022,HeM2023,TanBY2025,Yoneta2024}, as depicted in Fig.~\ref{fig:Figure1}(a).

The other crosscap state is obtained by applying the Kramers-Wannier duality transformation to $|\mathcal{C}^+_{\mathrm{latt}}\rangle$:
\begin{align}
|\mathcal{C}^-_{\mathrm{latt}}\rangle &\equiv U_{\mathrm{KW}}|\mathcal{C}^+_{\mathrm{latt}}\rangle   \nonumber \\
&= \prod_{j=1}^{N/2}\left(1+\mu_j\mu_{j+N/2}\right) \tfrac{1}{\sqrt{2}}(|\Rightarrow \rangle + |\Leftarrow\rangle)  \, ,
\label{eq:cross-latt-dual}
\end{align}
where $\mu_j = \prod_{l=1}^{j}\sigma_l^z$ is the Ising disorder operator (dual spin). $|\Rightarrow\rangle \equiv |\rightarrow_1 \rightarrow_2 \cdots \rightarrow_N\rangle$ and $|\Leftarrow\rangle \equiv |\leftarrow_1 \leftarrow_2 \cdots \leftarrow_N\rangle$ are fully polarized states in the $\sigma^x$-basis, i.e., $|\rightarrow\rangle = \frac{1}{\sqrt{2}}(|\uparrow\rangle + |\downarrow\rangle)$ and $|\leftarrow\rangle = \frac{1}{\sqrt{2}}(|\uparrow\rangle -|\downarrow\rangle)$. As $\sigma^z$ flips spins in the $\sigma^x$-basis, $\mu_j \mu_{j+N/2}$ creates two \emph{domain walls} at antipodal positions on top of $|\Rightarrow \rangle$ or $|\Leftarrow\rangle$, as illustrated in Fig.~\ref{fig:Figure1}(a). Thus, the lattice crosscap state $|\mathcal{C}^-_{\mathrm{latt}}\rangle$ identifies each pair of \emph{dual} spins (domain walls) at the antipodal sites.

The two lattice crosscap states satisfy $U_{\mathrm{KW}}|\mathcal{C}^{\pm}_{\mathrm{latt}}\rangle = |\mathcal{C}^{\mp}_{\mathrm{latt}}\rangle$, and therefore, are dual to each other under the Kramers-Wannier transformation. Remarkably, the dual state $|\mathcal{C}^-_{\mathrm{latt}}\rangle$ can be expressed as~\cite{Append}
\begin{align}
    |\mathcal{C}^-_{\mathrm{latt}}\rangle = \frac{1}{\sqrt{2}}(|\mathcal{C}^+_{\mathrm{latt}}\rangle + |\mathcal{C}'_{\mathrm{latt}}\rangle)\, ,
\label{eq:dual-cross-fusion}
\end{align}
which is an equal weight superposition of $|\mathcal{C}^+_{\mathrm{latt}}\rangle$ and another modified crosscap state, defined by
\begin{align}
    |\mathcal{C}'_{\mathrm{latt}}\rangle =  \prod_{j=1}^{N/2}(1-\sigma^x_j\sigma^x_{j+N/2}) |\Uparrow\rangle \, .
\label{eq:cross-latt-prime}
\end{align}
The meaning of these lattice crosscap states will become more transparent in the continuum limit, as we shall demonstrate below.

To identify their field theory counterparts, we adopt the strategy of computing the overlaps of $|\mathcal{C}_{\mathrm{latt}}^\pm\rangle$ with eigenstates of the critical Ising chain~\eqref{eq:Ising-Ham}. As the eigenstates of the critical Ising chain can be easily identified with primary and descendant states of the 2D Ising CFT, the overlaps yield the expansion coefficients of the crosscap states in the Ising CFT basis.

To compute the overlaps, we solve the critical Ising chain~\eqref{eq:Ising-Ham} by using the Jordan-Wigner transformation~\cite{Pfeuty1970}, $\sigma_j^{x} = (c_j^{\dagger} + c_j) \prod_{l=1}^{j-1} e^{i\pi c_l^\dagger c_l} $ and $\sigma^z_j = 2 c^{\dagger}_j c_j - 1$, and represent the eigenstates in the fermionic basis. As $Q|\mathcal{C}_{\mathrm{latt}}^\pm\rangle = |\mathcal{C}_{\mathrm{latt}}^\pm\rangle$, both crosscap states live in the NS sector, so it is sufficient to consider the fermionized Hamiltonian of Eq.~\eqref{eq:Ising-Ham} in the NS sector. After the Fourier transform $c_j = \frac{1}{\sqrt{N}}\sum_{k\in\mathrm{NS}} e^{ik j} c_k$ (lattice spacing set to one) and a Bogoliubov transformation, we arrive at
\begin{align}
    H^{\mathrm{NS}}_{\mathrm{latt}} = \sum_{k\in\mathrm{NS}} 4\cos \frac{k}{2}\left(d_k^\dagger d_k -\frac{1}{2} \right) ,
\label{eq:latt-NS-Hamil}
\end{align}
where $d_k = e^{i\pi/4}\sin\frac{k}{4} c_k + e^{-i\pi/4}\cos\frac{k}{4} c_{-k}^\dagger$ is the annihilation operator of the Bogoliubov mode and $k\in\mathrm{NS}$ denotes allowed single-particle momenta in the NS sector: $k = \pm[\pi -\frac{2\pi}{N}(n_{k}-\frac{1}{2})]$ with $n_k = 1,2,\ldots,N/2$ ($n_k \in \mathbb{Z}^{+}$ in the continuum limit). The ground state of Eq.~\eqref{eq:latt-NS-Hamil}, annihilated by all $d_k$, can be written as
\begin{align}
 |0\rangle_d = \prod_{k>0}\left(\sin\frac{k}{4}+i\cos\frac{k}{4}c_k^\dagger c_{-k}^\dagger\right) |0\rangle_c \, ,
\label{eq:latt-NS-GS}
\end{align}
where $|0\rangle_c$ is the vacuum of the Jordan-Wigner fermions, $c_j|0\rangle_c = 0 \; \forall j$ ($|0\rangle_c$ is just the fully polarized state $|\Downarrow\rangle \equiv |\downarrow_1 \downarrow_2 \cdots \downarrow_N\rangle$ in the spin basis). Excited states of Eq.~\eqref{eq:latt-NS-Hamil} are obtained by applying an even number of $d^{\dag}_k$'s (with distinct momenta) on top of the ground state $|0\rangle_d$.

A key observation enabling the overlap computation is that $|\mathcal{C}^+_{\mathrm{latt}}\rangle$ can be expressed as a sum of two fermionic Gaussian states
\begin{align}
|\mathcal{C}^+_{\mathrm{latt}}\rangle &= \frac{1-i}{2} \prod_{j=1}^{N/2} (1+ i c_j^\dagger c_{j+N/2}^\dagger) |0\rangle_c  \nonumber \\
    &\phantom{=} \;  + \frac{1+i}{2} \prod_{j=1}^{N/2} (1 - i c_j^\dagger c_{j+N/2}^\dagger) |0\rangle_c \, .
\label{eq:cross-latt-fermion}
\end{align}
After transforming Eq.~\eqref{eq:cross-latt-fermion} into momentum space, its overlaps with the eigenstates of Eq.~\eqref{eq:latt-NS-Hamil} can be calculated analytically. We find that only eigenstates of the form $|\psi_{k_1 \cdots k_M}\rangle = \prod_{\alpha=1}^{M} (id_{-k_\alpha}^\dagger d_{k_\alpha}^\dagger ) |0\rangle_d$ ($0<k_1<\cdots< k_M<\pi$) have nonvanishing overlaps with $|\mathcal{C}^+_{\mathrm{latt}}\rangle$~\cite{Append}:
\begin{align}
    \langle \psi_{k_1 \cdots k_M} |\mathcal{C}^+_{\mathrm{latt}}\rangle  = \begin{cases}
    (-1)^{\sum_{\alpha=1}^{M} n_{k_\alpha}}\sqrt{\frac{2+\sqrt{2}}{2}} & M\; \mathrm{even}\\
    i(-1)^{\sum_{\alpha=1}^{M} n_{k_\alpha}}\sqrt{\frac{2-\sqrt{2}}{2}} & M\; \mathrm{odd} \\
    \end{cases} \, .
\label{eq:cross-ovlp}
\end{align}
Most remarkably, the overlaps in Eq.~\eqref{eq:cross-ovlp} are free of any finite-size corrections and valid already for $N\geq 4$~\footnote{Jicheol Kim and Dong-Hee Kim independently obtained the crosscap overlap for the ground state $|0\rangle_d$ and proved that it is free of finite-size corrections.}.

For the other lattice crosscap state $|\mathcal{C}^-_{\mathrm{latt}}\rangle$, the overlaps with the eigenstates of Eq.~\eqref{eq:latt-NS-Hamil} can be computed with the help of the Kramers-Wannier duality. Using $U_{\mathrm{KW}}d_k^\dagger U_{\mathrm{KW}}^\dagger = ie^{-ik/2} d_k^\dagger$ and $U_{\mathrm{KW}}|0\rangle_d = |0\rangle_d$~\cite{Append}, we obtain $\langle \psi_{k_1 \cdots k_M} |\mathcal{C}^-_{\mathrm{latt}}\rangle = (-1)^{M} \langle \psi_{k_1 \cdots k_M}|\mathcal{C}^+_{\mathrm{latt}}\rangle$.

In the continuum limit, the low-energy effective Hamiltonian for the lattice model~\eqref{eq:latt-NS-Hamil} is just the Ising CFT Hamiltonian in the NS sector~\cite{Francesco-Book}
\begin{align}
    H^{\mathrm{NS}}_{0} = \frac{2\pi}{L}\left[\sum_{n=1}^{\infty} (n-\frac{1}{2}) \left(b^{\dag}_{n-\frac{1}{2}}b_{n-\frac{1}{2}} + \bar{b}^{\dag}_{n-\frac{1}{2}}\bar{b}_{n-\frac{1}{2}} \right)-\frac{1}{24}\right]
\label{eq:IsingCFT-ham}
\end{align}
with $b^{\dag}_{n-\frac{1}{2}}$ ($b_{n-\frac{1}{2}}$) being creation (annihilation) operators of the left-moving Majorana fermion (right-moving ones are similar). The connection of Eq.~\eqref{eq:IsingCFT-ham} with the lattice model is through the identification of low-energy modes in Eq.~\eqref{eq:latt-NS-Hamil}: $(d_k, d_k^\dagger) \Leftrightarrow (b_{n_k-\frac{1}{2}},b^{\dag}_{n_k-\frac{1}{2}})$ and $(id_{-k}, -id_{-k}^\dagger) \Leftrightarrow (\bar{b}_{n_k-\frac{1}{2}},\bar{b}^{\dag}_{n_k-\frac{1}{2}})$. This is valid for $k$ close to $\pm\pi$, where the dispersion of the lattice model~\eqref{eq:latt-NS-Hamil} can be linearized. The Kramers-Wannier duality of Eq.~\eqref{eq:IsingCFT-ham} is inherited from the lattice model: $U_{\mathrm{KW}} b_{n-\frac{1}{2}}U^{\dag}_{\mathrm{KW}} = b_{n-\frac{1}{2}}$ and $U_{\mathrm{KW}} \bar{b}_{n-\frac{1}{2}}U^{\dag}_{\mathrm{KW}} = -\bar{b}_{n-\frac{1}{2}}$.

In the continuum limit, the eigenstates $|\psi_{k_1 \cdots k_M}\rangle$ which have nonvanishing overlaps with crosscap states [Eq.~\eqref{eq:cross-ovlp}] become $\prod_{\alpha=1}^{M}b^{\dag}_{n_{k_\alpha}-\frac{1}{2}}\bar{b}^{\dag}_{n_{k_\alpha}-\frac{1}{2}} |0\rangle_{\mathrm{NS}}$ in the Ising CFT, where $|0\rangle_{\mathrm{NS}}$ is the vacuum in the NS sector. By using Eq.~\eqref{eq:cross-ovlp}, the continuum counterparts of the crosscap states are expressed in the Ising CFT basis as
\begin{align}
    |\mathcal{C}_\pm\rangle &= \frac{e^{i\pi/8}}{\sqrt{2}}\exp\left[\pm\sum_{n=1}^\infty (-1)^{n}b^{\dag}_{n-\frac{1}{2}}\bar{b}^{\dag}_{n-\frac{1}{2}}\right]|0\rangle_{\mathrm{NS}}\nonumber\\
    &\phantom{=} + \frac{e^{- i\pi/8}}{\sqrt{2}}\exp\left[\mp\sum_{n=1}^\infty (-1)^{n} b^{\dag}_{n-\frac{1}{2}}\bar{b}^{\dag}_{n-\frac{1}{2}}\right]|0\rangle_{\mathrm{NS}} \, .
\label{eq:cross-field}
\end{align}
The crosscap Ishibashi states of the 2D Ising CFT are labeled by the primary fields as $|a\rangle\rangle_{\mathcal{C}}$ with $a = \mathbbm{1}, \sigma, \varepsilon$~\cite{Ishibashi1989}. The crosscap states $|\mathcal{C}_\pm\rangle$ obtained in Eq.~\eqref{eq:cross-field} are just linear combinations of two Ishibashi states~\cite{Append}
\begin{align}
|\mathcal{C}_\pm\rangle = \sqrt{\frac{2+\sqrt{2}}{2}}|\mathbbm{1}\rangle\rangle_\mathcal{C} \pm \sqrt{\frac{2-\sqrt{2}}{2}}|\varepsilon\rangle\rangle_\mathcal{C} \, .
\label{eq:Ising-crosscaps}
\end{align}
Comparing with the PSS crosscap states for the Ising CFT~\cite{Pradisi1995,Huiszoon1999}, $|\mathcal{C}_{\mathbbm{1}}\rangle = \sqrt{2}\cos\frac{\pi}{8}|\mathbbm{1}\rangle\rangle_\mathcal{C} + \sqrt{2}\sin\frac{\pi}{8}|\varepsilon\rangle\rangle_\mathcal{C}$ and $|\mathcal{C}_{\varepsilon}\rangle = \sqrt{2}\sin\frac{\pi}{8}|\mathbbm{1}\rangle\rangle_\mathcal{C} - \sqrt{2}\cos\frac{\pi}{8}|\varepsilon\rangle\rangle_\mathcal{C}$, it becomes clear that $|\mathcal{C}_+\rangle = |\mathcal{C}_{\mathbbm{1}}\rangle$ and $|\mathcal{C}_-\rangle = \frac{1}{\sqrt{2}}(|\mathcal{C}_{\mathbbm{1}}\rangle + |\mathcal{C}_{\varepsilon}\rangle)$~\cite{Harada2025}. Interestingly, Eqs.~\eqref{eq:dual-cross-fusion} and \eqref{eq:cross-latt-prime} provide a perfect lattice correspondence of the latter relation. It becomes evident once we notice that, in the continuum limit, $|\mathcal{C}'_{\mathrm{latt}}\rangle$ corresponds to $|\mathcal{C}_{\varepsilon}\rangle$~\cite{Append}, by the same type of calculation that identifies $|\mathcal{C}^+_{\mathrm{latt}}\rangle$ with $|\mathcal{C}_{\mathbbm{1}}\rangle$.

{\em Crosscap correlators} --- For the Ising CFT, the two crosscap states $|\mathcal{C}_\pm\rangle$ are indistinguishable from the partition function (with crosscap boundaries): $\langle \mathcal{C}_+|e^{-\beta H_0}|\mathcal{C}_+\rangle = \langle \mathcal{C}_-|e^{-\beta H_0}|\mathcal{C}_-\rangle$. To distinguish them, we calculate the crosscap correlators, which are defined as the conformal correlation functions on the semi-infinite cylinder with a crosscap state at the boundary, in the complex coordinate $z = \tau + ix$ ($\tau \in [0,\infty)$ and $x\in [0,L)$). Geometrically, it is more convenient to perform a conformal transformation $w=e^{\frac{2\pi z}{L}}$ and interpret the crosscap correlators as the correlation functions on the real projective plane ($\mathbb{RP}^2$), as depicted in Fig.~\ref{fig:Figure1}(b). 

We extend bosonization techniques~\cite{Zuber1977,Francesco-Book} to the Ising CFT with crosscap boundaries. In this approach, the Ising crosscap correlators can be expressed in terms of those of the $\mathbb{Z}_2$-orbifold compactified boson CFT, enabling a systematic calculation. For instance, the crosscap correlators of two Ising spin fields are
\begin{align}
{}_{\mathrm{NS}}\langle 0|\sigma (w_1,\bar{w}_1)\sigma (w_2,\bar{w}_2)|\mathcal{C}_\pm\rangle =\sqrt{\frac{2+\sqrt{2}}{2}} \frac{G_\pm (\eta)}{|w_1-w_2|^{\frac{1}{4}}}
\label{eq:cross-spin-correlator}
\end{align}
with
\begin{align}
G_\pm (\eta) = \frac{\frac{\sqrt{2}}{2}\sqrt{1+\sqrt{1-\eta}} \pm \frac{2-\sqrt{2}}{2} \sqrt{1-\sqrt{1-\eta}} }{(1-\eta)^{\frac{1}{8}}} \, ,
\label{eq:crosscap-Gfactor}
\end{align} 
where $w_1$ and $w_2$ are the complex coordinates on $\mathbb{RP}^2$ and $\eta = \frac{|w_1-w_2|^2}{(1+|w_1|^2)(1+|w_2|^2)}$ is the crosscap cross ratio. The correlator with the boundary state $|\mathcal{C}_+\rangle$ is indeed consistent with that presented in Refs.~\cite{Fioravanti1994,Nakayama2016a,Nakayama2016b}. Within the bosonization framework, we also obtain multi-point crosscap correlators for the 2D Ising CFT~\cite{Append}.

{\em Universal scaling functions} --- While expressed in the Ising CFT basis, the validity of Ising crosscap states in Eq.~\eqref{eq:cross-field} extends to the 2D Ising field theory when relevant perturbations are present. The resulting crosscap overlap, defined as the overlap of crosscap states with perturbed ground states, is a universal scaling function of the dimensionless couplings~\cite{ZhangYS2023}. Despite its great potential for identifying critical theories in numerics, there was no systematic method to compute it analytically. Here we take the first step by developing a conformal perturbation theory that allows for a perturbative expansion of the crosscap overlap in powers of the coupling constants, applicable to general 2D CFTs with perturbations.

We consider the Hamiltonian $H = H_0 + H_1$, where $H_0$ is the Hamiltonian of a unitary CFT defined on a circle of length $L$. The perturbation is given by $H_1 = -g \int_0^L \mathrm{d}x \, \varphi(x)$, where $\varphi$ is a primary operator of conformal dimension $(h,\bar{h})$ (we take $h=\bar{h}$ for simplicity and $h<1$ so that the perturbation is relevant), normalized as $\lim_{x\rightarrow \infty}\lim_{L\rightarrow \infty} x^{4h} \langle \varphi(0) \varphi(x) \rangle = 1$ with the expectation value taken in the CFT vacuum. For notational simplicity we consider a single relevant perturbation; the extension to multiple perturbations is straightforward. The perturbed ground state, denoted as $|\psi_0(s)\rangle$, depends only on dimensionless coupling $s = g L^{2-2h}$~\cite{Saleur1987}. Denoting the crosscap state as $|\mathcal{C}\rangle$, our goal is to compute the crosscap overlap $\langle \psi_0(s)|\mathcal{C}\rangle$, which depends solely on $s$ and is thus a universal scaling function~\cite{ZhangYS2023}.

To perform a perturbative analysis, we factorize the crosscap overlap into two parts:
\begin{align}
    \langle\psi_0(s)|\mathcal{C}\rangle = Z(s) \, \exp\left[\frac{1}{2}W(s)\right]
\end{align}
with $Z(s) = \langle \psi_0 (s)|\mathcal{C}\rangle/\langle\psi_0(s ) |\psi_0 (0)\rangle$ and $\exp [\frac{1}{2}W(s)] = \langle\psi_0(s ) |\psi_0(0)\rangle$. For practical calculations, $Z(s)$ and $W(s)$ are written as
\begin{align}
    Z(s ) &= \lim_{\beta\to \infty}\frac{\langle \psi_0(0)|\mathcal{T} e^{-\int_0^\beta \mathrm{d}\tau \, H_1(\tau)}|\mathcal{C}\rangle}{\langle \psi_0(0)|\mathcal{T} e^{-\int_0^\beta \mathrm{d}\tau  \, H_1(\tau)} |\psi_0(0)\rangle}\,,\nonumber\\
    W(s) &= \lim_{\beta\to\infty}\left[\left\langle \mathcal{T} e^{-\int_0^\beta \mathrm{d}\tau \, H_1(\tau)}\right\rangle_\mathrm{c} -\frac{\pi\beta}{6L} \delta c(s) \right] \, ,
\label{eq:perturbation-expan}
\end{align}
where $\mathcal{T}$ stands for time-ordering and $H_1 (\tau)$ is the perturbation term in the interaction picture, $\langle \mathcal{T}e^{-\int_0^\beta \mathrm{d}\tau \, H_1(\tau)}\rangle_\mathrm{c}$ denotes the \emph{connected} contribution of $\langle \psi_0(0)|\mathcal{T}e^{-\int_0^\beta \mathrm{d}\tau \, H_1(\tau)}|\psi_0(0)\rangle$, and $\delta c(s) = \lim_{\beta\to \infty} \frac{6L}{\pi\beta}\langle \mathcal{T}e^{-\int_0^\beta \mathrm{d}\tau \, H_1(\tau)}\rangle_{\mathrm{c}}$ is the change of the ``running'' central charge $c(s) \equiv \delta c(s) + c$. With the knowledge of the CFT correlators (on $\mathbb{RP}^2$ and the plane), the expansion of the time-ordered exponential $\mathcal{T}e^{-\int_0^\beta \mathrm{d}\tau \, H_1(\tau)}$ (in powers of the coupling $g$) allows us to calculate the crosscap overlap in a perturbative way. 

We obtain a general expression for the first-order correction to the crosscap overlap~\cite{Append}:
\begin{align}
    \delta\langle\psi_0(s)|\mathcal{C}\rangle
    &= \mathcal{A}\cdot \frac{_2F_1(h,2h,h +1;-1)}{2(2\pi)^{1-2h}\, h} \cdot s + \cdots\, ,
\label{eq:1st-perturb}
\end{align}
where $\mathcal{A} = \lim_{w\to 0}\langle \psi_0(0)|\varphi (w)|\mathcal{C}\rangle$ is the crosscap coefficient in the one-point conformal crosscap correlator, and $_2F_1(a,b,c;x)$ is the hypergeometric function. Since $_2F_1(h,2h,h+1;-1)>0$ for $h<1$, the sign of the first-order correction is solely determined by $\mathcal{A}$.
As a concrete example, consider the $A$-series unitary minimal models~\cite{Francesco-Book}, with central charge $c= 1 - \frac{6}{m(m+1)}$, $m\geq 3$. The primary fields are labeled by integer pairs $(r,s)$ with $1 \leq r \leq m-1$ and $1 \leq s \leq m$, subject to $r+s=\mathrm{even}$. The crosscap state associated with the identity field $(1,1)$ can be expressed in terms of the Ishibashi states as
\begin{align}
    |\mathcal{C}_{(1,1)}\rangle = \sum_{1\leq r\leq m-1\,, 1\leq s\leq m}^{r+s = \mathrm{even}} \mathcal{A}_{(r,s)} |(r,s)\rangle\rangle_{\mathcal{C}} \, ,
\label{eq:minimal-model-crosscap}
\end{align}
where $\mathcal{A}_{(r,s)}$ are the crosscap coefficients of the primary fields $\varphi_{(r,s)}$~\cite{Bianchi1991,Pradisi1995}:
\begin{align}
    \mathcal{A}_{(r,s)} = O_r O_s \sqrt{\frac{\sqrt{2}\cot \left(\frac{\pi r}{2m}\right) \cot \left(\frac{\pi s}{2(m+1)}\right)}{\sqrt{m(m+1)}}}
\end{align}
with $O_n = \frac{1-(-1)^{n}}{2}$. Since $\mathcal{A}_{(r,s)} \geq 0$ for all $(r,s)$, Eq.~\eqref{eq:1st-perturb} provides a perturbative proof of the monotonicity conjecture~\cite{ZhangYS2023} for the crosscap overlap (and hence Klein bottle entropy) under the renormalization group flow in $A$-series unitary minimal models whenever the crosscap coefficient of the perturbing field is non-vanishing.

The conformal perturbation theory outlined above is directly applicable to the Ising field theory in Eq.~\eqref{eq:IsingFT}, where $|\mathcal{C}_{(1,1)}\rangle$ in Eq.~\eqref{eq:minimal-model-crosscap} reduces to $|\mathcal{C}_+\rangle$. The results are summarized as follows: (i) For the thermal perturbation, the crosscap overlap $\langle \psi_0 (s_1)|\mathcal{C}_+\rangle$, with $s_1 = g_1 L$, has been computed non-perturbatively in Ref.~\cite{ZhangYS2023}. Our conformal perturbation theory calculations agree with the non-perturbative result up to second order~\cite{Append}. (ii) For the magnetic perturbation, conformal perturbation theory shows that the crosscap overlap is an even function of the dimensionless coupling: $\langle \psi_0 (s_2)|\mathcal{C}_+\rangle =\sum_{n=0}^\infty\mathcal{C}_{2n} s_2^{2n}$, with $s_2 = g_2 L^{15/8}$, since crosscap correlators with odd number of $\sigma$-fields vanish. Utilizing the two-point crosscap correlator [Eq.~\eqref{eq:cross-spin-correlator}], we obtain the leading-order correction $\mathcal{C}_2 \approx -1.63528$~\cite{Append}. This shows that the crosscap overlap reaches its maximum at $s_2=0$, thereby supporting the monotonicity conjecture in Ref.~\cite{ZhangYS2023}. An important future direction is to extend this perturbative proof to other perturbed CFTs in which the first-order correction of the crosscap overlap vanishes.

{\em Summary and outlook} --- In conclusion, we have thoroughly investigated two distinct crosscap states in the 2D Ising field theory, connected by Kramers-Wannier duality. By developing a conformal perturbation theory for the crosscap overlap, we extended its definition beyond CFT and established a perturbative expansion in the coupling constant. This framework not only deepens our understanding of field theories on non-orientable manifolds, but also enriches the numerical toolbox for identifying critical theories in lattice models~\cite{LiZQ2020,KimJC2024,TanBY2025,DongJM2025}. The identification of lattice crosscap states and the bosonization techniques developed here also provide a natural route to the study of nonequilibrium dynamics involving crosscap states~\cite{Chalas2025a,WeiZX2024,ChenHH2025,BaiC2025,Dulac2025,Chalas2025b}.

Beyond the Ising model, many field theories exhibit dualities (e.g., $\mathbb{Z}_N$ parafermion CFTs~\cite{Zamolodchikov1985}), and it is natural to ask how crosscap states transform under such dualities. Exploring this interplay may reveal new structures of non-orientable CFTs. A second intriguing direction is to clarify the link between Klein bottle entropy and renormalization-group flow~\cite{ZhangYS2023}, in analogy with the $c$-theorem~\cite{Zamolodchikov1986} and $g$-theorem~\cite{Affleck1991a,Friedan2004}. Finally, crosscap overlaps may provide a direct route to extract crosscap coefficients in 3D CFTs~\cite{DongJM2025}, thereby complementing bootstrap results where only ratios of such coefficients are accessible~\cite{Nakayama2016a}.

{\em Acknowledgments} --- We are grateful to Meng Cheng, Davide Fioravanti, Anton Hulsch, Dong-Hee Kim, Jicheol Kim, Shota Komatsu, Wei Tang, and Hua-Chen Zhang for stimulating discussions. H.-H.T. extends his gratitude to Thomas Quella for many inspiring discussions and hospitality during a research visit to School of Mathematics and Statistics at the University of Melbourne, where this work is completed. Y.S.Z. is supported by the Sino-German (CSC-DAAD) Postdoc Scholarship Program. Y.H.W. is supported by the National Natural Science Foundation of China under Grant No. 12174130. L.W. is supported by National Natural Science Foundation of China under Grant No. T2225018.

{\em Data availability} --- The data that support the findings of this article are openly available~\cite{ZhangYS-Data}.

\bibliographystyle{apsrev4-1}
\bibliography{refs}

\clearpage

\appendix

\begin{widetext}

\begin{center}
\textbf{Supplemental Material for ``Crosscap states and duality of Ising field theory in two dimensions''}
\end{center}

\setcounter{table}{0}
\renewcommand{\thetable}{S\arabic{table}}
\setcounter{figure}{0}
\renewcommand{\thefigure}{S\arabic{figure}}
\setcounter{equation}{0}
\renewcommand{\theequation}{S\arabic{equation}}

This Supplemental Material provides derivation details of some results in the main text. In Sec.~I, we briefly review the exact solution of the transverse field Ising chain (TFIC) and its Kramers-Wannier duality. In Sec.~II, we introduce the lattice crosscap states and show how they are related by the Kramers-Wannier duality transformation. In Sec.~III, we derive the fermionic representation of the lattice crosscap states and calculate their overlaps with the eigenstates of the TFIC. In Sec.~IV, we identify the continuum counterparts of the lattices crosscap states and relate them to the Pradisi-Sagnotti-Stanev crosscap states in the Ising conformal field theory (CFT). The exact crosscap overlaps are derived for the Ising CFT in the presence of thermal perturbation. In Sec.~V, we discuss the bosonization of the conformal crosscap states and use it to calculate the crosscap correlators in the Ising CFT. In Sec.~VI, we develop the conformal perturbation theory to calculate the crosscap overlaps as universal scaling functions. We use it to provide a perturbative proof for the monotonicity of the Klein bottle entropy for the perturbed $A$-series unitary minimal models. We also apply the conformal perturbation theory to calculate the leading-order corrections to the crosscap overlaps in the Ising field theory and the $\mathbb{Z}_3$ parafermion CFT with thermal perturbation.

\tableofcontents

\section{I. Transverse field Ising chain}

In this section, we briefly review the exact solution of the transverse field Ising chain (TFIC) and its Kramers-Wannier duality.

\subsection{A. Exact solution}

The Hamiltonian of the TFIC is given by
\begin{align}
    H_{\mathrm{latt}} = -\sum_{j=1}^N \sigma^x_j\sigma^x_{j+1} - h\sum_{j=1}^N\sigma^z_j \, ,
\label{eq:ham-TFIC-SM}
\end{align}
where we consider even $N$ and $h>0$. The Hamiltonian has $\mathbb{Z}_2$ symmetry, $[H_{\mathrm{latt}}, Q]=0$ with $Q= \prod_{j=1}^N\sigma_j^z$. In the following, we focus on the $\mathbb{Z}_2$ even ($Q=1$) subspace, which is called the Neveu-Schwarz (NS) sector following the CFT convention. 

After performing the Jordan-Wigner transformation
\begin{align}
    \sigma_j^x = \prod_{l=1}^{j-1} e^{i\pi c_l^\dagger c_l} (c_j+c_j^\dagger),\quad
    \sigma_j^z = 2c_j^\dagger c_j-1 \, ,
\label{eq:JW-trans-SM}
\end{align}
the Fourier transform $c_j = \frac{1}{\sqrt{N}}\sum_{k\in\mathrm{NS}}e^{i k j}c_k$, and the Bogoliubov transformation, the Hamiltonian is diagonalized in the NS sector:
\begin{align}
    H^{\mathrm{NS}}_{\mathrm{latt}} = \sum_{k\in\mathrm{NS}}\varepsilon_k \left(d_k^\dagger d_k -\frac{1}{2}\right) \, ,
\end{align}
where the allowed single-particle momenta in the NS sector are $k= \pm\frac{2\pi}{N}(n_k-\frac{1}{2})$ with $n_k = 1,2,\ldots,N/2$. The Bogoliubov single-particle spectrum is
\begin{align}
    \varepsilon_k (h) = 2\sqrt{(h-1)^2+4h\cos^2(k/2)} \, ,
\end{align}
and the annihilation operator of the Bogoliubov mode is
\begin{align}
    d_k (h) =  e^{i\pi/4}\sin(\theta_k/2) c_k + e^{-i\pi/4}\cos(\theta_k/2) c_{-k}^\dagger
    \label{eq:TFIC-dk-SM}
\end{align}
with the Bogoliubov phase $\theta_k (h)$ being determined as
\begin{align}
    \cos\theta_k  = (2h+2\cos k)/\varepsilon_k,\quad \sin\theta_k = 2\sin k/\varepsilon_k \, .
\label{eq:TFIC-Bog-phase-SM}
\end{align}

The ground state of the TFIC in the NS sector is a fermionic Gaussian state
\begin{align}
    |\psi_0 (h)\rangle  =\prod_{k>0}\left(\sin\frac{\theta_k}{2}+i\cos\frac{\theta_k}{2}c_k^\dagger c_{-k}^\dagger\right)|0\rangle_c \, ,
\label{eq:TFIC-GS-SM}
\end{align}
where $|0\rangle_c$, the vacuum of the Jordan-Wigner fermions, satisfies $c_j|0\rangle_c =0 \; \forall j$ and is the fully polarized state in the original $\sigma^z$-basis: $|0\rangle_c = |\Downarrow\rangle \equiv  |\downarrow_1\downarrow_2\cdots\downarrow_N\rangle$. To fix the phase of the ground state, we require that its overlap with $|\Downarrow\rangle$ is positive: $\langle \Downarrow |\psi_0(h)\rangle = \prod_{k>0}\sin\frac{\theta_k}{2}>0$, and $|\psi_0(h)\rangle $ is then a \emph{real positive} wave function in the $\sigma^z$-basis -- this agrees with the Perron-Frobenius theorem, as the off-diagonal matrix elements of the Hamiltonian~\eqref{eq:ham-TFIC-SM} are non-positive in the $\sigma^z$-basis. All excited states in the NS sector are obtained by acting an even number of $d^{\dag}_k$'s (with distinct momenta) on top of the ground state $|\psi_0 (h)\rangle$.

\subsection{B. Kramers-Wannier duality}

To construct the unitary operator for the Kramers-Wannier duality transformation, it is convenient to consider the Majorana representation of the TFIC [Eq.~\eqref{eq:ham-TFIC-SM}]:
\begin{align}
    H_{\mathrm{latt}} = \frac{i}{2}\sum_{j=1}^N (\chi_j-\bar{\chi}_j)(\chi_{j+1}+\bar{\chi}_{j+1})-ih\sum_{j=1}^N \chi_j\bar{\chi}_j\,,
    \label{eq:ham-TFIC-Maj-SM}
\end{align}
where 
\begin{align}
    \chi_j = (-1)^j \left(e^{i\pi/4}c_j + e^{-i\pi/4}c_j^\dagger\right),\quad \bar{\chi}_j = (-1)^j \left(e^{-i\pi/4}c_j + e^{i\pi/4}c_j^\dagger\right)\,,~\label{eq:MajField-latt-SM}
\end{align}
are the lattice Majorana operators satisfying $\{\chi_j,\chi_l\}=\{\bar{\chi}_j,\bar{\chi}_l\} = 2\delta_{jl}$ and $\{\chi_j,\bar{\chi}_l\}=0$. As will be shown later, the lattice Majorana operators correspond to the Majorana fields in the Ising CFT in the continuum limit.

After the basis rotation
\begin{align}
    \chi'_j = \frac{1}{\sqrt{2}}(\chi_j+\bar{\chi}_j)=(-1)^j(c_j+c_j^\dagger),\quad \bar{\chi}'_j = \frac{1}{\sqrt{2}}(\chi_j-\bar{\chi}_j)=i(-1)^j (c_j-c_j^\dagger)\,,~\label{eq:MajField-rotat-latt-SM}
\end{align}
the Hamiltonian [Eq.~\eqref{eq:ham-TFIC-Maj-SM}] becomes
\begin{align}
    H_{\mathrm{latt}} = i\sum_{j=1}^N \bar{\chi}'_j\chi'_{j+1} +  ih\sum_{j=1}^N \chi'_j\bar{\chi}'_j \, .
    \label{eq:ham-TFIC-Maj-rotat-SM}
\end{align}
The Kramers-Wannier duality can be revealed by defining the unitary operator~\cite{ChenBB2022}
\begin{align}
    U_{\mathrm{KW}} = e^{i\frac{\pi}{4}N}\prod_{j=1}^{N-1}e^{-\frac{\pi}{4}\chi'_j\bar{\chi}'_j} e^{-\frac{\pi}{4}\bar{\chi}'_j\chi'_{j+1}} e^{-\frac{\pi}{4}\chi'_N\bar{\chi}'_N} \, ,
\label{eq:KW-fermion-SM}
\end{align}
whose action on the (new) lattice Majorana operators is given by
\begin{align}
    U_{\mathrm{KW}}\chi'_j U_{\mathrm{KW}}^\dagger &= \bar{\chi}'_j\,,\nonumber\\
    U_{\mathrm{KW}}\bar{\chi}'_j U_{\mathrm{KW}}^\dagger &= \begin{cases}
    \chi'_{j+1} &\quad j=1,2,\ldots,N-1\, ,\\
    -\chi'_1   &\quad j=N\, .
    \end{cases} 
    \label{eq:KW-trans-Maj-SM}
\end{align}
In the NS sector, the lattice Majorana operators $\chi'_j$ satisfy the anti-periodic boundary condition, i.e., $\chi'_{N+1} = -\chi'_1$. Thus, we have $U_{\mathrm{KW}}\bar{\chi}'_j U_{\mathrm{KW}}^\dagger = \chi'_{j+1} \; \forall j$, and applying the Kramers-Wannier duality transformation to the Hamiltonian [Eq.~\eqref{eq:ham-TFIC-Maj-rotat-SM}] gives
\begin{align}
    U_{\mathrm{KW}} H^{\mathrm{NS}}_{\mathrm{latt}}(h) U_{\mathrm{KW}}^\dagger = hH^{\mathrm{NS}}_{\mathrm{latt}}(1/h) \, ,
\end{align}
which transforms the TFIC with transverse field $h$ to the same Hamiltonian but with transverse field $1/h$.

The spin operators $\sigma_j^\alpha$ ($\alpha =x,z$) are related to the Majorana operators $\chi'_j$ and $\bar{\chi}'_j$ [Eq.~\eqref{eq:MajField-rotat-latt-SM}] via the Jordan-Wigner transformation [Eq.~\eqref{eq:JW-trans-SM}]:
\begin{align}
    \sigma_j^z &= -i\chi'_j\bar{\chi}'_j\,,\quad \sigma^x_{j} = (-1)^j\prod_{l=1}^{j-1}(i\chi'_l\bar{\chi}'_l)\chi'_j \, , \nonumber\\
    \sigma_j^x\sigma_{j+1}^x &= - \chi'_j (i\chi'_j\bar{\chi}'_j )\chi'_{j+1} = -i \bar{\chi}'_j\chi'_{j+1} \, .
\label{eq:spin-MajRep-SM}
\end{align}
Using the above relations, we can write down the Kramers-Wannier unitary operator [Eq.~\eqref{eq:KW-fermion-SM}] in the spin basis:
\begin{align}
    U_{\mathrm{KW}} = e^{i\frac{\pi}{4}N}\prod_{j=1}^{N-1} \left(e^{-i\frac{\pi}{4}\sigma_j^z} e^{-i\frac{\pi}{4}\sigma^x_{j}\sigma^x_{j+1}}\right) e^{-i\frac{\pi}{4}\sigma_N^z}\,.
    \label{eq:KW-spin-SM}
\end{align}

When viewing $h$ as a continuous parameter, the Kramers-Wannier duality transformation maps the ground state $|\psi_0(h)\rangle$ [Eq.~\eqref{eq:TFIC-GS-SM}] to the dual ground state $|\psi_0(1/h)\rangle$, possibly up to a phase:
\begin{align}
    U_{\mathrm{KW}}|\psi_0 (h)\rangle = e^{i\gamma (h)} |\psi_0(1/h)\rangle\,,
    \label{eq:KW-trans-GS-phase-SM}
\end{align}
where $e^{i\gamma (h)}$ is a \emph{continuous} function of $h\in (0,+\infty)$ with modulus one: $|e^{i\gamma (h)}|=1$. We assert that the phase factor is trivial $e^{i\gamma (h)}= 1 \; \forall h\in (0,+\infty)$, so that
\begin{align}
    U_{\mathrm{KW}}|\psi_0 (h)\rangle = |\psi_0(1/h)\rangle \, .
\label{eq:KW-trans-GS-SM}
\end{align}
Specifically, for the limiting case $h\to +\infty$, we require
\begin{align}
    U_{\mathrm{KW}}|\Uparrow\rangle = \frac{1}{\sqrt{2}}\left(|\Rightarrow\rangle + |\Leftarrow\rangle\right)\,,
\end{align}
where $|\Uparrow\rangle \equiv |\uparrow_1 \uparrow_2 \cdots \uparrow_N\rangle$ is the fully polarized state in the $\sigma^z$-basis, and $|\Rightarrow\rangle \equiv |\rightarrow_1 \rightarrow_2 \cdots \rightarrow_N\rangle$  and $|\Leftarrow\rangle \equiv |\leftarrow_1 \leftarrow_2 \cdots \leftarrow_N\rangle$ are fully polarized states in the $\sigma^x$-basis, with $|\rightarrow\rangle = \frac{1}{\sqrt{2}}(|\uparrow\rangle + |\downarrow\rangle )$ and $ |\leftarrow\rangle = \frac{1}{\sqrt{2}}(|\uparrow\rangle - |\downarrow\rangle )$.

To prove the about assertion, we first consider the limiting case $h\to +\infty$. In this limit, the ground states of $H^{\mathrm{NS}}_{\mathrm{latt}}(h)$ and $H^{\mathrm{NS}}_{\mathrm{latt}}(1/h)$ are given by
\begin{align}
    \lim_{h\to +\infty}|\psi_0 (h)\rangle = |\Uparrow\rangle \, , \qquad \lim_{h\to +\infty}|\psi_0 (1/h)\rangle = \frac{1}{\sqrt{2}}\left(|\Rightarrow\rangle + |\Leftarrow\rangle\right) \, .
\end{align}
These states live in the NS sector and have \emph{real positive} wave function coefficients in the $\sigma^z$-basis. The phase factor $e^{i\gamma (h)}$ [Eq.~\eqref{eq:KW-trans-GS-phase-SM}] in this limit can be determined via
\begin{align}
    e^{i\gamma (+\infty)} = \lim_{h\to +\infty}\frac{\langle \Uparrow|U_{\mathrm{KW}}|\psi_0(h)\rangle}{\langle \Uparrow |\psi_0(1/h)\rangle} = \frac{\langle \Uparrow|U_{\mathrm{KW}}|\Uparrow\rangle}{\langle \Uparrow |\cdot \frac{1}{\sqrt{2}}(|\Rightarrow\rangle + |\Leftarrow\rangle)} = 1 \, ,
\end{align}
where we used
\begin{align}
    \langle \Uparrow|U_{\mathrm{KW}}|\Uparrow\rangle = e^{i\frac{\pi}{4}N}\langle \Uparrow|\prod_{j=1}^{N-1}\left[ e^{-i\frac{\pi}{4}\sigma_j^z} \left(\frac{1-i \sigma^x_{j}\sigma^x_{j+1}}{\sqrt{2}}\right)\right]e^{-i\frac{\pi}{4}\sigma_N^z}|\Uparrow\rangle = e^{i\frac{\pi}{4}N}\cdot 2^{\frac{1-N}{2}}\langle \Uparrow|\prod_{j=1}^{N} e^{-i\frac{\pi}{4}\sigma_j^z} |\Uparrow\rangle  = 2^{\frac{1-N}{2}}\,.
\end{align}

To determine the phase factor $e^{i\gamma (h)}$ [Eq.~\eqref{eq:KW-trans-GS-phase-SM}] for general $h$, we consider
\begin{align}
    e^{-i\gamma (h)} =\frac{\langle \Uparrow|U^\dag_{\mathrm{KW}}|\psi_0(1/h)\rangle}{\langle \Uparrow |\psi_0(h)\rangle } = \frac{\frac{1}{\sqrt{2}}(\langle\Rightarrow| + \langle \Leftarrow|)\cdot |\psi_0(1/h)\rangle}{\langle \Uparrow|\psi_0(h)\rangle } >0\,,
\end{align}
since $|\psi_0(h)\rangle$ is a \emph{real positive} wave function in the $\sigma^z$-basis by definition [Eq.~\eqref{eq:TFIC-GS-SM}]. Therefore, as a \emph{continuous} function with modulus one, $e^{i\gamma (h)}$ must be a constant which equals one: 
\begin{align}
    e^{i\gamma (h)} =1\,,\quad \forall h\in (0,+\infty) \, .
\end{align}
This proves our assertion [Eq.~\eqref{eq:KW-trans-GS-SM}].

Lastly, we determine how Bogoliubov modes change under the Kramers-Wannier duality transformation. 

We first determine the transformation rule of the Jordan-Wigner fermion via Eq.~\eqref{eq:KW-trans-Maj-SM}:
\begin{align}
    U_{\mathrm{KW}}c_j U_{\mathrm{KW}}^\dagger = \frac{(-1)^j}{2}U_{\mathrm{KW}} (\chi'_j-i\bar{\chi}'_j) U^\dag_{\mathrm{KW}} =\frac{(-1)^j}{2}(\bar{\chi}'_j-i\chi'_{j+1}) = \frac{i}{2}(c_j+c_{j+1}-c_j^\dagger +c_{j+1}^\dagger)\,.
\end{align}
When going to momentum space, it reads
\begin{align}
    U_{\mathrm{KW}} c_k U^\dag_{\mathrm{KW}} = \frac{1}{\sqrt{N}}\sum_{j=1}^N e^{-ik j}U_{\mathrm{KW}} c_j U^\dag_{\mathrm{KW}} =\frac{i}{2}(c_{k} +e^{ik}c_{k} - c_{-k}^\dagger +e^{ik}c_{-k}^\dagger)= ie^{i k/2}(\cos\frac{k}{2}c_{k} +i\sin\frac{k}{2}c_{-k}^\dagger)
\end{align}
and $U_{\mathrm{KW}} c^\dagger_k U^\dag_{\mathrm{KW}} = -ie^{-i k/2}(\cos\frac{k}{2}c_{k}^\dagger -i\sin\frac{k}{2}c_{-k})$. Therefore, the Bogoliubov mode $d_k^\dagger (h)$ [Eq.~\eqref{eq:TFIC-dk-SM}] transforms as
\begin{align}
    U_{\mathrm{KW}}d_k^\dagger (h) U_{\mathrm{KW}}^\dagger &= U_{\mathrm{KW}}\left[e^{-i\pi/4}\sin\frac{\theta_k (h)}{2}c_k^\dagger  +e^{i\pi/4}\cos\frac{\theta_k (h)}{2} c_{-k}\right]U_{\mathrm{KW}}^\dagger \nonumber\\
    &=-ie^{-i k/2} e^{-i\pi/4}\sin\frac{\theta_k}{2} (\cos\frac{k}{2}c_{k}^\dagger -i\sin\frac{k}{2}c_{-k}) + ie^{-i k/2}e^{i\pi/4}\cos\frac{\theta_k}{2}(\cos\frac{k}{2}c_{-k} -i\sin\frac{k}{2}c_{k}^\dagger)\nonumber\\
    &= ie^{-ik/2}\left[e^{-i\pi/4}\sin\left(\frac{k-\theta_k (h)}{2}\right)c_k^\dagger + e^{i\pi/4}\cos\left(\frac{k-\theta_k (h)}{2}\right)c_{-k}\right]\nonumber\\
    &= ie^{-ik/2} d_k^\dagger (1/h)\,,\quad k\in\mathrm{NS}\,,~\label{eq:KW-trans-dk-SM}
\end{align}
where we used $\theta_k (1/h) = k-\theta_k(h)$ with $\theta_k(h)$ defined in Eq.~\eqref{eq:TFIC-Bog-phase-SM}.

\section{II. Kramers-Wannier duality of the lattice crosscap states}

In this section, we consider the lattice crosscap states and discuss how they are related via the Kramers-Wannier duality. We also derive an alternative representation of the Kramers-Wannier dual lattice crosscap state, which is expressed as an equal weight superposition of the original lattice crosscap state and another modified lattice crosscap state.

\subsection{A. Lattice crosscap state and its dual}

The lattice crosscap state $|\mathcal{C}_{\mathrm{latt}}^+\rangle$ is the product of Bell pairs identifying Ising spins at two antipodal sites:
\begin{align}
|\mathcal{C}_{\mathrm{latt}}^+\rangle = \prod_{j=1}^{N/2} \left(|\uparrow_j\uparrow_{j+N/2}\rangle + |\downarrow_j\downarrow_{j+N/2}\rangle\right) \, .
\label{eq:cross-latt-SM}
\end{align}
For revealing its Kramers-Wannier dual state, it is convenient to rewrite it as
\begin{align}
|\mathcal{C}^+_{\mathrm{latt}}\rangle =  \prod_{j=1}^{N/2}(1+\sigma^x_j\sigma^x_{j+N/2}) |\Uparrow\rangle \, .
\end{align}

Using the Majorana fermion representation of the spin operators [Eq.~\eqref{eq:spin-MajRep-SM}], it is easy to obtain via Eq.~\eqref{eq:KW-trans-Maj-SM} how they change under the Kramers-Wannier duality transformation:
\begin{align}
    U_{\mathrm{KW}}\sigma_j^z U_{\mathrm{KW}}^\dagger = -i (U_{\mathrm{KW}} \chi'_j U_{\mathrm{KW}}^\dagger ) (U_{\mathrm{KW}} \bar{\chi}'_j U_{\mathrm{KW}}^\dagger )= -i \bar{\chi}'_j\chi'_{j+1}  = \sigma^x_{j}\sigma^x_{j+1} \, ,
\end{align}
and
\begin{align}
    U_{\mathrm{KW}}\sigma_j^x  U_{\mathrm{KW}}^\dagger 
    = -\prod_{l=1}^{j-1}\left[ U_{\mathrm{KW}}\left(-i\chi'_l\bar{\chi}'_l\right)U_{\mathrm{KW}}^\dagger \right](U_{\mathrm{KW}} \chi'_j U_{\mathrm{KW}}^\dagger )
    = -\bar{\chi}'_1 \prod_{l=2}^j (-i\chi'_l\bar{\chi}'_l)= \sigma_1^y \prod_{l=2}^j \sigma_l^z\,.
\end{align}

Since the lattice crosscap state $|\mathcal{C}_{\mathrm{latt}}^+\rangle$ lives in the NS sector, $Q|\mathcal{C}_{\mathrm{latt}}^+\rangle = |\mathcal{C}_{\mathrm{latt}}^+\rangle$, it is natural to consider its Kramers-Wannier dual state, denoted as $|\mathcal{C}_{\mathrm{latt}}^-\rangle \equiv U_{\mathrm{KW}}|\mathcal{C}_{\mathrm{latt}}^+\rangle$, which also lives in the NS sector. The explicit form of the dual lattice crosscap state $|\mathcal{C}_{\mathrm{latt}}^-\rangle$ reads
\begin{align}
    |\mathcal{C}_{\mathrm{latt}}^-\rangle \equiv U_{\mathrm{KW}}|\mathcal{C}_{\mathrm{latt}}^+\rangle &= 
    \prod_{j=1}^{N/2}\left(1+U_{\mathrm{KW}}\sigma_j^x U_{\mathrm{KW}}^\dagger \cdot U_{\mathrm{KW}} \sigma_{j+1}^x  U_{\mathrm{KW}}^\dagger \right) \left(U_{\mathrm{KW}}|\Uparrow\rangle\right)\nonumber\\
    &=\prod_{j=1}^{N/2}
    \left(1+\mu_j\mu_{j+N/2}\right) \frac{1}{\sqrt{2}}\left(|\Rightarrow\rangle + |\Leftarrow\rangle\right) \, ,
\label{eq:cross-latt-dual-SM}
\end{align}
where $\mu_j = \prod_{l=1}^{j}\sigma_l^z$ is the Ising disorder operator (dual spin). This corresponds to Eq.~(4) in the main text.

Moreover, since $U_{\mathrm{KW}}\mu_j\mu_{j+N/2} U^\dag_{\mathrm{KW}} = \sigma^x_{j+1}\sigma^x_{j+1+N/2}$ and $U_{\mathrm{KW}}\, \frac{1}{\sqrt{2}} \left(|\Rightarrow\rangle + |\Leftarrow\rangle\right) = |\Uparrow\rangle$, imposing periodic boundary conditions one finds
\begin{align}
    U_{\mathrm{KW}}|\mathcal{C}_{\mathrm{latt}}^-\rangle = |\mathcal{C}_{\mathrm{latt}}^+\rangle\, .
    \label{eq:cross-latt-dual-inv-SM}
\end{align}
Thus, the two lattice crosscap states, $|\mathcal{C}_{\mathrm{latt}}^+\rangle$ and $|\mathcal{C}_{\mathrm{latt}}^-\rangle$, are dual to each other under the Kramers-Wannier transformation, as shown in Eqs.~\eqref{eq:cross-latt-dual-SM} and ~\eqref{eq:cross-latt-dual-inv-SM}. 

\subsection{B. Alternative representation of the dual lattice crosscap state}

In the main text [Eq.~(5)], we mentioned that the dual lattice crosscap state $|\mathcal{C}_{\mathrm{latt}}^-\rangle$ can be expressed as
\begin{align}
    |\mathcal{C}^-_{\mathrm{latt}}\rangle = \frac{1}{\sqrt{2}} \left(|\mathcal{C}^+_{\mathrm{latt}}\rangle + |\mathcal{C}'_{\mathrm{latt}}\rangle \right)\, .
\label{eq:dual-cross-fusion-SM}
\end{align}
Here, $|\mathcal{C}^+_{\mathrm{latt}}\rangle$ is the original lattice corsscap state [Eq.~\eqref{eq:cross-latt-SM}], and $|\mathcal{C}'_{\mathrm{latt}}\rangle$ is a modified lattice crosscap state (with a different type of Bell pairs between antipodal sites):
\begin{align}
    |\mathcal{C}'_{\mathrm{latt}}\rangle
    = \prod_{j=1}^{N/2} \left(|\uparrow_j\uparrow_{j+N/2}\rangle - |\downarrow_j\downarrow_{j+N/2}\rangle\right)\, .
\end{align}

To prove Eq.~\eqref{eq:dual-cross-fusion-SM}, we compare both sides explicitly in the $\sigma^z$-basis. We start with the definition of $|\mathcal{C}^-_{\mathrm{latt}}\rangle$ in Eq.~\eqref{eq:cross-latt-dual-SM} and decompose it as
\begin{align}
    |\mathcal{C}^-_{\mathrm{latt}}\rangle
    = \frac{1}{\sqrt{2}} \left(|D_+\rangle + |D_-\rangle\right)
\end{align}
with
\begin{align}
    |D_\pm\rangle = \frac{1}{2^{N/2}}\prod_{j=1}^{N/2} \left(1+ \prod_{l=j+1}^{j+N/2}\sigma_l^z\right)  \prod_{j=1}^N (1\pm\sigma_j^x)|\Uparrow\rangle\, .
\end{align}
Since $\sigma^z |s\rangle = (-1)^{\frac{s-1}{2}} |s\rangle$ for $s=\pm 1$, one finds
\begin{align}
    |D_\pm\rangle 
    &= \frac{1}{2^{N/2}}\sum_{s_1,\cdots,s_N=\pm 1} 
    \prod_{j=1}^{N/2} \left[1+ \prod_{l=j+1}^{j+N/2} (-1)^{\frac{s_l-1}{2}}\right] (\pm 1)^{\sum_{j=1}^N\frac{s_j-1}{2}} |s_1\cdots s_N\rangle\nonumber\\
    &= \sum_{s_1,\cdots,s_N=\pm 1} \prod_{j=1}^{N/2}\delta^{(2)}_{\sum_{l=j+1}^{j+N/2}\frac{s_l-1}{2},0} \, (\pm 1)^{\sum_{j=1}^N\frac{s_j-1}{2}} |s_1\cdots s_N\rangle \, ,
\end{align}
where $\delta^{(2)}_{n,0} \equiv \frac{1+(-1)^n}{2}$ ($n\in\mathbb{Z}$) is the Kronecker-$\delta$ function with period $2$. The relation
\begin{align}
    \delta^{(2)}_{\sum_{l=j}^{j-1+N/2}\frac{s_l-1}{2},0} \, \delta^{(2)}_{\sum_{l=j+1}^{j+N/2}\frac{s_l-1}{2},0} = \delta^{(2)}_{\sum_{l=j}^{j-1+N/2}\frac{s_l-1}{2},0} \delta_{s_j,s_{j+N/2}}\, ,\quad j=2,\cdots,N/2\,
\end{align}
allows us to simplify the expression of $|D_\pm\rangle$,
\begin{align}
    |D_\pm\rangle 
    &= \sum_{s_1,\cdots,s_N=\pm 1} 
    \delta^{(2)}_{\sum_{j=2}^{N/2 +1}\frac{s_j-1}{2},0}\, \prod_{j=2}^{N/2}\delta_{s_j,s_{j+N/2}} (\pm 1)^{\sum_{j=1}^N\frac{s_j-1}{2}}|s_1\cdots s_N\rangle \nonumber\\
    &= \sum_{s_1,\cdots,s_{N/2+1}=\pm 1} \delta^{(2)}_{\sum_{j=2}^{N/2 +1}\frac{s_j-1}{2},0}\, (\pm 1)^{\frac{s_1-1}{2} - \frac{s_{N/2+1}-1}{2} + 2\sum_{j=2}^{N/2+1}\frac{s_j-1}{2}} |s_1s_2\cdots s_{N/2} s_{N/2 +1} s_2 \cdots s_{N/2} \rangle\nonumber\\
    &= \sum_{s_1,\cdots,s_{N/2+1}=\pm 1} \delta^{(2)}_{\sum_{j=2}^{N/2 +1}\frac{s_j-1}{2},0}\, (\pm 1)^{\frac{1}{2}(s_1-s_{N/2+1})} |s_1s_2\cdots s_{N/2} s_{N/2 +1} s_2 \cdots s_{N/2} \rangle\, .
\end{align}
Summing $|D_+\rangle$ and $|D_-\rangle$ gives
\begin{align}
    |D_+\rangle + |D_-\rangle &= \sum_{s_1,\cdots,s_{N/2+1}=\pm 1} \delta^{(2)}_{\sum_{j=2}^{N/2 +1}\frac{s_j-1}{2},0}\, \left[1+ (- 1)^{\frac{1}{2}(s_1-s_{N/2+1})}\right] |s_1s_2\cdots s_{N/2} s_{N/2 +1} s_2 \cdots s_{N/2} \rangle\nonumber\\
    &= 2 \sum_{s_1,\cdots,s_{N/2+1}=\pm 1} \delta^{(2)}_{\sum_{j=2}^{N/2 +1}\frac{s_j-1}{2},0} \delta_{s_1,s_{N/2+1}} |s_1s_2\cdots s_{N/2} s_{N/2 +1} s_2 \cdots s_{N/2} \rangle\nonumber\\
    &= 2 \sum_{s_1,\cdots,s_{N/2}=\pm 1} \delta^{(2)}_{\sum_{j=1}^{N/2}\frac{s_j-1}{2},0} |s_1\cdots s_{N/2} s_1 \cdots s_{N/2} \rangle\, .
\end{align}
On the other hand, the right-hand side of Eq.~\eqref{eq:dual-cross-fusion-SM} expands as
\begin{align}
    |\mathcal{C}^+_{\mathrm{latt}}\rangle + |\mathcal{C}'_{\mathrm{latt}}\rangle
    &= \sum_{s_1,\cdots,s_{N/2}=\pm 1} \left[1+ (-1)^{\sum_{j=1}^{N/2} (s_j-1)/2}\right] |s_1\cdots s_{N/2} s_1 \cdots s_{N/2}\rangle\nonumber\\
    &= 2\sum_{s_1,\cdots,s_{N/2}=\pm 1} \delta^{(2)}_{\sum_{j=1}^{N/2} \frac{s_j-1}{2}, 0}\, |s_1\cdots s_{N/2} s_1\cdots s_{N/2}\rangle\, ,
\end{align}
Since the two expansions coincide, Eq.~\eqref{eq:dual-cross-fusion-SM} is proved.

\section{III. Exact lattice crosscap overlaps}

In this section, we calculate the overlaps of the lattice crosscap states $|\mathcal{C}_{\mathrm{latt}}^\pm\rangle$, and the modified crosscap state $|\mathcal{C}'_{\mathrm{latt}}\rangle$, with eigenstates of the TFIC. 

We first derive the fermionic representation of the lattice crosscap state $|\mathcal{C}_{\mathrm{latt}}^+\rangle$, which is a superposition of two fermionic Gaussian states in the NS sector. This crucial observation allows us to calculate its overlap with the ground state $|\psi_0(h)\rangle$ of the TFIC as well as all excited states. The overlaps for the dual state $|\mathcal{C}_{\mathrm{latt}}^-\rangle$ then follow directly from Kramers-Wannier duality. Moreover, the modified lattice crosscap state $|\mathcal{C}'_{\mathrm{latt}}\rangle$ admits an analogous fermionic representation, allowing its overlaps to be computed as well. The expansions of all three lattice crosscap states in the TFIC eigenbasis provide an alternative verification of Eq.~\eqref{eq:dual-cross-fusion-SM}.

\subsection{A. Fermionic representation of the lattice crosscap state}

Our important result is that the lattice crosscap state $|\mathcal{C}_{\mathrm{latt}}^+\rangle$ [Eq.~\eqref{eq:cross-latt-SM}] is the following equal weight superposition of certain states in the Jordan-Wigner fermion basis:
\begin{align}
|\mathcal{C}_{\mathrm{latt}}^+\rangle = \prod_{j=1}^{N/2}\left(1 + \sigma_j^+\sigma_{j+N/2}^+\right)|\Downarrow\rangle
    =\sum_{n=0}^{N/2}\sum_{1\leq j_1 <\cdots j_n\leq N/2} c_{j_1}^\dagger \cdots c_{j_n}^\dagger c_{j_1+N/2}^\dagger \cdots c_{j_n+N/2}^\dagger |0\rangle_c\,,
\end{align}
which can be divided into the ``even'' and ``odd'' parts by introducing the half-chain fermion parity $P_{\mathrm{half}} = (-1)^{\sum_{j=1}^{N/2}c_j^\dagger c_j}$,
\begin{align}
|\mathcal{C}_{\mathrm{latt}}^+\rangle = \frac{1+P_{\mathrm{half}}}{2}|\mathcal{C}_{\mathrm{latt}}^+\rangle +\frac{1-P_{\mathrm{half}}}{2}|\mathcal{C}_{\mathrm{latt}}^+\rangle
\label{eq:half-chain-parity-SM}
\end{align}
with 
\begin{align}
    \frac{1+P_{\mathrm{half}}}{2}|\mathcal{C}_{\mathrm{latt}}^+\rangle 
    &= \sum_{n=0 \,(n \; \mathrm{even})}^{N/2} \sum_{1\leq j_1 <\cdots< j_n\leq N/2} c_{j_1}^\dagger \cdots c_{j_n}^\dagger c_{j_1+N/2}^\dagger \cdots c_{j_n+N/2}^\dagger |0\rangle_c \, , \nonumber \\
    \frac{1-P_{\mathrm{half}}}{2}|\mathcal{C}_{\mathrm{latt}}^+\rangle 
    &= \sum_{n=1 \,(n \; \mathrm{odd})}^{N/2} \sum_{1\leq j_1 <\cdots< j_n\leq N/2} c_{j_1}^\dagger \cdots c_{j_n}^\dagger c_{j_1+N/2}^\dagger \cdots c_{j_n+N/2}^\dagger |0\rangle_c \, .
\end{align}
The key observation is that the even and odd parts can be expressed via a pair of fermionic Gaussian states
\begin{align}
    |B_{\mathrm{latt}}^\pm\rangle = \prod_{j=1}^{N/2}(1\pm i c_j^\dagger c_{j+N/2}^\dagger)|0\rangle_c\,.
    \label{eq:Gaussian-Bpm-SM}
\end{align}
To see this, we consider the expansion $|B_{\mathrm{latt}}^\pm\rangle = \sum_{n=0}^{N/2} \sum_{1\leq j_1  <\cdots< j_n\leq N/2} (\pm i)^n  c_{j_1}^\dagger c_{j_1+N/2}^\dagger \cdots c_{j_n}^\dagger c_{j_n+N/2}^\dagger |0\rangle_c$ and notice 
\begin{align}
    (\pm i)^n c_{j_1}^\dagger c_{j_1+N/2}^\dagger \cdots c_{j_n}^\dagger c_{j_n+N/2}^\dagger =
    \begin{cases}
    c_{j_1}^\dagger \cdots c_{j_n}^\dagger c_{j_1+N/2}^\dagger \cdots c_{j_n+N/2}^\dagger &\quad n\;\mathrm{even} \\
    \pm i \; c_{j_1}^\dagger \cdots c_{j_n}^\dagger c_{j_1+N/2}^\dagger \cdots c_{j_n+N/2}^\dagger &\quad n\;\mathrm{odd}
    \end{cases} ,
\end{align}
which allows us to relate the fermionic Gaussian states $|B_{\mathrm{latt}}^\pm\rangle$ to the even and odd parts of $|\mathcal{C}_{\mathrm{latt}}^+\rangle$:
\begin{align}
    |B_{\mathrm{latt}}^\pm\rangle = \frac{1+P_{\mathrm{half}}}{2}|\mathcal{C}_{\mathrm{latt}}\rangle \pm i \frac{1-P_{\mathrm{half}}}{2}|\mathcal{C}_{\mathrm{latt}}\rangle \, .
\end{align}
Thus, the lattice crosscap state $|\mathcal{C}_{\mathrm{latt}}^+\rangle$ [Eq.~\eqref{eq:half-chain-parity-SM}] can be expressed in terms of the fermionic Gaussian states $|B^\pm_{\mathrm{latt}}\rangle$ as
\begin{align}
|\mathcal{C}_{\mathrm{latt}}^+\rangle =\frac{1-i}{2}|B_{\mathrm{latt}}^+\rangle +\frac{1+i}{2}|B_{\mathrm{latt}}^-\rangle \, ,
\label{eq:cross-latt-fermi-SM}
\end{align}
which corresponds to Eq.~(9) in the main text.

The fermionic Gaussian states $|B_{\mathrm{latt}}^\pm\rangle =\exp\left(\pm \frac{i}{2} \sum_{j=1}^{N}c_j^\dagger c_{j+N/2}^\dagger\right)|0\rangle_c$ are translationally invariant. After performing the Fourier transform $c_j = \frac{1}{\sqrt{N}}\sum_{k\in\mathrm{NS}}e^{i k j}c_k$, we obtain
\begin{align}
    |B_{\mathrm{latt}}^\pm\rangle
    &= \exp \left(\pm i\sum_{k>0}e^{ik N/2}c_k^\dagger c_{-k}^\dagger\right)|0\rangle_c = \prod_{k>0}\left[1\mp  (-1)^{n_k+N/2}c_k^\dagger c_{-k}^\dagger\right]|0\rangle_c \, ,
\label{eq:fermi-Gauss-latt-SM}
\end{align}
where we used $k = \pi - \frac{2\pi}{N}(n_k-\frac{1}{2})$ for $k>0$.

\subsection{B. Lattice crosscap overlap for the ground state}
\label{sec:lattice-ovlp-GS-SM}

To obtain the lattice crosscap overlap for the ground state $|\psi_0 (h)\rangle$ [Eq.~\eqref{eq:TFIC-GS-SM}], we just need to calculate the overlap $\langle \psi_0(h)|B_{\mathrm{latt}}^\pm\rangle$ by using the fermionic representation of the lattice crosscap state [Eq.~\eqref{eq:cross-latt-fermi-SM}]. The calculation is straightforward since both $|B_{\mathrm{latt}}^\pm\rangle$ [Eq.~\eqref{eq:fermi-Gauss-latt-SM}] and $|\psi_0(h)\rangle$ are fermionic Gaussian states:
\begin{align}
    \langle \psi_0 (h)|B_{\mathrm{latt}}^\pm\rangle &= {}_c\langle 0|\prod_{k>0}\left(\sin\frac{\theta_k}{2}-i\cos\frac{\theta_k}{2}c_{-k}c_k\right)\left(1\mp (-1)^{n_k+N/2}c_k^\dagger c_{-k}^\dagger\right)|0\rangle_c \nonumber\\
    &= \prod_{k>0}\left[\sin\frac{\theta_k}{2}\pm i(-1)^{n_k +N/2}\cos\frac{\theta_k}{2}\right]\nonumber\\
    &= \exp\left[\pm i \sum_{k>0}(-1)^{n_k+N/2}\left(\frac{\pi}{2}-\frac{\theta_k}{2}\right)\right] \nonumber\\
    &\equiv \exp\left[\pm i \left(\frac{\pi}{4}-(-1)^{N/2}\Theta (h)\right)\right]\,,
\end{align}
where we defined the angle variable 
\begin{align}
    \Theta (h)= \frac{\pi}{4}+\frac{1}{2}\sum_{k>0}(-1)^{n_k}\theta_k(h)\,.
    \label{eq:cross-latt-angle-variable-SM}
\end{align}
Therefore, the overlap of the lattice crosscap state $|\mathcal{C}_{\mathrm{latt}}^+\rangle$ [Eq.~\eqref{eq:cross-latt-fermi-SM}] with the ground state $|\psi_0(h)\rangle$ is
\begin{align}
    \langle \psi_0 (h)|\mathcal{C}_{\mathrm{latt}}^+\rangle &= \frac{1-i}{2}\langle \psi_0 (h)|B_{\mathrm{latt}}^+\rangle + \frac{1+i}{2}\langle \psi_0 (h)|B_{\mathrm{latt}}^-\rangle \nonumber\\
    &= \sin\left[\frac{\pi}{4}-(-1)^{N/2}\Theta (h)\right] + \cos\left[\frac{\pi}{4}-(-1)^{N/2}\Theta (h)\right]\nonumber\\
    &=\sqrt{2}\cos\Theta (h)\,.
\end{align}

\subsection{C. Lattice crosscap overlaps for the excited states}
\label{sec:lattice-ovlp-excited-SM}

It is not difficult to see that only ``paired'' eigenstates of the TFIC
\begin{align}
    |\psi_{k_1 \cdots k_M} (h)\rangle = \prod_{\alpha=1}^{M} \left[id_{-k_\alpha}^\dagger (h) d_{k_\alpha}^\dagger (h)\right] |\psi_0(h)\rangle~\label{eq:eigenstate-TFIC-SM}
\end{align}
with $0<k_1<\cdots< k_M<\pi$ have non-vanishing overlaps with the fermionic Gaussian states $|B_{\mathrm{latt}}^\pm\rangle$ [Eq.~\eqref{eq:fermi-Gauss-latt-SM}]. Thus, these are also the eigenstates which may have nontrivial overlaps with the lattice crosscap state $|\mathcal{C}_{\mathrm{latt}}^+\rangle$.

The overlap calculation is similar to that of the ground state. We have
\begin{align}
    \langle \psi_{k_1 \cdots k_M} (h)|B_{\mathrm{latt}}^\pm \rangle &=(-i)^M \prod_{\alpha =1}^M\left[\cos\frac{\theta_{k_\alpha}}{2}\mp i(-1)^{n_{k_\alpha}+N/2}\sin\frac{\theta_{k_\alpha}}{2}\right]\prod_{k>0,k\neq \{k_\alpha\}}\left[\sin\frac{\theta_k}{2}\pm i(-1)^{n_k+N/2}\cos\frac{\theta_k}{2}\right]\nonumber\\
    &= (-i)^M \prod_{\alpha =1}^M\frac{\cos\frac{\theta_{k_\alpha}}{2}\mp i(-1)^{n_{k_\alpha}+N/2}\sin\frac{\theta_{k_\alpha}}{2}}{\sin\frac{\theta_{k_\alpha}}{2}\pm i(-1)^{n_{k_\alpha}+N/2}\cos\frac{\theta_{k_\alpha}}{2}} \cdot \prod_{k>0}\left[\sin\frac{\theta_k}{2}\pm i(-1)^{n_k+N/2}\cos\frac{\theta_k}{2}\right]\nonumber\\
    &=(\mp)^M (-1)^{\frac{MN}{2} +\sum_{\alpha=1}^M n_{k_\alpha}} \cdot\langle\psi_0 (h)|B_{\mathrm{latt}}^\pm\rangle\nonumber\\
    &= \begin{cases}
     (-1)^{\sum_{\alpha=1}^M n_{k_\alpha}} \cdot\langle\psi_0 (h)|B_{\mathrm{latt}}^\pm\rangle &\quad M\; \mathrm{even} \\
     \mp (-1)^{N/2} (-1)^{\sum_{\alpha=1}^M n_{k_\alpha}} \cdot\langle\psi_0 (h)|B_{\mathrm{latt}}^\pm\rangle &\quad M\; \mathrm{odd} 
    \end{cases} .
\end{align}
Using Eq.~\eqref{eq:cross-latt-fermi-SM}, we obtain
\begin{align}
    \langle \psi_{k_1 \cdots k_M} (h)|\mathcal{C}_{\mathrm{latt}}^+ \rangle = (-1)^{\sum_{\alpha=1}^M n_{k_\alpha}} \langle \psi_0 (h)|\mathcal{C}_{\mathrm{latt}}^+ \rangle = (-1)^{\sum_{\alpha=1}^M n_{k_\alpha}} \sqrt{2}\cos\Theta (h)
\end{align}
for $M$ even, and
\begin{align}
    \langle \psi_{k_1 \cdots k_M} (h)|\mathcal{C}_{\mathrm{latt}}^+ \rangle &= (-1)^{N/2} (-1)^{\sum_{\alpha=1}^M n_{k_\alpha}} \left[-\frac{1-i}{2}e^{i\left(\frac{\pi}{4}-(-1)^{N/2}\Theta \right)} +\frac{1+i}{2}e^{-i\left(\frac{\pi}{4}-(-1)^{N/2}\Theta \right)} \right]\nonumber\\
    &=  i (-1)^{\sum_{\alpha=1}^M n_{k_\alpha}} \, (-1)^{N/2} \sqrt{2}\sin\left[(-1)^{N/2}\Theta\right]\nonumber\\
    &= i (-1)^{\sum_{\alpha=1}^M n_{k_\alpha}} \sqrt{2}\sin \Theta (h)
\end{align}
for $M$ odd.

In summary, the lattice crosscap overlaps for $|\mathcal{C}_{\mathrm{latt}}^+\rangle$ are given by
\begin{align}
    \langle \psi_{k_1 \cdots k_M} (h)|\mathcal{C}_{\mathrm{latt}}^+ \rangle = \begin{cases}
     (-1)^{\sum_{\alpha=1}^M n_{k_\alpha}} \sqrt{2}\cos\Theta (h) &\quad M\; \mathrm{even} \\
     i (-1)^{\sum_{\alpha=1}^M n_{k_\alpha}} \sqrt{2}\sin\Theta (h) &\quad M\; \mathrm{odd} 
    \end{cases} .
\label{eq:cross-latt-ovlp-standard-SM}
\end{align}

\subsection{D. Overlaps for the dual and modified lattice crosscap states}

The crosscap overlaps of the dual lattice crosscap state $|\mathcal{C}_{\mathrm{latt}}^-\rangle$ [Eq.~\eqref{eq:cross-latt-dual-SM}] can be obtained via the Kramers-Wannier duality. Using Eqs.~\eqref{eq:KW-trans-GS-SM} and \eqref{eq:KW-trans-dk-SM}, we find that the eigenstate $|\psi_{k_1 \cdots k_M} (h)\rangle$ transforms under the Kramers-Wannier duality transformation as
\begin{align}
    U_{\mathrm{KW}}|\psi_{k_1 \cdots k_M} (h)\rangle &= \prod_{\alpha=1}^{M} \left[i U_{\mathrm{KW}}d_{-k_\alpha}^\dagger (h) U_{\mathrm{KW}}^\dag U_{\mathrm{KW}}d_{k_\alpha}^\dagger (h) U_{\mathrm{KW}}^\dag\right] U_{\mathrm{KW}} |\psi_0(h)\rangle \nonumber\\
    &= \prod_{\alpha=1}^{M} \left[-id_{-k_\alpha}^\dagger (1/h) d_{k_\alpha}^\dagger (1/h)\right] |\psi_0(1/h)\rangle \nonumber\\
    &= (-1)^M |\psi_{k_1 \cdots k_M} (1/h)\rangle \, ,
\end{align}
which gives
\begin{align}
    \langle \psi_{k_1 \cdots k_M} (h)|\mathcal{C}_{\mathrm{latt}}^- \rangle = \langle \psi_{k_1 \cdots k_M} (h)|U_{\mathrm{KW}}|\mathcal{C}_{\mathrm{latt}}^+ \rangle = (-1)^M \langle \psi_{k_1 \cdots k_M} (1/h)|\mathcal{C}_{\mathrm{latt}}^+ \rangle\,.
\end{align}

For the modified lattice crosscap state $|\mathcal{C}'_{\mathrm{latt}}\rangle$, we have the analogous fermionic representation:
\begin{align}
    (-1)^{N/2}|\mathcal{C}'_{\mathrm{latt}}\rangle = \frac{1 + i}{2} |B_{\mathrm{latt}}^+\rangle + \frac{1 - i}{2} |B_{\mathrm{latt}}^-\rangle\, ,
\end{align}
where $|B_{\mathrm{latt}}^\pm\rangle$ are the fermionic Gaussian states defined in Eq.~\eqref{eq:Gaussian-Bpm-SM}. This enables analytic computation of overlaps of $|\mathcal{C}'_{\mathrm{latt}}\rangle$ with TFIC eigenstates:
\begin{align}
    \langle \psi_{k_1 \cdots k_M} (h)|\mathcal{C}_{\mathrm{latt}}' \rangle = \begin{cases}
     (-1)^{\sum_{\alpha=1}^M n_{k_\alpha}} \sqrt{2}\sin\Theta (h) &\quad M\; \mathrm{even} \\
     -i (-1)^{\sum_{\alpha=1}^M n_{k_\alpha}} \sqrt{2}\cos\Theta (h) &\quad M\;\mathrm{odd} 
    \end{cases} .
\label{eq:cross-latt-ovlp-modified-SM}
\end{align}

\subsection{E. Expansion of crosscap states in the critical Ising chain eigenbasis}

Specifically, at the critical point of TFIC [Eq.~\eqref{eq:ham-TFIC-SM} with $h=1$], we have $\theta_k = \frac{k}{2}$ [Eq.~\eqref{eq:TFIC-Bog-phase-SM}] and the angle variable $\Theta (h=1)$ [Eq.~\eqref{eq:cross-latt-angle-variable-SM}] is
\begin{align}
    \Theta (h=1) = \frac{\pi}{4}+\frac{1}{4}\sum_{k>0}(-1)^{n_k}k = \frac{\pi}{8} \, ,
\end{align}
from which we obtain the non-vanishing overlaps for the lattice crosscap states
\begin{align}
    \langle \psi_{k_1 \cdots k_M} (h=1)|\mathcal{C}_{\mathrm{latt}}^+ \rangle = (-1)^M\langle \psi_{k_1 \cdots k_M} (h=1)|\mathcal{C}_{\mathrm{latt}}^- \rangle 
    = \begin{cases}
     (-1)^{\sum_{\alpha=1}^M n_{k_\alpha}} \sqrt{\frac{2+\sqrt{2}}{2}} &\quad M\; \mathrm{even} \\
     i (-1)^{\sum_{\alpha=1}^M n_{k_\alpha}} \sqrt{\frac{2-\sqrt{2}}{2}} &\quad M\; \mathrm{odd} 
\label{eq:cross-latt-ovlp-critical-SM}
    \end{cases} ,
\end{align}
and for the modified crosscap state
\begin{align}
    \langle \psi_{k_1 \cdots k_M} (h=1)|\mathcal{C}_{\mathrm{latt}}' \rangle 
    = \begin{cases}
     (-1)^{\sum_{\alpha=1}^M n_{k_\alpha}} \sqrt{\frac{2-\sqrt{2}}{2}} &\quad M\; \mathrm{even} \\
     -i (-1)^{\sum_{\alpha=1}^M n_{k_\alpha}} \sqrt{\frac{2+\sqrt{2}}{2}} &\quad M\; \mathrm{odd} 
    \end{cases} .
\label{eq:cross-latt-ovlp-modified-critical-SM}
\end{align}
Note that Eq.~\eqref{eq:cross-latt-ovlp-critical-SM} corresponds to Eq.~(10) in the main text.

According to Eqs.~\eqref{eq:cross-latt-ovlp-critical-SM} and ~\eqref{eq:cross-latt-ovlp-modified-critical-SM}, the lattice crosscap states $|\mathcal{C}_{\mathrm{latt}}^\pm \rangle$ are expanded in the eigenbasis of the \emph{critical} Ising chain as
\begin{align}
 |\mathcal{C}^\pm_{\mathrm{latt}}\rangle &= \frac{e^{ i \pi/8 }}{\sqrt{2}}\exp\left[\pm i\sum_{k>0}(-1)^{n_k}d_{-k}^\dagger (h=1)d_k^\dagger (h=1)\right] |\psi_0(h=1)\rangle \nonumber\\
    &\phantom{=} \; + \frac{e^{-i \pi/8 }}{\sqrt{2}}\exp\left[\mp i\sum_{k>0}(-1)^{n_k}d_{-k}^\dagger (h=1) d_k^\dagger (h=1)\right] |\psi_0(h=1)\rangle\, ,
\label{eq:cross-latt-expan-SM}
\end{align}
and the modified crosscap state $|\mathcal{C}'_{\mathrm{latt}} \rangle$ is expanded as
\begin{align}
     |\mathcal{C}'_{\mathrm{latt}}\rangle 
     &= \frac{-ie^{ i \pi/8 }}{\sqrt{2}}\exp\left[ i\sum_{k>0}(-1)^{n_k}d_{-k}^\dagger (h=1)d_k^\dagger (h=1)\right] |\psi_0(h=1)\rangle \nonumber\\
    & \phantom{=} \; + \frac{ie^{-i \pi/8 }}{\sqrt{2}}\exp\left[- i\sum_{k>0}(-1)^{n_k}d_{-k}^\dagger (h=1) d_k^\dagger (h=1)\right] |\psi_0(h=1)\rangle\, .
\label{eq:cross-latt-expan-Z2-SM}
\end{align}

\section{IV. Ising conformal crosscap states}

In the continuum limit, the critical Ising chain is described by the Ising CFT. In this section, we first review the Pradisi-Sagnotti-Stanev (PSS) conformal crosscap states in the Ising CFT and then derive the associated dual state via the Kramers-Wannier topological defect line (TDL). Using the Majorana free field representation, we identify the continuum counterparts of the lattice crosscap states $|\mathcal{C}_{\mathrm{latt}}^\pm\rangle$, denoted as $|\mathcal{C}_\pm\rangle$, and determine their relation to the PSS crosscap states.

\subsection{A. Pradisi-Sagnotti-Stanev crosscap states}

The Ising CFT is the simplest example of the rational CFT with diagonal modular invariant. In such theories, conformal crosscap states can be systematically constructed from the bulk conformal data. 

Let the modular $\mathcal{S}$-matrix be $S_{ab}$ and the $\mathcal{T}$-matrix be $T_{ab} = \delta_{ab}\, e^{2\pi i (h_a - \tfrac{c}{24})}$, where $h_a$ is the conformal weight of the chiral primary field, and $c$ is the central charge. The so-called PSS crosscap states can then be formally defined, each labeled by a primary field $a$~\cite{Fioravanti1994,Pradisi1995,Huiszoon1999},
\begin{align}
    |\mathcal{C}_a\rangle 
    = \sum_b \frac{P_{ab}}{\sqrt{S_{0b}}}\, | b \rangle\rangle_{\mathcal{C}}\, ,
\label{eq:gPSS-RCFT-SM}
\end{align}
where the $P_{ab}$ is a symmetric and unitary matrix, defined as
\begin{align}
    P = \sqrt{T}\, S\, T^2\, S\, \sqrt{T}\, ,
    \label{eq:P-matrix-SM}
\end{align}
and $|a\rangle\rangle_{\mathcal{C}} \equiv e^{i\pi (L_0 -h_a)} |a\rangle\rangle$ are crosscap Ishibashi states, with $L_0$ the zeroth Virasoro generator and $|a\rangle\rangle$ the usual boundary CFT Ishibashi states. If the primary field $a$ is a simple current of the theory, the corresponding PSS crosscap state $|\mathcal{C}_a\rangle$ is physical~\cite{Huiszoon1999}. In particular, the identity field is a trivial simple current, and the associated state $|\mathcal{C}_0\rangle$ is known as the ``standard'' PSS crosscap state.

Verlinde TDLs are defined as~\cite{Petkova2001}
\begin{align}
    \mathcal{D}_a = \sum_b \frac{S_{ab}}{S_{0b}} \mathcal{P}_b\, ,
\end{align}
where $\mathcal{P}_{b}$ projects onto the sector labeled by the primary field $b$. Acting with $\mathcal{D}_a$ on a PSS crosscap state yields
\begin{align}
    \mathcal{D}_a |\mathcal{C}_b\rangle 
    = \sum_c Y_{ab}^c \,|\mathcal{C}_c\rangle\, ,
    \label{eq:gPSS-fusion-SM}
\end{align}
where 
\begin{align}
    Y_{ab}^c \equiv (Y_a)_{bc} =
    \sum_{d} \frac{S_{ad}\,P_{bd}\,P^*_{cd}}{S_{0d}}
\end{align}
is an integer-valued tensor~\cite{Pradisi1995,Bantay1997,Huiszoon1999,Gannon2000}.

The Ising CFT, with central charge $c=\frac{1}{2}$, has three (chiral) primary fields $\mathbbm{1}\,,\varepsilon\,,\sigma$, with conformal weights $0\,,\frac{1}{2}\,,\frac{1}{16}$, respectively. Its modular $\mathcal{S}$-matrix is
\begin{align}
    S = \frac{1}{2}
    \begin{pmatrix}
        1 & 1 & \sqrt{2} \\
        1 & 1 & -\sqrt{2} \\
        \sqrt{2} & -\sqrt{2} & 0
    \end{pmatrix}\, ,
\end{align}
from which the $P$-matrix and the $Y$-tensor can be computed explicitly:
\begin{align}
    P = \begin{pmatrix}
        \cos(\pi/8) & \sin(\pi/8) & 0 \\
        \sin (\pi/8) & -\cos (\pi/8) & 0 \\
        0 & 0 & 1
    \end{pmatrix}\, ,\quad
    Y_{\mathbbm{1}} = \begin{pmatrix}
        1 & 0 & 0 \\ 0 & 1 & 0 \\ 0 & 0 & 1
    \end{pmatrix}\, ,\quad
    Y_\varepsilon = \begin{pmatrix}
        1 & 0 & 0 \\ 0 & 1 & 0 \\ 0 & 0 & -1
    \end{pmatrix}\, ,\quad
    Y_\sigma = \begin{pmatrix}
        1 & 1 & 0 \\ 1 & -1 & 0 \\ 0 & 0 & 0
    \end{pmatrix}\, .
\end{align}

From Eq.~\eqref{eq:gPSS-RCFT-SM}, two PSS crosscap states, corresponding to the simple currents $\mathbbm{1}$ and $\varepsilon$, are given by~\cite{Fioravanti1994,Pradisi1995} 
\begin{align}
    |\mathcal{C}_{\mathbbm{1}}\rangle
    &= \sqrt{\frac{2+\sqrt{2}}{2}} |\mathbbm{1}\rangle\rangle_{\mathcal{C}} + \sqrt{\frac{2-\sqrt{2}}{2}} |\varepsilon\rangle\rangle_{\mathcal{C}}\, ,\nonumber\\
    |\mathcal{C}_\varepsilon\rangle
    &= \sqrt{\frac{2-\sqrt{2}}{2}} |\mathbbm{1}\rangle\rangle_{\mathcal{C}} - \sqrt{\frac{2+\sqrt{2}}{2}} |\varepsilon\rangle\rangle_{\mathcal{C}}\, .
    \label{eq:gPSS-Ising-SM}
\end{align}
Although these two PSS crosscap states are both physical, their crosscap partition functions $\langle \mathcal{C}_a|e^{-\beta H_0}|\mathcal{C}_a\rangle$ ($a=\mathbbm{1},\varepsilon$) differ. Under a modular-$\mathcal{S}$ transformation, the one for $|\mathcal{C}_{\mathbbm{1}}\rangle$ gives the standard Klein bottle partition function, while that for $|\mathcal{C}_\varepsilon\rangle$ yields the non-standard Klein bottle partition function~\cite{Huiszoon1999}.

The Verlinde line $\mathcal{D}_\sigma = \sqrt{2}(\mathcal{P}_{\mathbbm{1}} -\mathcal{P}_\varepsilon)$ realizes Kramers-Wannier duality in the Ising CFT. Acting with $\mathcal{D}_\sigma$ on the (standard) PSS crosscap state $|\mathcal{C}_{\mathbbm{1}}\rangle$ yields a dual conformal crosscap state, which by Eq.~\eqref{eq:gPSS-fusion-SM} is a superposition of $|\mathcal{C}_{\mathbbm{1}}\rangle$ and $|\mathcal{C}_{\varepsilon}\rangle$~\cite{Harada2025}:
\begin{align}
    \frac{\mathcal{D}_\sigma}{\langle \mathcal{D}_\sigma\rangle} |\mathcal{C}_{\mathbbm{1}}\rangle 
    = \sqrt{\frac{2+\sqrt{2}}{2}} |\mathbbm{1}\rangle\rangle_{\mathcal{C}} - \sqrt{\frac{2-\sqrt{2}}{2}} |\varepsilon\rangle\rangle_{\mathcal{C}}
    = \frac{1}{\sqrt{2}} \left(|\mathcal{C}_{\mathbbm{1}}\rangle + |\mathcal{C}_{\varepsilon}\rangle \right)\, ,
\label{eq:dual-cross-fusion-CFT-SM}
\end{align}
where $\langle \mathcal{D}_\sigma\rangle = \sqrt{2}$ denotes the vacuum expectation value of the TDL. The crosscap states $|\mathcal{C}_{\mathbbm{1}}\rangle$ and $\frac{\mathcal{D}_\sigma}{\langle \mathcal{D}_\sigma\rangle} |\mathcal{C}_{\mathbbm{1}}\rangle$ share the same crosscap partition function and are mapped into each other under the Kramers-Wannier duality transformation.

\subsection{B. Majorana free field representation}

The operator formalism of the Ising CFT can be formulated in terms of the free Majorana Hamiltonian:
\begin{align}
    H^{\mathrm{NS}}_0 = \frac{i}{4\pi}\int_0^L \mathrm{d}x \left[\chi\partial_x\chi -\bar{\chi}\partial_x\bar{\chi}\right] \, ,
\label{eq:ham-Ising-CFT-SM}
\end{align}
where we restrict ourselves in the NS sector for our purpose. Here $\chi$ and $\bar{\chi}$ are Majorana fields, with the mode expansion: 
\begin{align}
    \chi (z) = \sqrt{\frac{2\pi}{L}}\sum_{n\in\mathbb{Z}} b_{n-1/2} e^{-\frac{2\pi}{L}(n-\frac{1}{2})z},\quad
    \bar{\chi} (\bar{z}) = \sqrt{\frac{2\pi}{L}}\sum_{n\in\mathbb{Z}} \bar{b}_{n-1/2} e^{-\frac{2\pi}{L}(n-\frac{1}{2})\bar{z}} \, ,
\label{eq:mode-expan-CFT-SM}
\end{align}
where $z=\tau +ix$ is the complex coordinate. The modes in momentum space satisfy $b^\dag_{n-1/2}= b_{-n+1/2}$ and $\bar{b}^\dag_{n-1/2}= \bar{b}_{-n+1/2}$ as well as the following anticommutation relations: $\{b_{n-1/2},b^\dag_{m-1/2}\}=\{\bar{b}_{n-1/2},\bar{b}^\dag_{m-1/2}\}=\delta_{n,m}$ and $\{b_{n-1/2},\bar{b}_{m-1/2}\}=0$.

The Virasoro generators can be expressed in terms of these modes as
\begin{align}
    L_n = \frac{1}{2}\sum_{m\in \mathbb{Z}} (m-1/2):b_{n-m+1/2}b_{m-1/2}:\,,\quad \bar{L}_n = \frac{1}{2}\sum_{m\in \mathbb{Z}} (m-1/2):\bar{b}_{n-m+1/2}\bar{b}_{m-1/2}:\, ,
\label{eq:Vir-Ln-SM}
\end{align}
where $:\cdots:$ denotes normal ordering. The primary states $|\mathbbm{1}\rangle$ and $|\varepsilon\rangle$ are
\begin{align}
    |\mathbbm{1}\rangle = |0\rangle_{\mathrm{NS}}\, ,\quad 
    |\varepsilon\rangle = -i b_{-1/2}\bar{b}_{-1/2}|0\rangle_{\mathrm{NS}}\, ,
\label{eq:primary-fermion-SM}
\end{align}
corresponding to the vacuum and the first excited state in the NS sector, respectively.

The crosscap Ishibashi states associated with the primary fields $\mathbbm{1}$ and $\varepsilon$ are given by~\cite{Ishibashi1989}
\begin{align}
    |\alpha\rangle\rangle_\mathcal{C} = \prod_{n=1}^\infty \left[\sum_{p_n=0}^\infty (-1)^{np_n}L_{-n}^{p_n}\bar{L}_{-n}^{p_n}\right]|\alpha\rangle \, , \quad \alpha = \mathbbm{1},\varepsilon\, .
\label{eq:cross-Ishibashi-SM}
\end{align}
Utilizing the fermionic representation of the Virasoro generators [Eq.~\eqref{eq:Vir-Ln-SM}] and the associated primary states [Eq.~\eqref{eq:primary-fermion-SM}], these Ishibashi states can be expanded in the fermionic basis as
\begin{align}
    |\mathbbm{1}\rangle\rangle_\mathcal{C} &= \sum_{M\; \mathrm{even}} \sum_{0<n_1<\cdots <n_M } (-1)^{\sum_{\alpha=1}^M (n_\alpha -1/2)} \prod_{\alpha =1}^M b_{-n_\alpha +1/2} \prod_{\alpha =1}^M \bar{b}_{-n_\alpha +1/2} |0\rangle_{\mathrm{NS}} \, , \nonumber \\
    |\varepsilon\rangle\rangle_\mathcal{C} &= -i \sum_{M\; \mathrm{odd}} \sum_{0<n_1<\cdots <n_M } (-1)^{-\frac{1}{2} +\sum_{\alpha=1}^M (n_\alpha -1/2)} \prod_{\alpha =1}^M b_{-n_\alpha +1/2} \prod_{\alpha =1}^M \bar{b}_{-n_\alpha +1/2} |0\rangle_{\mathrm{NS}} \, .
    \label{eq:Ishibashi-fermion-SM}
\end{align}

\subsection{C. Continuum counterparts of the lattice crosscap states}

The Ising CFT is the effective theory of the critical Ising chain [Eq.~\eqref{eq:ham-TFIC-SM} with $h=1$], where the lattice Majorana fermion operators $\chi_j$ and $\bar{\chi}_j$ in Eq.~\eqref{eq:MajField-latt-SM} are identified with the Majorana fields $\chi (x)$ and $\bar{\chi} (x)$ in Eq.~\eqref{eq:mode-expan-CFT-SM}, and the Bogoliubov modes near $\pm \pi$ are identified with the momentum modes of the Majorana fields:
\begin{align}
    b_{n-1/2} \leftrightarrow \begin{cases}
    d_{k=\pi - \frac{2\pi}{N} (n-1/2)}(h=1) &\quad n>0\\
    d_{k=\pi + \frac{2\pi}{N} (n-1/2)}^\dagger (h=1) &\quad n\leq 0\\
    \end{cases}\,,\quad
    \bar{b}_{n-1/2} \leftrightarrow \begin{cases}
    id_{k=-\pi + \frac{2\pi}{N} (n-1/2)} (h=1) &\quad n>0\\
    -id_{k=-\pi - \frac{2\pi}{N}(n-1/2)}^\dagger (h=1) &\quad n\leq 0\\
    \end{cases}\,,
\end{align}
which is valid for the low-energy mode with $|n|/N \ll 1$ in the continuum limit. Therefore, the low-energy ``paired'' eigenstate [Eq.~\eqref{eq:eigenstate-TFIC-SM}] of the critical Ising chain is identified as
\begin{align}
    |\psi_{k_1\cdots k_M}(h=1)\rangle \leftrightarrow \prod_{\alpha=1}^M b_{-n_{k_\alpha}+1/2}\bar{b}_{-n_{k_\alpha}+1/2}|0\rangle_{\mathrm{NS}} \, ,
\end{align}
and the continuum counterparts of the lattice crosscap states $|\mathcal{C}_{\mathrm{latt}}^\pm\rangle$ [Eq.~\eqref{eq:cross-latt-expan-SM}] are identified as 
\begin{align}
    |\mathcal{C}_\pm\rangle = \frac{e^{i\pi/8}}{\sqrt{2}} |B_\pm\rangle + \frac{e^{- i\pi/8}}{\sqrt{2}} |B_\mp\rangle 
\label{eq:cross-field-SM}
\end{align}
with 
\begin{align}
    |B_\pm \rangle = \exp\left[\pm\sum_{n=1}^\infty (-1)^{n}b_{-n+\frac{1}{2}}\bar{b}_{-n+\frac{1}{2}}\right]|0\rangle_{\mathrm{NS}} \, .
\label{eq:Bpm-Ising-CFT-SM}
\end{align}
This corresponds to Eq.~(12) in the main text.

To identify $|\mathcal{C}_\pm\rangle$ as the conformal crosscap states, we notice that the crosscap Ishibashi states $|\mathbbm{1}\rangle\rangle_\mathcal{C}$ and $|\varepsilon\rangle\rangle_\mathcal{C}$ [Eq.~\eqref{eq:Ishibashi-fermion-SM}] can be expressed in terms of $|B_\pm\rangle$ [Eq.~\eqref{eq:Bpm-Ising-CFT-SM}]:
\begin{align}
    |\mathbbm{1}\rangle\rangle_\mathcal{C} 
    &= \sum_{M\; \mathrm{even}} \sum_{0<n_1\cdots <n_M } (-1)^{\sum_{\alpha=1}^M (n_\alpha -1/2)}\, (-1)^{M/2} \prod_{\alpha =1}^M b_{-n_\alpha +1/2} \bar{b}_{-n_\alpha +1/2} |0\rangle_{\mathrm{NS}}\nonumber\\
    &=  \sum_{M\; \mathrm{even}} \sum_{0<n_1\cdots <n_M } (-1)^{\sum_{\alpha=1}^M n_\alpha }\prod_{\alpha =1}^M b_{-n_\alpha +1/2} \bar{b}_{-n_\alpha +1/2} |0\rangle_{\mathrm{NS}}\nonumber\\
    &= \frac{1}{2}\left(|B_+\rangle + |B_-\rangle\right)
\label{eq:Ishibashi-identity-fermion-rep-SM}
\end{align}
and
\begin{align}
    |\varepsilon\rangle\rangle_\mathcal{C} &= -i \sum_{M\; \mathrm{odd}} \sum_{0<n_1<\cdots <n_M } (-1)^{-1/2+\sum_{\alpha=1}^M (n_\alpha -1/2)} \, (-1)^{\frac{M-1}{2}} \prod_{\alpha =1}^M b_{-n_\alpha +1/2} \bar{b}_{-n_\alpha +1/2} |0\rangle_{\mathrm{NS}} \nonumber \\
    &= i \sum_{M\; \mathrm{odd}} \sum_{0<n_1<\cdots <n_M } (-1)^{\sum_{\alpha=1}^M n_\alpha }\prod_{\alpha =1}^M b_{-n_\alpha +1/2} \bar{b}_{-n_\alpha +1/2} |0\rangle_{\mathrm{NS}}\nonumber\\
    &= \frac{i}{2}\left(|B_+\rangle - |B_-\rangle\right)\, ,
\label{eq:Ishibashi-energy-fermion-rep-SM}
\end{align}
where we used
\begin{align}
    \prod_{\alpha =1}^M b_{-n_\alpha +1/2} \prod_{\alpha =1}^M \bar{b}_{-n_\alpha +1/2} = \begin{cases}
     (-1)^{\frac{M}{2}} \prod_{\alpha =1}^M b_{-n_\alpha +1/2} \bar{b}_{-n_\alpha +1/2}  &\quad M\; \mathrm{even} \\
     (-1)^{\frac{M-1}{2}} \prod_{\alpha =1}^M b_{-n_\alpha +1/2} \bar{b}_{-n_\alpha +1/2}  &\quad M\; \mathrm{odd} 
    \end{cases} .
\end{align}
Therefore, $|\mathcal{C}_\pm\rangle$ are the following linear combinations of the crosscap Ishibashi states:
\begin{align}
    |\mathcal{C}_\pm\rangle = \sqrt{\frac{2+\sqrt{2}}{2}}|\mathbbm{1}\rangle\rangle_\mathcal{C} \pm \sqrt{\frac{2-\sqrt{2}}{2}}|\varepsilon\rangle\rangle_\mathcal{C} \, .
\label{eq:cross-Ishibashi-expan-SM}
\end{align}
This corresponds to Eq.~(13) in the main text.
Comparing with Eqs.~\eqref{eq:gPSS-Ising-SM} and \eqref{eq:dual-cross-fusion-CFT-SM}, we find that $|\mathcal{C}_\pm\rangle$ are the (standard) PSS conformal crosscap state $|\mathcal{C}_{\mathbbm{1}}\rangle$ and its dual $\frac{\mathcal{D}_\sigma}{\langle\mathcal{D}_\sigma\rangle}|\mathcal{C}_{\mathbbm{1}}\rangle$, respectively.

Moreover, from Eq.~\eqref{eq:cross-latt-expan-Z2-SM}, we can derive the continuum counterpart of the modified lattice crosscap states $|\mathcal{C}'_{\mathrm{latt}}\rangle$:
\begin{align}
    |\mathcal{C}'\rangle &= \frac{-ie^{i\pi/8}}{\sqrt{2}} |B_+\rangle + \frac{ie^{- i\pi/8}}{\sqrt{2}} |B_-\rangle  \nonumber \\
    &= \sqrt{\frac{2-\sqrt{2}}{2}} |\mathbbm{1}\rangle\rangle_{\mathcal{C}} - \sqrt{\frac{2+\sqrt{2}}{2}} |\varepsilon\rangle\rangle_{\mathcal{C}} \, ,
\end{align}
where we used Eqs.~\eqref{eq:Ishibashi-identity-fermion-rep-SM} and \eqref{eq:Ishibashi-energy-fermion-rep-SM}. This proves $|\mathcal{C}'\rangle = |\mathcal{C}_{\varepsilon}\rangle$, where $|\mathcal{C}_{\varepsilon}\rangle$ is one of the PSS crosscap states in Eq.~\eqref{eq:gPSS-Ising-SM}.

Having identified the continuum counterparts of $|\mathcal{C}_{\mathrm{latt}}^\pm\rangle$ and $|\mathcal{C}'_{\mathrm{latt}}\rangle$ in the Ising CFT, we find that Eq.~\eqref{eq:dual-cross-fusion-SM} is the perfect lattice correspondence of Eq.~\eqref{eq:dual-cross-fusion-CFT-SM}, which can be interpreted as the interplay of the Kramers-Wannier TDL and the PSS crosscap states.

\subsection{D. Exact crosscap overlaps in the presence of thermal perturbation}

For the Ising CFT [Eq.~\eqref{eq:ham-Ising-CFT-SM}] with thermal perturbation ($\varepsilon = -i\chi\bar{\chi}$), the perturbed Hamiltonian (in the NS sector) is still quadratic in the fermionic representation:
\begin{align}
    H^{\mathrm{NS}} = H^{\mathrm{NS}}_0 -g_1\int_0^L \mathrm{d}x \,\varepsilon (x)=\frac{i}{4\pi}\int_0^L \mathrm{d}x \left[\chi (x)\partial_x\chi(x)-\bar{\chi}(x)\partial_x\bar{\chi}(x)\right] +\frac{im}{2\pi}\int_0^L \mathrm{d}x \, \chi(x)\bar{\chi}(x) \,,
\end{align}
where $g_1=m/2\pi$ and $z=\tau + ix$. The mode expansions of the Majorana fields $\chi$ and $\bar{\chi}$ are given in Eq.~\eqref{eq:mode-expan-CFT-SM}.

The perturbed Hamiltonian can be diagonalized with a Bogoliubov transformation~\cite{ZhangYS2023}
\begin{align}
    H^{\mathrm{NS}} 
    &= \sum_{n=1}^\infty \frac{2\pi}{L}(n-\frac{1}{2})(b_{-n+1/2}b_{n-1/2}-\bar{b}_{n-1/2}\bar{b}_{-n+1/2}) + im (b_{-n+1/2}\bar{b}_{-n+1/2}-\bar{b}_{n-1/2} b_{n-1/2}) \nonumber\\
    &= \frac{2\pi}{L}\sum_{n\in\mathbb{Z}}\sqrt{\left(n-\frac{1}{2}\right)^2+s_1^2}\left(\eta_{n-1/2}^\dagger\eta_{n-1/2} -\frac{1}{2}\right) \, ,
\label{eq:thermal-perturbed-hamil-SM}
\end{align}
where $s_1 = g_1 L = \frac{mL}{2\pi}$ is the dimensionless coupling. The annihilation operator of the Bogoliubov mode is given by
\begin{align}
    \eta_{n-1/2} (s_1)  = \cos\left(\frac{\pi}{4}-\frac{\theta_{n-1/2}(s_1)}{2}\right) b_{n-1/2} + i\sin\left(\frac{\pi}{4}-\frac{\theta_{n-1/2}(s_1)}{2}\right) \bar{b}_{-n+1/2} 
\end{align}
with
\begin{align}
    \cos\theta_{n-1/2} (s_1) = \frac{s_1}{\sqrt{\left(n-\frac{1}{2}\right)^2+s_1^2}}\,,\quad \sin\theta_{n-1/2}(s_1) = \frac{n-1/2}{\sqrt{\left(n-\frac{1}{2}\right)^2+s_1^2}}\,,\quad \theta_{n-1/2} \in (-\pi,\pi) \, .
\end{align}

The perturbed ground state of Eq.~\eqref{eq:thermal-perturbed-hamil-SM} is given by
\begin{align}
    |\psi_0 (s_1)\rangle 
    =\prod_{n=1}^\infty\left[\cos\left(\frac{\pi}{4}-\frac{\theta_{n-1/2}}{2}\right) + i \sin\left(\frac{\pi}{4}-\frac{\theta_{n-1/2}}{2}\right)\bar{b}_{-n+1/2}b_{-n+1/2}\right]|0\rangle_{\mathrm{NS}} \, ,
\end{align}
where $|0\rangle_{\mathrm{NS}}$ is the Ising CFT ground state.

The crosscap overlap calculation is similar to the lattice case (see Sec.~\ref{sec:lattice-ovlp-GS-SM}):
\begin{align}
    \langle\psi_0 (s_1)|\exp\left[\pm\sum_{n=1}^\infty (-1)^{n}b_{-n+1/2}\bar{b}_{-n+1/2}\right]|0\rangle_{\mathrm{NS}} = \exp\left[\pm i\Theta (s_1)\right] \, ,
\end{align}
and thus
\begin{align}
    \langle \psi_0 (s_1)|\mathcal{C}_\pm\rangle =\frac{e^{i\pi/8}}{\sqrt{2}}e^{\pm i\Theta (s_1)} + \frac{e^{-i\pi/8}}{\sqrt{2}}e^{\mp i\Theta (s_1)}=\sqrt{2}\cos\left[\frac{\pi}{8}\pm \Theta (s_1)\right] \, ,
\label{eq:perturbed-crosscap-overlap-SM}
\end{align}
where
\begin{align}
    \Theta (s_1) &= \sum_{n=1}^\infty (-1)^{n}\left(\frac{\pi}{4}-\frac{\theta_{n-1/2}(s_1)}{2}\right)\nonumber\\
    &= \sum_{n=1}^\infty (-1)^n \left[\frac{\pi}{4} - \arctan \left(\frac{n-\frac{1}{2}}{s_1+\sqrt{(n-\frac{1}{2})^2+s_1^2}}\right)\right] \, .
\end{align}
As a side remark, we note that the above expression of $\Theta (s_1)$ is expanded in orders of the energy levels. This indicates that the truncated conformal space approach may serve as an effective \emph{non-perturbative} method for calculating the crosscap overlap. In fact, by retaining about one hundred terms in the expansion of $\Theta (s_1)$, one can already obtain the universal scaling function for the crosscap overlap with very high precision.

The series expansion shows that $\Theta (s_1)$ is an odd function of $s_1$:
\begin{align}
    \Theta (s_1) = \sum_{k=0}^\infty \left[\sum_{n=1}^\infty \frac{(-1)^n}{\left(n-\frac{1}{2}\right)^{2k+1}}\right] \cdot \frac{s_1^{2k+1}}{4k+2} = -\Theta (-s_1) \, .
\end{align}
Actually, the crosscap overlap in Eq.~\eqref{eq:perturbed-crosscap-overlap-SM} is in perfect agreement with the exact solution of the Klein bottle entropy obtained in Ref.~\cite{ZhangYS2023} if one uses
\begin{align}
    \sqrt{2}\cos\left[\frac{\pi}{8}\pm \Theta (s_1)\right] =\sqrt{1+ \frac{1}{\sqrt{1+e^{\mp 2\pi s_1}}}} \, .
    \label{eq:Theta-s-SM}
\end{align}

When approaching the critical point ($s_1\to 0$), the Bogoliubov modes reduce to Majorana modes in the Ising CFT
\begin{align}
    \eta_{n-1/2} (s_1\to 0) =\begin{cases}
    b_{n-1/2} &\quad n>0\\
    i\bar{b}_{-n+1/2} &\quad n\leq 0\\
    \end{cases} \, .
\end{align}
Viewed as a continuous deformation of the (unperturbed) CFT eigenstates $|\psi_{n_1\cdots n_M}(0)\rangle = \prod_{\alpha =1}^M b_{-n+1/2}\bar{b}_{-n+1/2}|0\rangle_{\mathrm{NS}}$, the perturbed eigenstates
\begin{align}
    |\psi_{n_1\cdots n_M}(s_1)\rangle = \prod_{\alpha =1}^M \left[i\;\eta_{n-1/2}^\dagger (s_1)\eta_{-n+1/2}^\dagger (s_1) \right]|\psi_0(s_1)\rangle
    \label{eq:Ising-deformed-eigenstate-SM}
\end{align}
have the following non-vanishing overlaps with the crosscap states:
\begin{align}
    \langle \psi_{n_1 \cdots n_M} (s_1)|\mathcal{C}_\pm \rangle 
    = \begin{cases}
    (-1)^{\sum_{\alpha=1}^M n_{\alpha}} \sqrt{2}\cos \left[\frac{\pi}{8}\pm\Theta (s_1)\right] = (-1)^{\sum_{\alpha=1}^M n_{\alpha}} \sqrt{1+ \frac{1}{\sqrt{1+e^{\mp 2\pi s_1}}}} &\quad M\; \mathrm{even}\\
    \pm i (-1)^{\sum_{\alpha=1}^M n_{\alpha}} \sqrt{2}\sin \left[\frac{\pi}{8}\pm \Theta (s_1)\right] = \pm i (-1)^{\sum_{\alpha=1}^M n_{\alpha}} \sqrt{1- \frac{1}{\sqrt{1+e^{\mp 2\pi s_1}}}} &\quad M\; \mathrm{odd}\\
    \end{cases} \, ,
\end{align}
where the calculation is similar to the lattice case (see Sec.~III C). This is the continuum counterpart of the lattice crosscap overlaps in Eq.~\eqref{eq:cross-latt-ovlp-standard-SM}. However, we note that in the off-critical region of the lattice model ($h\neq 1$), finite-size corrections are present, unlike the critical point  ($h = 1$) where finite-size corrections are absent.

In summary, we have the expansion of the crosscap state $|\mathcal{C}_\pm\rangle$ in the eigenbasis of the Ising CFT with the thermal perturbation:
\begin{align}
    |\mathcal{C}_\pm\rangle = \frac{e^{i\left[\frac{\pi}{8} \pm \Theta (s_1)\right]} }{\sqrt{2}}\exp\left[\pm i\sum_{n=1}^\infty (-1)^{n}\eta^\dagger_{n-\frac{1}{2}}\eta^\dagger_{-n+\frac{1}{2}}\right]|\psi_0(s_1)\rangle + \frac{e^{-i\left[\frac{\pi}{8} \pm \Theta (s_1)\right]} }{\sqrt{2}}\exp\left[\mp i\sum_{n=1}^\infty (-1)^{n}\eta^\dagger_{n-\frac{1}{2}}\eta^\dagger_{-n+\frac{1}{2}}\right]|\psi_0(s_1)\rangle\, ,
\end{align}
where $\Theta (s_1)$ has been defined in Eq.~\eqref{eq:Theta-s-SM}. More specifically, the squared norms of the crosscap overlaps take the simple form,
\begin{align}
    |\langle\psi_{n_1 \cdots n_M}(s_1)|\mathcal{C}_\pm\rangle|^2 = 1 + \frac{(-1)^{M}}{\sqrt{1+ e^{\mp 2\pi s_1}}} \, ,
    \label{eq:Ising-crossovlp-thermal-SM}
\end{align}
where $|\psi_{n_1 \cdots n_M}(s_1)\rangle$ is the eigenstate deformed from the unperturbed state $|\psi_{n_1 \cdots n_M} (0)\rangle \equiv \prod_{\alpha=1}^M b_{-n_\alpha + 1/2}\bar{b}_{-n_\alpha +1/2}|0\rangle_{\mathrm{NS}}$ in the Ising CFT basis.

\section{V. Crosscap correlators for the Ising CFT}

In this section, we extend the bosonization approach for calculating the crosscap correlators of the Ising CFT, i.e., the conformal correlation functions on the real projective plane ($\mathbb{RP}^2$). The general multi-point crosscap correlators for both $\varepsilon$ and $\sigma$ fields are derived in this framework.

\subsection{A. Bosonization of crosscap states}

To develop the bosonization formalism of the Ising crosscap correlators, we begin with two copies of the Ising CFTs [Eq.~\eqref{eq:ham-Ising-CFT-SM}], which together form the Dirac fermion CFT. Let $\chi_1 (\bar{\chi}_1)$ and $\chi_2 (\bar{\chi}_2)$ denote the (anti-)chiral Majorana fields of the two Ising CFTs, respectively. The corresponding (anti-)chiral Dirac fermion fields are then given by
\begin{align}
    \Psi (z) &=\frac{1}{\sqrt{2}}\left(\chi_1(z) + i\chi_2(z) \right)\,,\quad \Psi^\dagger (z) =\frac{1}{\sqrt{2}}\left(\chi_1(z) - i\chi_2(z) \right)\,,\nonumber\\
    \bar{\Psi} (\bar{z}) &=\frac{1}{\sqrt{2}}\left(\bar{\chi}_1(\bar{z}) + i \bar{\chi}_2(\bar{z}) \right)\,,\quad \bar{\Psi}^\dagger (\bar{z}) =\frac{1}{\sqrt{2}}\left(\bar{\chi}_1(\bar{z}) - i \bar{\chi}_2(\bar{z}) \right)\,,
\end{align}
where $z=\tau + ix$ is the complex coordinate. The Dirac fermion fields satisfy the anti-periodic boundary condition in the NS sector.

The Dirac fermion CFT is equivalent to the compactified boson CFT via the following bosonization identity~\cite{Francesco-Book}:
\begin{align}
    \Psi(z)&=\sqrt{\frac{2\pi w}{L}}:e^{i\phi(w)}:\,,\quad
    \Psi^\dagger (z)=\sqrt{\frac{2\pi w}{L}}:e^{-i\phi(w)}:\,,\nonumber\\
    \bar{\Psi}(\bar{z})&=\sqrt{\frac{2\pi \bar{w}}{L}}:e^{i\bar{\phi}(\bar{w})}:\,,\quad
    \bar{\Psi}^\dagger(\bar{z})=\sqrt{\frac{2\pi \bar{w}}{L}}:e^{-i\bar{\phi}(\bar{w})}:
\label{eq:bosonization-id-SM}
\end{align}
with $w=e^{\frac{2\pi}{L}z}$. The prefactor $\sqrt{\frac{2\pi w}{L}}$ arises from the conformal transformation $z = \frac{L}{2\pi}\ln w$, which maps the plane to the cylinder.

To ensure consistency with the boundary conditions of the Dirac fermions, the compactified boson field $\varphi (x,\tau) = \phi (w) + \bar{\phi} (\bar{w})$ has the radius $R=2$: $\varphi (x+L,\tau) \sim \varphi (x,\tau) + 2\pi m R$, where $m\in\mathbb{Z}$ is referred to as the winding number. For convenience, we divide the (anti-)chiral boson field $\phi$ ($\bar{\phi}$) into the zero-mode part and the oscillatory part $\phi'$ ($\bar{\phi}'$):
\begin{align}
    \phi (w) =  x_0-i a_0\ln w + \phi'(w) \,,\quad
    \bar{\phi} (\bar{w}) &= \bar{x}_0-i\bar{a}_0\ln\bar{w} + \bar{\phi}' (\bar{w})\,,
\end{align}
where the oscillatory part can be further decomposed into the positive and negative mode parts, given by the mode expansion:
\begin{align}
    \phi'(w) &= \phi'_+ (w) + \phi'_- (w) = -i\sum_{k=1}^\infty \frac{1}{k} w^k a_{-k} + i\sum_{k=1}^\infty \frac{1}{k} w^{-k} a_{k}\,,\nonumber\\
    \bar{\phi}'(\bar{w}) & = \bar{\phi}'_+(\bar{w}) + \bar{\phi}'_-(\bar{w}) = -i\sum_{k=1}^\infty \frac{1}{k}\bar{w}^k \bar{a}_{-k} + i\sum_{k=1}^\infty \frac{1}{k}\bar{w}^{-k} \bar{a}_{k} \, .
\label{eq:mode-expan-boson-SM}
\end{align}
The momentum modes satisfy commutation relations $[x_0,a_0]=[\bar{x}_0,\bar{a}_0]=i$ and $[a_k,a_l]=[\bar{a}_k,\bar{a}_l]=k\delta_{k+l} \; \forall k,l\in\mathbb{Z}$. The Virasoro primary states of the compactified boson CFT are labeled by the eigenvalues of $a_0$ and $\bar{a}_0$:
\begin{align}
    a_0 |n,m\rangle = \left(\frac{n}{2}+ m\right)|n,m\rangle \, , \quad \bar{a}_0 |n,m\rangle = \left(\frac{n}{2}- m\right)|n,m\rangle \, ,
\end{align}
which can be generated from the ground state $|0,0\rangle$ via the ladder operator, sometimes referred to as the  \emph{Klein factor} in the literature~\cite{vonDelft1998}:
\begin{align}
    e^{i(x_0+\bar{x}_0)/2}|n,m\rangle = |n+1,m\rangle \,,\quad e^{i(x_0-\bar{x}_0)} |n,m\rangle = |n,m+1\rangle \,.
\end{align}
The normal-ordered vertex operators are defined as follows
\begin{align}
    :e^{i\phi (w)}: &= e^{i x_0} e^{a_0\ln w} :e^{i\phi' (w)}: = e^{i x_0} e^{a_0\ln w} e^{i\phi'_+ (w)} e^{i\phi'_- (w)} \,,\nonumber\\
    :e^{i\bar{\phi} (\bar{w})}: &= e^{i \bar{x}_0} e^{\bar{a}_0\ln\bar{w}} :e^{i\bar{\phi}' (\bar{w})}: = e^{i \bar{x}_0} e^{\bar{a}_0\ln\bar{w}} e^{i\bar{\phi}'_+ (\bar{w})} e^{i\bar{\phi}'_- (\bar{w})} \,.~\label{eq:vertex-normal-order-SM}
\end{align}

The key observation for bosonizing the crosscap states $|\mathcal{C}_\pm\rangle = \frac{e^{i\pi/8}}{\sqrt{2}}|B_+\rangle + \frac{e^{-i\pi/8}}{\sqrt{2}}|B_-\rangle$ is that the fermionic Gaussian states $|B_\pm\rangle$ [Eq.~\eqref{eq:Bpm-Ising-CFT-SM}] satisfy the following constraints:
\begin{align}
    [b_{n-1/2}\pm (-1)^n \bar{b}_{-n+1/2}]|B_\pm\rangle =0\,,\quad \forall n\in\mathbb{Z} \, .
\end{align}
This is equivalent to the condition $[\chi (x) \mp i \bar{\chi}(x+L/2)]|B_\pm\rangle = 0$, where $\chi$ and $\bar{\chi}$ are the Majorana fields [Eq.~\eqref{eq:mode-expan-CFT-SM}] in the Ising CFT. Consequently, for two copies of such fermionic Gaussian states $|B_\pm\rangle$, we obtain the constraints on the Dirac fields:
\begin{align}
    \left[\Psi (x) \mp i\bar{\Psi} (x+L/2)\right]|B_\pm\rangle^{(1)}|B_\pm\rangle^{(2)} &= 0\,,\nonumber\\
    \left[\Psi (x) \mp i\bar{\Psi}^\dagger (x+L/2)\right]|B_\pm\rangle^{(1)}|B_\mp\rangle^{(2)} &= 0\,.
\end{align}
These constraints can be translated into the bosonic language using the bosonization identity [Eq.~\eqref{eq:bosonization-id-SM}]:
\begin{align}
    \left[e^{i\frac{\pi}{L}x} :e^{i\phi(x)}: \mp\; i e^{-i\frac{\pi }{L}(x+L/2)}:e^{i\bar{\phi}(x+L/2)}:\right]|B_\pm\rangle^{(1)}|B_\pm\rangle^{(2)}&=0\,,\nonumber\\
    \left[e^{i\frac{\pi}{L}x} :e^{i\phi(x)}: \mp\; i e^{-i\frac{\pi }{L}(x+L/2)}:e^{-i\bar{\phi}(x+L/2)}:\right]|B_\pm\rangle^{(1)}|B_\mp\rangle^{(2)}&=0\,,~\label{eq:boson-constraint-BB-SM}
\end{align}
which completely determine the expansions of $|B_\pm\rangle^{(1)}|B_\pm\rangle^{(2)}$ and $|B_\pm\rangle^{(1)}|B_\mp\rangle^{(2)}$ in the bosonic basis.

We first analyse the bosonization of $|B_\pm\rangle^{(1)}|B_\pm\rangle^{(2)}$. Inserting the definition of the normal-ordered vertex operator [Eq.~\eqref{eq:vertex-normal-order-SM}] into the bosonic constraint [Eq.~\eqref{eq:boson-constraint-BB-SM}], we have
\begin{align}
    \left[e^{i\frac{\pi}{L}x} e^{i x_0} e^{i\frac{2\pi x}{L}a_0} :e^{i\phi'(x)}: \mp \; i e^{-i\frac{\pi }{L}(x+L/2)} e^{i \bar{x}_0} e^{-i\frac{2\pi}{L} (x+L/2)\bar{a}_0} :e^{i\bar{\phi}'(x+L/2)}:\right]|B_\pm\rangle^{(1)}|B_\pm\rangle^{(2)}=0\,.
\end{align}
Since the constraint is valid for all $x\in [0,L)$, we obtain the constraint for the oscillatory part:
\begin{align}
    \left[:e^{i\phi'(x)}: - :e^{i\bar{\phi}'(x+L/2)}:\right]|B_\pm\rangle^{(1)}|B_\pm\rangle^{(2)}=0\,, \quad x\in [0,L)\,,
\end{align}
which is equivalent to 
\begin{align}
    \left[a_k+(-1)^k\bar{a}_{-k}\right]|B_\pm\rangle^{(1)}|B_\pm\rangle^{(2)} = 0\,,\quad \forall k\in\mathbb{Z}\,,
\end{align}
by comparing with the mode expansions [Eq.~\eqref{eq:mode-expan-boson-SM}] of boson fields $\phi'$ and $\bar{\phi}'$.
This constraint completely determines the oscillatory part of $|B_\pm\rangle^{(1)}|B_\pm\rangle^{(2)}$:
\begin{align}
    |B_\pm\rangle^{(1)}|B_\pm\rangle^{(2)} = \exp\left[-\sum_{k=1}^\infty\frac{(-1)^k}{k}a_{-k}\bar{a}_{-k}\right]\sum_{n,m\in\mathbb{Z}} c^\pm_{n,m} |n,m\rangle \, ,
\end{align}
where the zero mode part remains to be fixed.

The zero mode part satisfies the following constraint:
\begin{align}
    \left[e^{i\frac{\pi}{L}x} e^{i x_0} e^{i\frac{2\pi x}{L}a_0} \mp i e^{-i\frac{\pi }{L}(x+L/2)} e^{i \bar{x}_0} e^{-i\frac{2\pi}{L} (x+L/2)\bar{a}_0} \right]\sum_{n,m\in\mathbb{Z}} c^\pm_{n,m} |n,m\rangle =0\,,
\end{align}
which simplifies to:
\begin{align}
    &\quad \sum_{n,m\in\mathbb{Z}} c^\pm_{n,m} \left[e^{i\frac{\pi}{L}x} e^{i x_0} e^{i\frac{2\pi x}{L}(n/2 +m)} \mp i e^{-i\frac{\pi }{L}(x+L/2)} e^{i \bar{x}_0} e^{-i\frac{2\pi}{L} (x+L/2)(n/2 -m)} \right] |n,m\rangle\nonumber\\
    &= \sum_{n,m\in\mathbb{Z}}  c^\pm_{n,m} \left[ e^{i\frac{2\pi x}{L}(\frac{n+1}{2} +m)} e^{i x_0}|n,m\rangle \mp (-i)^n (-1)^m e^{-i\frac{2\pi x}{L} (\frac{n+1}{2} -m)}e^{ix_0}|n,m-1\rangle\right]\nonumber\\
    &= \sum_{n,m\in\mathbb{Z}} \left[e^{i\frac{2\pi x}{L}(m+\frac{1}{2}+\frac{n}{2})} c^\pm_{n,m} \pm (-i)^n (-1)^m e^{i\frac{2\pi x}{L} (m+\frac{1}{2}-\frac{n}{2})} c^\pm_{n,m+1} \right]e^{i x_0}|n,m\rangle\nonumber\\
    &= 0 \, ,
\end{align}
where we used the relation $e^{i(\bar{x}_0-x_0)}|n,m\rangle = |n,m-1\rangle$. Therefore, the coefficients are determined as $c^\pm_{n,m+1} = \delta_{n,0}c^\pm_{0,m+1}$ and $c^\pm_{0,m} \pm (-1)^m c^\pm_{0,m+1} = 0$. Consequently, we determine the bosonic representation of $|B_\pm\rangle^{(1)}|B_\pm\rangle^{(2)}$ completely:
\begin{align}
    |B_\pm\rangle^{(1)}|B_\pm\rangle^{(2)}=\exp\left[-\sum_{k=1}^\infty\frac{(-1)^k}{k}a_{-k}\bar{a}_{-k}\right] \sum_{m\in\mathbb{Z}} (\pm)^m (-1)^{\frac{m(m+1)}{2}} |0,m\rangle\,,
\end{align}
where the normalization factor is fixed by $\langle 0,0|(|B_\pm\rangle^{(1)}|B_\pm\rangle^{(2)}) = ({}_{\mathrm{NS}}\langle 0|B_\pm\rangle)^2  = 1$, since the boson vacuum is identified as two copies of the Ising CFT ground states: $|0,0\rangle = |0\rangle_{\mathrm{NS}}^{(1)}|0\rangle_{\mathrm{NS}}^{(2)}$ in the bosonization.

The bosonization of $|B_\pm\rangle^{(1)}|B_\mp\rangle^{(2)}$ is similar to the derivation for $|B_\pm\rangle^{(1)}|B_\pm\rangle^{(2)}$. The oscillatory part is determined by
\begin{align}
    \left[a_k-(-1)^k\bar{a}_{-k}\right]|B_\pm\rangle^{(1)}|B_\mp\rangle^{(2)} = 0\,,\quad \forall k\in\mathbb{Z}\,,
\end{align}
and the zero-mode part is fixed as
\begin{align}
    \sum_{n\in\mathbb{Z}} (\pm)^n (-1)^{\frac{n(n+1)}{2}}|2n,0\rangle \, .
\end{align}
Therefore, the bosonic representation of $|B_\pm\rangle^{(1)}|B_\mp\rangle^{(2)}$ is given by
\begin{align}
    |B_\pm\rangle^{(1)}|B_\mp\rangle^{(2)} = \exp\left[\sum_{k=1}^\infty\frac{(-1)^k}{k}a_{-k}\bar{a}_{-k}\right] \sum_{n\in\mathbb{Z}} (\pm)^n (-1)^{\frac{n(n+1)}{2}}|2n,0\rangle
\end{align}
with the normalization $\langle 0,0|(|B_\pm\rangle^{(1)}|B_\mp\rangle^{(2)}) = {}_{\mathrm{NS}}\langle 0|B_+\rangle \cdot {}_{\mathrm{NS}}\langle 0|B_-\rangle = 1$.

However, there is a subtle point in the bosonization of the Ising CFT, where the operator content of two copies of Ising CFTs is not the ordinary Dirac CFT due to the constraints imposed by the fermion parity-dependent space-time boundary conditions. In fact, the bosonization of two copies of Ising CFTs is the $\mathbb{Z}_2$ \emph{orbifold} compactified boson CFT. Any physical bosonic crosscap state should be consistent with the $\mathbb{Z}_2$ orbifold constraint, i.e., invariant under the reflection of the boson field: $\varphi \leftrightarrow -\varphi$, which is realized via the operator $G$, with the action: $G|n,m\rangle = |-m,-n\rangle$ and $G (a_k, \bar{a}_k) G^{-1} = (-a_k,-\bar{a}_k)$, $\forall n,m,k\in\mathbb{Z}$~\cite{Francesco-Book}.

We find that the two copies of the Ising Ishibashi states $|\alpha\rangle\rangle_\mathcal{C}^{(1)}|\alpha\rangle\rangle_\mathcal{C}^{(2)}$, with $\alpha =\mathbbm{1},\varepsilon$, are suitable crosscap states for the $\mathbb{Z}_2$ orbifold boson CFT:
\begin{align}
|\mathbbm{1}\rangle\rangle_\mathcal{C}^{(1)}|\mathbbm{1}\rangle\rangle_\mathcal{C}^{(2)} &= \left[\frac{1}{2}(|B_+\rangle^{(1)}+|B_-\rangle^{(1)})\right] \left[\frac{1}{2}(|B_+\rangle^{(2)}+|B_-\rangle^{(2)})\right]
    =\frac{1}{2}\left(|\mathcal{O}_+\rangle + |\mathcal{O}_-\rangle\right)\,,\nonumber\\
    |\varepsilon\rangle\rangle_\mathcal{C}^{(1)}
    |\varepsilon\rangle\rangle_\mathcal{C}^{(2)} 
    &= \left[\frac{i}{2}(|B_+\rangle^{(1)}-|B_-\rangle^{(1)})\right] \left[\frac{i}{2}(|B_+\rangle^{(2)}-|B_-\rangle^{(2)})\right]
    =\frac{1}{2}\left(|\mathcal{O}_+\rangle -|\mathcal{O}_-\rangle\right)\,,
    \label{eq:bosonization-Ishibashi-SM}
\end{align}
with
\begin{align}
    |\mathcal{O}_+\rangle &= \exp\left[\sum_{k=1}^\infty\frac{(-1)^k}{k}a_{-k}\bar{a}_{-k}\right]\sum_{n\in\mathbb{Z}} (-1)^n |4n,0\rangle\,,\nonumber\\
    |\mathcal{O}_-\rangle &= \exp\left[-\sum_{k=1}^\infty\frac{(-1)^k}{k}a_{-k}\bar{a}_{-k}\right] \sum_{m\in\mathbb{Z}} (-1)^m |0,2m\rangle \, ,
    \label{eq:cross-orbifold-boson-SM}
\end{align}
satisfying $G|\mathcal{O}_\pm\rangle = |\mathcal{O}_\pm\rangle$. Here, we used the fermionic representation of the Ishibashi states in Eqs.~\eqref{eq:Ishibashi-identity-fermion-rep-SM} and ~\eqref{eq:Ishibashi-energy-fermion-rep-SM}.

The crosscap correlators can be calculated in each Ishibashi sector separately.

\subsection{B. Wick's theorem with crosscap states involoved}

The bosonization dictionary for two copies of $\varepsilon$ fields and $\sigma$ fields is given by~\cite{Francesco-Book}
\begin{align}
    \varepsilon_1(w,\bar{w})\varepsilon_2(w,\bar{w}) = \partial\phi (w)\bar{\partial}\bar{\phi}(\bar{w})\,,\quad \sigma_1(w,\bar{w})\sigma_2(w,\bar{w}) = \sqrt{2} :\cos \left[\vartheta (w,\bar{w})/2\right]:\,,
    \label{eq:bosonization-primary-fields-SM}
\end{align}
where $\vartheta (w,\bar{w}) = \phi (w) -\bar{\phi}(\bar{w})$ is the \emph{dual} compactified boson field with radius $R' = 2/R = 1$, and $w= e^{\frac{2\pi}{L}z}$ is the plane coordinate.

To effectively calculate the multipoint crosscap correlators of $\varepsilon$ and $\sigma$ fields, we apply Wick's theorems. For the U(1) current, Wick's theorem gives
\begin{align}
    \prod_{j=1}^n \partial\phi (w_j) &= \sum_{0\leq m \leq \left[\frac{n}{2}\right]} \sum_{1\leq k_1 < \cdots k_{2m}\leq n} :\prod_{j\neq k_1,\cdots k_{2m}} \partial\phi (w_j): \frac{1}{2^m m!}\sum_{\sigma \in S_{2m}} \prod_{\alpha =1}^m\langle \partial\phi (w_{k_{\sigma (2\alpha -1)}})\partial\phi (w_{k_{\sigma (2\alpha)}})\rangle\,,\nonumber\\
    \prod_{j=1}^n \bar{\partial}\bar{\phi} (\bar{w}_j) &= \sum_{0\leq m \leq \left[\frac{n}{2}\right]} \sum_{1\leq k_1 < \cdots k_{2m}\leq n} :\prod_{j\neq k_1,\cdots k_{2m}} \bar{\partial}\bar{\phi} (\bar{w}_j): \frac{1}{2^m m!}\sum_{\sigma \in S_{2m}} \prod_{\alpha =1}^m\langle \bar{\partial}\bar{\phi} (\bar{w}_{k_{\sigma (2\alpha -1)}})\bar{\partial}\bar{\phi} (\bar{w}_{k_{\sigma (2\alpha)}})\rangle \, ,
\label{eq:Wick-1-SM}
\end{align}
where $\langle\partial\phi (w)\partial\phi (w')\rangle = -\frac{1}{(w-w')^2}$ and $\langle\bar{\partial}\bar{\phi} (\bar{w})\bar{\partial}\bar{\phi} (\bar{w}')\rangle = -\frac{1}{(\bar{w}-\bar{w}')^2}$ are the (anti-)chiral current-current correlators on the plane, respectively.
For the vertex operators, the theorem gives
\begin{align}
    \prod_{j=1}^n  :e^{i\alpha_j\vartheta (w_j,\bar{w}_j)}: = \prod_{1\leq j< k\leq n} |w_j-w_k|^{2\alpha_j\alpha_k}:e^{i\sum_{j=1}^n\alpha_j\vartheta (w_j,\bar{w}_j)}: \, .
\label{eq:Wick-2-SM}
\end{align}
The proofs of these theorems can be found in standard textbooks, such as Ref.~\cite{Francesco-Book}.

The theorems stated above are the ordinary Wick's theorems. To account for crosscap states, we also need the following two generalized Wick's theorems:

\emph{Theorem 1}:
\begin{align}
    \prod_{j=1}^{n}\partial\phi'_-(w_j)\prod_{l=1}^{m}\bar{\partial}\bar{\phi}'_-(\bar{w}_l) e^{\pm K} &= e^{\pm K} \prod_{j=1}^{n}\left[\partial\phi'_-(w_j)\mp \frac{1}{w_j^2}\bar{\partial}\bar{\phi}'_+ (-1/w_j)\right]\prod_{l=1}^{m}
    \left[\bar{\partial}\bar{\phi}'_-(\bar{w}_l)\mp \frac{1}{\bar{w}_l^2}\partial\phi'_+(-1/\bar{w}_l)\right] \, ,
\label{eq:cross-Wick-1-SM}   
\end{align}

\emph{Theorem 2}:
\begin{align}
    :e^{i\sum\limits_{j=1}^n\alpha_j \vartheta' (w_j,\bar{w}_j)}:\; e^{\pm K} &=\prod_{1\leq j,l\leq n}\left(1+\frac{1}{w_j\bar{w}_l}\right)^{\pm\alpha_j\alpha_l}  e^{\pm K \pm i\sum\limits_{j=1}^n\alpha_j\vartheta'_+(-1/\bar{w}_j,-1/w_j)}\; :e^{i\sum\limits_{j=1}^n\alpha_j \vartheta' (w_j,\bar{w}_j)}: \, ,
\label{eq:cross-Wick-2-SM}
\end{align}
where 
\begin{align}
    K = \sum_{k=1}^\infty \frac{(-1)^k}{k}a_{-k}\bar{a}_{-k}
\end{align}
is the quadratic form of boson modes, and $\phi'$ ($\bar{\phi}'$) represents the oscillatory part of the (anti-)chiral boson field [Eq.~\eqref{eq:mode-expan-boson-SM}], while $\vartheta'$ denotes the oscillatory part of the dual boson field:
\begin{align}
    \vartheta' (w_j,\bar{w}_j) = \phi' (w) -\bar{\phi}'(\bar{w})\equiv \vartheta'_+ (w_j,\bar{w}_j) + \vartheta'_- (w_j,\bar{w}_j) \,,
\end{align}
which can be divided into positive and negative energy contributions.

To prove the above generalized Wick's theorems, we first introduce the following lemma:

\emph{Lemma 1}:
\begin{align}
    [\phi'_-(w),K] &= -\bar{\phi}'_+ (-1/w)\,,\quad [\bar{\phi}'_-(\bar{w}),K] =-\phi'_+(-1/\bar{w})\,.
    \label{eq:lemma-1-SM}
\end{align}
This lemma can be verified straightforwardly by comparing the mode expansions [Eq.~\eqref{eq:mode-expan-boson-SM}]: 
\begin{align}
    [\phi'_-(w),K] &= i\sum_{k=1}^\infty \frac{(-1)^k}{k}w^{-k}\bar{a}_{-k} = i\sum_{k=1}^\infty\frac{1}{k}(-1/w)^k\bar{a}_{-k} =-\bar{\phi}'_+(-1/w)\,,\nonumber\\
    [\bar{\phi}'_-(\bar{w}),K] &= i\sum_{k=1}^\infty \frac{(-1)^k}{k}\bar{w}^{-k}a_{-k} = i\sum_{k=1}^\infty\frac{1}{k}(-1/\bar{w})^k a_{-k} =-\phi'_+(-1/\bar{w})\,,
\end{align}
where we used the commutation relations $[\phi'_-(w),a_{-k}]=iw^{-k}$ and $[\bar{\phi}'_-(\bar{w}),a_{-k}]=i\bar{w}^{-k}$, $\forall k>0$. 

\emph{Theorem 1} [Eq.~\eqref{eq:cross-Wick-1-SM}] is a corollary of \emph{Lemma 1}. Since $[\partial\phi'_-(w),K] = -\frac{1}{w^2}\bar{\partial}\bar{\phi}'_+ (-1/w)$ and $[\bar{\partial}\bar{\phi}'_-(\bar{w}),K] =-\frac{1}{\bar{w}^2}\partial\phi'_+(-1/\bar{w})$, we have
\begin{align}
    \partial\phi'_-(w) e^{\pm K} = e^{\pm K}\left[\partial\phi'_-(w)\mp \frac{1}{w^2}\bar{\partial}\bar{\phi}'_+ (-1/w) \right]\,,\quad \bar{\partial}\bar{\phi}'_-(\bar{w}) e^{\pm K} = e^{\pm K }\left[\bar{\partial}\bar{\phi}'_-(\bar{w})\mp \frac{1}{\bar{w}^2}\partial\phi'_+(-1/\bar{w})\right]\,,
\end{align}
from which \emph{Theorem 1} is derived.

To prove \emph{Theorem 2} [Eq.~\eqref{eq:cross-Wick-2-SM}], an additional lemma is needed:

\emph{Lemma 2}:
\begin{align}
    e^{i\alpha \vartheta'_- (w,\bar{w})}\, e^{\pm K} &= \left(1+\frac{1}{w\bar{w}}\right)^{\pm\alpha^2} e^{\pm K \pm i \alpha\vartheta'_+(-1/\bar{w},-1/w)} e^{i\alpha \vartheta'_- (w,\bar{w})}\,,
    \label{eq:lemma-2-SM}
\end{align}
which can be proved as follows:
\begin{align}
    e^{i\alpha \vartheta'_- (w,\bar{w})}\, e^{\pm K} 
    &= e^{i\alpha \phi'_-(w)}\left(e^{-i\alpha\bar{\phi}'_-(\bar{w})}\, e^{\pm K}\right) \nonumber\\
    &= e^{i\alpha \phi'_-(w)} \left(e^{\pm K} e^{-i\alpha\bar{\phi}'_-(\bar{w})} e^{\mp i\alpha [\bar{\phi}'_-(\bar{w}),\, K]}\right)\nonumber\\
    &= \left(e^{i\alpha \phi'_-(w)} e^{\pm K \pm i \alpha\phi'_+ (-1/\bar{w})}\right) e^{-i\alpha\bar{\phi}'_-(\bar{w})}\nonumber\\
    &= \left(e^{\pm K \pm i \alpha\phi'_+ (-1/\bar{w})} e^{i\alpha \phi'_-(w)} e^{\pm i\alpha [\phi'_-(w),\, K + i\alpha\phi'_+ (-1/\bar{w})]}\right)e^{-i\alpha\bar{\phi}'_-(\bar{w})}\nonumber\\
    &= e^{\pm K \pm i\alpha [\phi'_+ (-1/\bar{w}) -\bar{\phi}'_+ (-1/w)]} e^{\mp \alpha^2 \langle\phi'(w)\phi'(-1/\bar{w}) \rangle}  e^{i\alpha \vartheta'_- (w,\bar{w})}\nonumber\\
    &= \left(1+\frac{1}{w\bar{w}}\right)^{\pm\alpha^2} e^{\pm K \pm i \alpha\vartheta'_+(-1/\bar{w},-1/w)} e^{i\alpha \vartheta'_- (w,\bar{w})} \, ,
    \label{eq:prove-lemma-2-SM}
\end{align}
where in the second and fourth equalities, we used the Baker-Campbell-Hausdorff formula: $e^{A}e^{B} =e^{A+B+\frac{1}{2}[A,B]}=e^Be^Ae^{[A,B]}$, assuming $[[A,B],A]=[[A,B],B]=0$; in the third and fifth equalities, we used the \emph{Lemma 1} [Eq.~\eqref{eq:lemma-1-SM}]; and in the last equality, we used the correlator $[\phi'_- (w),\phi'_+ (-1/\bar{w})]=\langle \phi' (w)\phi' (-1/\bar{w})\rangle = -\ln (1+\frac{1}{w\bar{w}})$. 

The simplest case ($n=1$) of \emph{Theorem 2} can be derived directly from \emph{Lemma 2} [Eq.~\eqref{eq:lemma-2-SM}]:
\begin{align}
    :e^{i\alpha \vartheta' (w,\bar{w})}:\, e^{\pm K} &= e^{i\alpha \vartheta'_+ (w,\bar{w})}\left(e^{i\alpha \vartheta'_- (w,\bar{w})}\, e^{\pm K}\right)\nonumber\\
    &= \left(1+\frac{1}{w\bar{w}}\right)^{\pm\alpha^2} e^{\pm K \pm i \alpha\vartheta'_+(-1/\bar{w},-1/w)} :e^{i\alpha \vartheta' (w,\bar{w})}:\,.
    \label{eq:cross-Wick-2-simplest-case-SM}
\end{align}
The general case ($n \geq 1$) can be proved by induction. For instance, we prove the $n=2$ case below:
\begin{align}
    :e^{i\alpha_1 \vartheta' (w_1,\bar{w}_1) + i\alpha_2 \vartheta' (w_2,\bar{w}_2)}:\, e^{\pm K} 
    &= e^{i\alpha_1 \vartheta'_+ (w_1,\bar{w}_1) + i\alpha_2 \vartheta'_+ (w_2,\bar{w}_2)} e^{i\alpha_2 \vartheta'_- (w_2,\bar{w}_2)}\left(e^{i\alpha_1 \vartheta'_- (w_1,\bar{w}_1)} e^{\pm K}\right)\nonumber\\
    &= e^{i\sum\limits_{j=1,2}\alpha_j \vartheta'_+ (w_j,\bar{w}_j)} e^{i\alpha_2 \vartheta'_- (w_2,\bar{w}_2)} \left(1+\frac{1}{w_1\bar{w}_1}\right)^{\pm\alpha_1^2} e^{\pm K \pm i \alpha_1\vartheta'_+(-1/\bar{w}_1,-1/w_1)} e^{i\alpha_1 \vartheta'_- (w_1,\bar{w}_1)}\nonumber\\
    &= \left(1+\frac{1}{w_1\bar{w}_1}\right)^{\pm\alpha_1^2} e^{i\sum\limits_{j=1,2}\alpha_j \vartheta'_+ (w_j,\bar{w}_j) } \left(1+\frac{1}{w_2\bar{w}_2}\right)^{\pm\alpha_2^2} e^{\pm K \pm i \alpha_2\vartheta'_+(-1/\bar{w}_2,-1/w_2)} \nonumber\\
    &\quad \times  e^{i\alpha_2 \vartheta'_- (w_2,\bar{w}_2)} e^{\pm i \alpha_1\vartheta'_+(-1/\bar{w}_1,-1/w_1)} e^{i\alpha_1 \vartheta'_- (w_1,\bar{w}_1)}\nonumber\\
    &= \left(1+\frac{1}{w_1\bar{w}_1}\right)^{\pm\alpha_1^2}\left(1+\frac{1}{w_2\bar{w}_2}\right)^{\pm\alpha_2^2} e^{\pm K \pm i \alpha_2\vartheta'_+(-1/\bar{w}_2,-1/w_2)}e^{i\alpha_1 \vartheta'_+ (w_1,\bar{w}_1) + i\alpha_2 \vartheta'_+ (w_2,\bar{w}_2)}\nonumber\\
    &\quad\times \left(e^{\pm i \alpha_1\vartheta'_+(-1/\bar{w}_1,-1/w_1)} e^{i\alpha_2 \vartheta'_- (w_2,\bar{w}_2)} e^{\mp \alpha_1\alpha_2 \langle \vartheta' (w_2,\bar{w}_2) \vartheta'(-1/\bar{w}_1,-1/w_1)\rangle}\right) e^{i\alpha \vartheta'_- (w_1,\bar{w}_1)}\nonumber\\
    &= \left(1+\frac{1}{w_1\bar{w}_1}\right)^{\pm\alpha_1^2}\left(1+\frac{1}{w_2\bar{w}_2}\right)^{\pm\alpha_2^2} \left(1+\frac{1}{w_1\bar{w}_2}\right)^{\pm\alpha_1\alpha_2} \left(1+\frac{1}{w_2\bar{w}_1}\right)^{\pm\alpha_1\alpha_2}\nonumber\\
    &\quad\times e^{\pm K \pm i \alpha_1\vartheta'_+(-1/\bar{w}_1,-1/w_1) \pm i \alpha_2\vartheta'_+(-1/\bar{w}_2,-1/w_2)} :e^{i\alpha_1 \vartheta' (w_1,\bar{w}_1) + i\alpha_2 \vartheta' (w_2,\bar{w}_2)}:\nonumber\\
    &= \prod_{1\leq j,l\leq 2}\left(1+\frac{1}{w_j\bar{w}_l}\right)^{\pm\alpha_j\alpha_l}  e^{\pm K \pm i\sum\limits_{j=1,2}\alpha_j\vartheta'_+(-1/\bar{w}_j,-1/w_j)}\; :e^{i\sum\limits_{j=1,2}\alpha_j \vartheta' (w_j,\bar{w}_j)}: \, ,
\end{align}
where, in the second equality, we used the result for the $n=1$ case [Eq.~\eqref{eq:cross-Wick-2-simplest-case-SM}] by induction. We also used the definition of the normal-ordered vertex operators [Eq.~\eqref{eq:vertex-normal-order-SM}] and the Baker-Campbell-Hausdorff formula [Below Eq.~\eqref{eq:prove-lemma-2-SM}], along with the correlator
\begin{align}
    \langle \vartheta' (w_2,\bar{w}_2) \vartheta'(-1/\bar{w}_1,-1/w_1)\rangle &= \langle\phi'(w_2)\phi'(-1/\bar{w}_1)\rangle + \langle\bar{\phi}'(\bar{w}_2)\bar{\phi}'(-1/w_1)\rangle\nonumber\\
    &= - \ln \left(1+\frac{1}{w_2\bar{w}_1}\right) -\ln \left(1+\frac{1}{w_1\bar{w}_2}\right)\,.
\end{align}

\subsection{C. $n$-point crosscap correlators of the $\varepsilon$ field}

The $n$-point crosscap correlator of the $\varepsilon$ field can be directly calculated in the fermionic basis. However, in this case, we will alternatively use the bosonization technique to perform the calculation, where a unified expression can be obtained.

Using Eq.~\eqref{eq:cross-Ishibashi-expan-SM}, we can calculate the $n$-point crosscap correlator of the $\varepsilon$ field in each Ishibashi sector respectively:
\begin{align}
    {}_{\mathrm{NS}}\langle 0| \prod_{j=1}^n \varepsilon (w_j,\bar{w}_j)|\mathcal{C}_\pm\rangle = \sqrt{\frac{2+\sqrt{2}}{2}} \; {}_{\mathrm{NS}}\langle 0| \prod_{j=1}^n \varepsilon (w_j,\bar{w}_j)|\mathbbm{1}\rangle\rangle_\mathcal{C} \pm \sqrt{\frac{2-\sqrt{2}}{2}} \; {}_{\mathrm{NS}}\langle 0| \prod_{j=1}^n \varepsilon (w_j,\bar{w}_j)|\varepsilon\rangle\rangle_\mathcal{C} \, ,
\end{align}
where the bosonization of the crosscap correlator in each Ishibashi sector is given by
\begin{align}
    {}_{\mathrm{NS}}\langle 0| \prod_{j=1}^{n} \varepsilon (w_j,\bar{w}_j)|\mathbbm{1}\rangle\rangle_\mathcal{C} &= \sqrt{\frac{1}{2}\langle 0,0|\prod_{j=1}^{n} \partial\phi (w_j)\bar{\partial}\bar{\phi}(\bar{w}_j)\left(|\mathcal{O}_+\rangle + |\mathcal{O}_-\rangle\right)}\,,\nonumber\\
    {}_{\mathrm{NS}}\langle 0| \prod_{j=1}^{n} \varepsilon (w_j,\bar{w}_j)|\varepsilon\rangle\rangle_\mathcal{C} &= \sqrt{\frac{1}{2}\langle 0,0|\prod_{j=1}^{n} \partial\phi (w_j)\bar{\partial}\bar{\phi}(\bar{w}_j)\left(|\mathcal{O}_+\rangle - |\mathcal{O}_-\rangle\right)}\,,
\end{align}
with
\begin{align}
    \langle 0,0|\prod_{j=1}^{n} \partial\phi (w_j)\bar{\partial}\bar{\phi}(\bar{w}_j)|\mathcal{O}_\pm\rangle = \langle 0,0|\prod_{j=1}^{n} \partial\phi (w_j)\bar{\partial}\bar{\phi}(\bar{w}_j)e^{\pm K}|0,0\rangle\,,
\end{align}
which are the crosscap correlators in the compactified boson CFT. We used the bosonization of two copies of $\varepsilon$ fields [Eq.~\eqref{eq:bosonization-primary-fields-SM}] and the bosonization of two copies of Ishibashi states [Eq.~\eqref{eq:bosonization-Ishibashi-SM}], where $|0,0\rangle$ is the boson ground state and $|\mathcal{O}_\pm\rangle$ are the $\mathbb{Z}_2$ orbifold bosonic crosscap states in Eq.~\eqref{eq:cross-orbifold-boson-SM}.

To calculate the bosonic crosscap correlators, we first use Wick's theorem [Eq.~\eqref{eq:Wick-1-SM}] to express the time-ordered operator product in a normal-ordered form:
\begin{align}
    &\phantom{=} \;\; \langle 0,0|\prod_{j=1}^{n} \partial\phi (w_j)\bar{\partial}\bar{\phi}(\bar{w}_j)e^{\pm K}|0,0\rangle \nonumber \\
    &= \sum_{0\leq p,q \leq \left[\frac{n}{2}\right]}  (-1)^{p+q}\sum_{\substack{1\leq k_1 < \cdots < k_{2p}\leq n \\ 1\leq k'_1 < \cdots < k'_{2q}\leq n } }\mathrm{Hf}\left[\frac{1}{(w_{k_\alpha}-w_{k_\beta})^2}\right]_{\alpha,\beta = 1,\ldots, 2p} \mathrm{Hf}\left[\frac{1}{(w_{k'_\alpha}-w_{k'_\beta})^2}\right]_{\alpha,\beta = 1,\ldots, 2q} \nonumber \\
    &\quad\times\langle 0,0|:\prod_{j\neq k_1,\ldots, k_{2p}} \partial\phi (w_j)\prod_{l\neq k'_1,\ldots, k'_{2q}} \bar{\partial}\bar{\phi} (\bar{w}_l): e^{\pm K}|0,0\rangle \, ,
\end{align}
where $\mathrm{Hf}(B)$ stands for the Hafnian of a symmetric $2n\times 2n$ matrix $B$:
\begin{align}
    \mathrm{Hf} (B) = \frac{1}{2^n n!}\sum_{\sigma \in S_{2n}} \prod_{\alpha =1}^n B_{\sigma (2\alpha -1),\sigma (2\alpha)}\, ,
\end{align}
and $[\frac{n}{2}]$ denotes the integer part of $\frac{n}{2}$. Next, using the generalized Wick's theorem in Eq.~\eqref{eq:cross-Wick-1-SM}, the expression simplifies further:
\begin{align}
    &\quad\langle 0,0|:\prod_{j\neq k_1,\ldots, k_{2p}} \partial\phi (w_j)\prod_{l\neq k'_1,\ldots, k'_{2q}} \bar{\partial}\bar{\phi} (\bar{w}_l): e^{\pm K}|0,0\rangle \nonumber\\
    &= \langle 0,0| \prod_{j\neq k_1,\ldots, k_{2p}} \partial\phi'_- (w_j)\prod_{l\neq k'_1,\ldots, k'_{2q}} \bar{\partial}\bar{\phi}'_- (\bar{w}_l) e^{\pm K}|0,0\rangle\nonumber\\
    &= \langle 0,0| \prod_{j\neq k_1,\ldots, k_{2p}} \left[\partial\phi'_- (w_j) \mp \frac{1}{w_j^2}\bar{\partial}\bar{\phi}'_+ (-1/w_j) \right]\prod_{l\neq k'_1,\ldots, k'_{2q}} \left[\bar{\partial}\bar{\phi}'_- (\bar{w}_l) \mp \frac{1}{\bar{w}_l^2}\partial\phi'_+ (-1/\bar{w}_l)\right]|0,0\rangle\,,
\end{align}
where the calculation reduces to the Wick contraction of boson fields, denoted as $\varphi_A (w)$ and $\varphi_B (\bar{w})$:
\begin{align}
    \varphi_A (w) = \partial\phi'_- (w) \mp \frac{1}{w^2}\bar{\partial}\bar{\phi}'_+ (-1/w)\,,\quad \varphi_B (\bar{w}) = \bar{\partial}\bar{\phi}'_- (\bar{w}) \mp \frac{1}{\bar{w}^2}\partial\phi'_+ (-1/\bar{w})
\end{align}
with their two-point correlators on the plane being
\begin{align}
    \langle \varphi_A (w_j)\varphi_B (\bar{w}_l) \rangle = \mp \frac{1}{\bar{w}_l^2}\langle  \partial\phi'_- (w_j)\partial\phi'_+ (-1/\bar{w}_l)\rangle = \mp \frac{1}{\bar{w}_l^2} \langle \partial\phi (w_j)\partial\phi (-1/\bar{w}_l)\rangle = \pm \frac{1}{\bar{w}_l^2} \frac{1}{(w_j+1/\bar{w}_l)^2} = \pm \frac{1}{(1+w_j\bar{w}_l)^2}\,,
\end{align}
and $\langle \varphi_A (w_j)\varphi_A (w_l) \rangle = \langle \varphi_B (\bar{w}_j)\varphi_B (\bar{w}_l) \rangle =0$. After performing the Wick contraction, we have
\begin{align}
    &\quad\langle 0,0| \prod_{j\neq k_1,\ldots, k_{2p}} \left[\partial\phi'_- (w_j) \mp \frac{1}{w_j^2}\bar{\partial}\bar{\phi}'_+ (-1/w_j) \right]\prod_{l\neq k'_1,\ldots, k'_{2q}} \left[\bar{\partial}\bar{\phi}'_- (\bar{w}_l) \mp \frac{1}{\bar{w}_l^2}\partial\phi'_+ (-1/\bar{w}_l)\right]|0,0\rangle\nonumber\\
    &=\delta_{p,q} \sum_{\sigma \in S_{n-2p}} \prod_{\substack{j\neq k_1,\ldots, k_{2p}\\ l\neq k'_1,\ldots, k'_{2p}} }\langle \varphi_A (w_j)\varphi_B (\bar{w}_{\sigma(l)}) \rangle =\delta_{p,q} (\pm)^{n-2p} \mathrm{Perm}\left[\frac{1}{(1+w_j\bar{w}_l)^2}\right]_{\substack{j\neq k_1,\ldots, k_{2p}\\ l\neq k'_1,\ldots, k'_{2p}} }\,,
\end{align}
where $\mathrm{Perm}(M)$ is the permanent of the $n\times n$ matrix $M$:
\begin{align}
    \mathrm{Perm}(M) = \sum_{\sigma \in S_n} \prod_{\alpha =1}^n M_{\alpha,\sigma (\alpha)}\,.
\end{align}

Finally, we arrive at the bosonic crosscap correlators:
\begin{align}
    &\quad \langle 0,0|\prod_{j=1}^{n} \partial\phi (w_j)\bar{\partial}\bar{\phi}(\bar{w}_j)|\mathcal{O}_\pm\rangle = \langle 0,0|\prod_{j=1}^{n} \partial\phi (w_j)\bar{\partial}\bar{\phi}(\bar{w}_j)e^{\pm K}|0,0\rangle \nonumber\\
    &= (\pm)^n\sum_{0\leq p \leq \left[\frac{n}{2}\right]}  \sum_{\substack{1\leq k_1 < \cdots < k_{2p}\leq n \\ 1\leq k'_1 < \cdots < k'_{2p}\leq n } }\mathrm{Hf}\left[\frac{1}{(w_{k_\alpha}-w_{k_\beta})^2}\right]\mathrm{Hf}\left[\frac{1}{(w_{k'_\alpha}-w_{k'_\beta})^2}\right]_{\alpha,\beta = 1,\ldots, 2p} \mathrm{Perm}\left[\frac{1}{(1+w_j\bar{w}_l)^2}\right]_{\substack{j\neq k_1,\ldots, k_{2p}\\ l\neq k'_1,\ldots, k'_{2p}} } \, .
\end{align}
Thus, the $n$-point crosscap correlator of the $\varepsilon$ field is given by
\begin{align}
    &\quad {}_{\mathrm{NS}}\langle 0| \prod_{j=1}^n \varepsilon (w_j,\bar{w}_j)|\mathcal{C}_\pm\rangle = (\pm)^n\sqrt{\frac{2 + (-1)^n \sqrt{2}}{2}}\nonumber\\
    &\times \sqrt{\sum_{0\leq p \leq \left[\frac{n}{2}\right]}  \sum_{\substack{1\leq k_1 < \cdots < k_{2p}\leq n \\ 1\leq k'_1 < \cdots < k'_{2p}\leq n } }\mathrm{Hf}\left[\frac{1}{(w_{k_\alpha}-w_{k_\beta})^2}\right]\mathrm{Hf}\left[\frac{1}{(w_{k'_\alpha}-w_{k'_\beta})^2}\right]_{\alpha,\beta = 1,\ldots, 2p} \mathrm{Perm}\left[\frac{1}{(1+w_j\bar{w}_l)^2}\right]_{\substack{j\neq k_1,\ldots, k_{2p}\\ l\neq k'_1,\ldots, k'_{2p}} }} \, .
\end{align}

In the end, we present the results for the one- and two-point correlators as examples.

For the one-point correlator, we have
\begin{align}
    {}_{\mathrm{NS}}\langle 0| \varepsilon (w,\bar{w})|\mathcal{C}_\pm \rangle 
    &= \pm\sqrt{\frac{2-\sqrt{2}}{2}}\frac{1}{1+w\bar{w}}\,,
    \label{eq:cross-correlator-epsilon-1-SM}
\end{align}

For the two-point correlator, the expression is
\begin{align}
    {}_{\mathrm{NS}}\langle 0| \varepsilon (w_1,\bar{w}_1)\varepsilon (w_2,\bar{w}_2)|\mathcal{C}_\pm\rangle
    &= \sqrt{\frac{2+\sqrt{2}}{2}}\sqrt{\frac{1}{(1+w_1\bar{w}_1)^2}\frac{1}{(1+w_2\bar{w}_2)^2} + \frac{1}{(1+w_1\bar{w}_2)^2}\frac{1}{(1+w_2 \bar{w}_1)^2} + \frac{1}{|w_1-w_2|^4} }\nonumber\\
    &= \sqrt{\frac{2+\sqrt{2}}{2}}\frac{1}{|w_1-w_2|^2}\sqrt{\eta^2 + \left(\frac{\eta}{1-\eta}\right)^2 +1}\nonumber\\
    &= \sqrt{\frac{2+\sqrt{2}}{2}} \frac{1}{|w_1-w_2|^2} \left(1 + \frac{\eta^2}{1-\eta}\right) \, ,
    \label{eq:cross-correlator-epsilon-2-SM}
\end{align}
where 
\begin{align}
    \eta = \frac{|w_1-w_2|^2}{(1+|w_1|^2)(1+|w_2|^2)}
    \label{eq:cross-ratio-SM}
\end{align}
is the crosscap cross ratio.

\subsection{D. $2n$-point crosscap correlators of the $\sigma$ field}

According to the fusion rule $\sigma \times\sigma \sim \mathbbm{1} + \varepsilon$, $(2n+1)$-point crosscap correlators of the $\sigma$ field vanish. For $2n$-point crosscap correlators, the calculation follows a similar logic to that of the $\varepsilon$ field. We express the full crosscap correlator as a combination of two Ishibashi crosscap correlators:
\begin{align}
    {}_{\mathrm{NS}}\langle 0|\prod_{j=1}^{2n} \sigma (w_j,\bar{w}_j) |\mathcal{C}_\pm\rangle = \sqrt{\frac{2+\sqrt{2}}{2}}  {}_{\mathrm{NS}}\langle 0|\prod_{j=1}^{2n} \sigma (w_j,\bar{w}_j) |\mathbbm{1}\rangle\rangle_\mathcal{C} \pm \sqrt{\frac{2-\sqrt{2}}{2}}  {}_{\mathrm{NS}}\langle 0|\prod_{j=1}^{2n} \sigma (w_j,\bar{w}_j) |\varepsilon\rangle\rangle_\mathcal{C}\,.
\end{align}
Each Ishibashi crosscap correlator is then expressed in terms of bosonic crosscap correlators via bosonization [Eqs.~\eqref{eq:bosonization-primary-fields-SM} and \eqref{eq:bosonization-Ishibashi-SM}]:
\begin{align}
    {}_{\mathrm{NS}}\langle 0| \prod_{j=1}^{2n} \sigma (w_j,\bar{w}_j)|\mathbbm{1}\rangle\rangle_\mathcal{C} &= \sqrt{2^{n-1}\langle 0,0|\prod_{j=1}^{2n} \cos\frac{\vartheta }{2}(w_j,\bar{w}_j)\left(|\mathcal{O}_+\rangle + |\mathcal{O}_-\rangle\right)}\,,\nonumber\\
    {}_{\mathrm{NS}}\langle 0| \prod_{j=1}^{2n} \sigma (w_j,\bar{w}_j)|\varepsilon\rangle\rangle_\mathcal{C} &= \sqrt{2^{n-1}\langle 0,0|\prod_{j=1}^{2n} \cos\frac{\vartheta}{2}(w_j,\bar{w}_j)\left(|\mathcal{O}_+\rangle - |\mathcal{O}_-\rangle\right)}
\label{eq:cross-sigma-Ishibashi-SM}
\end{align}
with
\begin{align}
    \langle 0,0|\prod_{j=1}^{2n} \cos\frac{\vartheta }{2}(w_j,\bar{w}_j)|\mathcal{O}_\pm\rangle =  \langle 0,0|\prod_{j=1}^{n} \cos\frac{\vartheta }{2}(w_j,\bar{w}_j) \; e^{\pm K}|\mathcal{O}^0_\pm \rangle\,,
\end{align}
where $|\mathcal{O}^0_\pm \rangle$ denotes the zero-mode parts of the bosonic crosscap states $|\mathcal{O}_\pm \rangle$ in Eq.~\eqref{eq:cross-orbifold-boson-SM}:
\begin{align}
    |\mathcal{O}^0_+ \rangle = \sum_{n\in\mathbb{Z}} (-1)^n |4n,0\rangle\,,\quad |\mathcal{O}^0_- \rangle = \sum_{m\in\mathbb{Z}} (-1)^m |0,2m\rangle\,.
\end{align}

The Wick's theorem [Eq.~\eqref{eq:Wick-2-SM}] turns the product of the vertex operators into the normal-ordered form:
\begin{align} 
     \langle 0,0|\prod_{j=1}^{2n} \cos\frac{\vartheta }{2}(w_j,\bar{w}_j)\; e^{\pm K}|\mathcal{O}^0_\pm \rangle &= \frac{1}{2^{2n}} \langle 0,0|\prod_{j=1}^{2n} \left[e^{i\vartheta (w_j,\bar{w}_j)/2} + e^{-i\vartheta (w_j,\bar{w}_j)/2}\right] e^{\pm K}|\mathcal{O}^0_\pm \rangle \nonumber\\
     &= \frac{1}{2^{2n}}\sum_{\substack{\epsilon_j =\pm 1,\\ j=1,\ldots,2n }} \prod_{1\leq j < l\leq 2n} |w_j-w_l|^{\epsilon_j\epsilon_l/2} \langle 0,0|:e^{i\sum_{j=1}^{2n}\epsilon_j\vartheta (w_j,\bar{w}_j)/2} : e^{\pm K}|\mathcal{O}^0_\pm \rangle \, ,
\end{align}
and the generalized Wick's theorem [Eq.~\eqref{eq:cross-Wick-2-SM}] further simplifies the contribution from the oscillatory part of the crosscap states:
\begin{align}
    \langle 0,0|:e^{i\sum_{j=1}^{2n}\epsilon_j\vartheta (w_j,\bar{w}_j)/2} : e^{\pm K}|\mathcal{O}^0_\pm \rangle &=\langle 0,0|e^{\frac{i}{2}\sum_{j=1}^{2n}\epsilon_j (x_0-\bar{x}_0)}e^{\sum_{j=1}^{2n}\epsilon_j (\ln w_j a_0-\ln \bar{w}_j\bar{a}_0)/2}
     :e^{i\sum_{j=1}^{2n}\epsilon_j\vartheta' (w_j,\bar{w}_j)/2} : e^{\pm K}|\mathcal{O}^0_\pm\rangle \nonumber\\
     &= \prod_{1\leq j,l\leq 2n}\left(1+\frac{1}{w_j\bar{w}_l}\right)^{\pm \epsilon_j\epsilon_l/4}\langle 0,0|e^{\frac{i}{2}\sum_{j=1}^{2n}\epsilon_j (x_0-\bar{x}_0)}e^{\sum_{j=1}^{2n}\epsilon_j (\ln w_j a_0-\ln \bar{w}_j\bar{a}_0)/2} |\mathcal{O}^0_\pm\rangle \, .
\end{align}
The remaining zero-mode part can be further simplified by a straightforward computation:
\begin{align}
    \langle 0,0|e^{\frac{i}{2}\sum_{j=1}^{2n}\epsilon_j (x_0-\bar{x}_0)}e^{\sum_{j=1}^{2n}\epsilon_j (\ln w_j a_0-\ln \bar{w}_j\bar{a}_0)/2} |\mathcal{O}^0_\pm\rangle &= \langle 0,-\frac{1}{2}\sum_{j=1}^{2n}\epsilon_j | e^{\sum_{j=1}^{2n}\epsilon_j (\ln w_j a_0-\ln \bar{w}_j\bar{a}_0)/2} |\mathcal{O}^0_\pm\rangle \nonumber\\
    &= \langle 0,-\frac{1}{2}\sum_{j=1}^{2n}\epsilon_j |\mathcal{O}^0_\pm\rangle\, \exp\left[-\frac{1}{4}\sum_{j=1}^{2n}\epsilon_j \, \sum_{l=1}^{2n}\epsilon_l \ln (w_l\bar{w}_l)\right]\,,
    \label{eq:sigma-orbifold-correlator-SM}
\end{align}
where we have used $e^{i\epsilon (x_0-\bar{x}_0)}|0,0\rangle = |0,m=\epsilon\rangle$. By using
\begin{align}
    \langle 0,-\frac{1}{2}\sum_{j=1}^{2n}\epsilon_j |\mathcal{O}^0_+\rangle = \delta_{\sum_{j=1}^{2n}\epsilon_j,0}\,,\quad
    \langle 0,-\frac{1}{2}\sum_{j=1}^{2n}\epsilon_j |\mathcal{O}^0_-\rangle = \sum_{m\in\mathbb{Z}} (-1)^m \delta_{\sum_{j=1}^{2n}\epsilon_j, 4m}\,,
\end{align}
Eq.~\eqref{eq:sigma-orbifold-correlator-SM} reduces to
\begin{align}
    \langle 0,-\frac{1}{2}\sum_{j=1}^{2n}\epsilon_j |\mathcal{O}^0_+\rangle\cdot \exp\left[-\frac{1}{4}\sum_{j=1}^{2n}\epsilon_j \cdot \sum_{l=1}^{2n}\epsilon_l \ln (w_l\bar{w}_l)\right] &= \delta_{\sum_{j=1}^{2n}\epsilon_j,0}\,,\nonumber\\
    \langle 0,-\frac{1}{2}\sum_{j=1}^{2n}\epsilon_j |\mathcal{O}^0_-\rangle\cdot \exp\left[-\frac{1}{4}\sum_{j=1}^{2n}\epsilon_j \cdot \sum_{l=1}^{2n}\epsilon_l \ln (w_l\bar{w}_l)\right] &= \sum_{m\in\mathbb{Z}} (-1)^m \delta_{\sum_{j=1}^{2n}\epsilon_j, 4m} \cdot \exp\left[-m \sum_{l=1}^{2n}\epsilon_l \ln (w_l\bar{w}_l)\right]\,.
\end{align}

Therefore, the bosonic crosscap correlators are given by
\begin{align}
    \langle 0,0|\prod_{j=1}^{2n} \cos\frac{\vartheta }{2}(w_j,\bar{w}_j)|\mathcal{O}_+ \rangle &=\langle 0,0|\prod_{j=1}^{2n} \cos\frac{\vartheta }{2}(w_j,\bar{w}_j)\; e^{ K}|\mathcal{O}^0_+ \rangle\nonumber\\
    &= \frac{1}{2^{2n}}\sum_{\substack{\epsilon_j =\pm 1,\\ j=1,\ldots,2n }}  \delta_{\sum_{j=1}^{2n}\epsilon_j,0}\prod_{1\leq j < l\leq 2n} |w_j-w_l|^{\epsilon_j\epsilon_l/2}\prod_{1\leq j,l\leq 2n}\left(1+\frac{1}{w_j\bar{w}_l}\right)^{ \epsilon_j\epsilon_l/4} \nonumber\\
    &= \frac{1}{2^{2n}}\sum_{\substack{\epsilon_j =\pm 1,\\ j=1,\ldots,2n }}  \delta_{\sum_{j=1}^{2n}\epsilon_j,0}\prod_{j=1}^{2n}\left(1+\frac{1}{w_j\bar{w}_j}\right)^{1/4}\prod_{1\leq j < l\leq 2n} \left|(w_j-w_l)\cdot\left(1+\frac{1}{w_j\bar{w}_l}\right)\right|^{\epsilon_j\epsilon_l/2}\,,
\end{align}
and
\begin{align}
    \langle 0,0|\prod_{j=1}^{2n} \cos\frac{\vartheta }{2}(w_j,\bar{w}_j)|\mathcal{O}_- \rangle &=\langle 0,0|\prod_{j=1}^{2n} \cos\frac{\vartheta }{2}(w_j,\bar{w}_j)\; e^{- K}|\mathcal{O}^0_- \rangle \nonumber\\
    &= \frac{1}{2^{2n}}\sum_{m\in\mathbb{Z}}\sum_{\substack{\epsilon_j =\pm 1,\\ j=1,\ldots,2n }} (-1)^m \delta_{\sum_{j=1}^{2n}\epsilon_j, 4m} \nonumber\\
    &\phantom{=} \;\times \prod_{j=1}^{2n} |w_j|^{-2m\epsilon_j}\prod_{1\leq j < l\leq 2n} |w_j-w_l|^{\epsilon_j\epsilon_l/2} \prod_{1\leq j, l\leq 2n}\left(1+\frac{1}{w_j\bar{w}_l}\right)^{- \epsilon_j\epsilon_l/4}\nonumber\\
    &= \frac{1}{2^{2n}}\sum_{m\in\mathbb{Z}}\sum_{\substack{\epsilon_j =\pm 1,\\ j=1,\ldots,2n }} (-1)^m \delta_{\sum_{j=1}^{2n}\epsilon_j, 4m} \nonumber\\
    &\phantom{=} \; \times \prod_{j=1}^{2n} \left[|w_j|^{-2m\epsilon_j} \left(1+\frac{1}{w_j\bar{w}_j}\right)^{-1/4}\right] \prod_{1\leq j < l\leq 2n} \left|\frac{1}{w_j-w_l}\cdot\left(1+\frac{1}{w_j\bar{w}_l}\right)\right|^{-\epsilon_j\epsilon_l/2}\,.
\end{align}
The general formula for the $2n$-point crosscap correlator of the $\sigma$ field is obtained by inserting the above expressions into Eq.~\eqref{eq:cross-sigma-Ishibashi-SM}. 

As an example, let us consider the two-point crosscap correlator. Starting with
\begin{align}
    \langle 0,0|\cos\frac{\vartheta }{2}(w_1,\bar{w}_1)\cos\frac{\vartheta }{2}(w_2,\bar{w}_2)|\mathcal{O}_+ \rangle &= \frac{1}{4}\cdot 2\prod_{j=1,2}\left(1+\frac{1}{w_j\bar{w}_j}\right)^{1/4} \cdot\left|(w_1-w_2)\cdot\left(1+\frac{1}{w_1\bar{w}_2}\right)\right|^{-1/2}\nonumber\\
    &= \frac{1}{2}\frac{1}{|w_1-w_2|^{1/2}} \left[\frac{|1+w_1\bar{w}_2|^2}{(1+|w_1|^2)(1+|w_2|^2)}\right]^{- 1/4}\nonumber\\
    &= \frac{1}{2}\frac{1}{|w_1-w_2|^{1/2}} \cdot (1-\eta)^{-1/4}\,,
\end{align}
and
\begin{align}
    \langle 0,0|\cos\frac{\vartheta }{2}(w_1,\bar{w}_1)\cos\frac{\vartheta }{2}(w_2,\bar{w}_2)|\mathcal{O}_- \rangle &= \frac{1}{4}\cdot 2\prod_{j=1,2}\left(1+\frac{1}{w_j\bar{w}_j}\right)^{-1/4} \cdot\left|\frac{1}{w_1-w_2}\cdot\left(1+\frac{1}{w_1\bar{w}_2}\right)\right|^{1/2}\nonumber\\
    &= \frac{1}{2}\frac{1}{|w_1-w_2|^{1/2}} \cdot (1-\eta)^{1/4}\,,
\end{align}
we find that the crosscap correlators in two Ishibashi sectors are given by
\begin{align}
    {}_{\mathrm{NS}}\langle 0| \sigma (w_1,\bar{w}_1)\sigma (w_2,\bar{w}_2) |\mathbbm{1}\rangle\rangle_\mathcal{C} 
    &= \frac{1}{|w_1-w_2|^{1/4}}\sqrt{\frac{1}{2}\left[(1-\eta)^{-1/4} + (1-\eta)^{1/4}\right]} \nonumber\\
    &= \frac{1}{|w_1-w_2|^{1/4}} \cdot \frac{1}{\sqrt{2}}\cdot (1-\eta)^{-1/8}\cdot \sqrt{1 +\sqrt{1-\eta}}\,,\nonumber\\
    {}_{\mathrm{NS}}\langle 0| \sigma (w_1,\bar{w}_1)\sigma (w_2,\bar{w}_2) |\varepsilon\rangle\rangle_\mathcal{C} &= \frac{1}{|w_1-w_2|^{1/4}}\sqrt{\frac{1}{2}\left[(1-\eta)^{-1/4} - (1-\eta)^{1/4}\right]} \nonumber\\
    &= \frac{1}{|w_1-w_2|^{1/4}} \cdot \frac{1}{\sqrt{2}}\cdot (1-\eta)^{-1/8}\cdot \sqrt{1 -\sqrt{1-\eta}} \, ,
\end{align}
so the two-point crosscap correlator of the $\sigma$ field reads
\begin{align}
    {}_{\mathrm{NS}}\langle 0| \sigma (w_1,\bar{w}_1)\sigma (w_2,\bar{w}_2) |\mathcal{C}_\pm\rangle 
    &= \sqrt{\frac{2+\sqrt{2}}{2}}\frac{1}{|w_1-w_2|^{1/4}}\, (1-\eta)^{-1/8}\left[\frac{\sqrt{2}}{2}\sqrt{1+\sqrt{1-\eta}} \pm \frac{2-\sqrt{2}}{2} \sqrt{1-\sqrt{1-\eta}}\right] \, ,
\label{eq:cross-correlator-sigma-SM}
\end{align}
where $\eta$ is the crosscap cross ratio [Eq.~\eqref{eq:cross-ratio-SM}]. This corresponds to Eq.~(14) in the main text.

\section{VI. Conformal perturbation theory for the crosscap overlap}

The conformal perturbation theory for computing \emph{crosscap overlap} is given in Eqs.~(16) and (17) in the main text. Here, we provide a simple derivation and then formulate the perturbation expansion for a general crosscap state $|\mathcal{C}\rangle$ up to second order. The application to the perturbed Ising CFT is specifically discussed.

\subsection{A. Formal perturbation series}

We consider a CFT on a circle with length $L$ with a relevant perturbation: $H = H_0 + H_1$, where $H_0$ is the CFT hamiltonian and $H_1= -g\int_0^L \varphi (x)$ is the perturbation term, $\varphi (x)$ is a primary field with conformal weight $h=\bar{h}<1$. Denoting the perturbed gorund state as $|\psi_0 (s)\rangle$, parameterized by the dimensionless coupling $s= gL^{2-2h}$, our aim is to formulate the conformal perturbation theory~\cite{Saleur1987} for the universal crosscap overlap $\langle \psi_0 (s)|\mathcal{C}\rangle$. For practical calculation, we split the crosscap overlap into two parts:
\begin{align}
    \langle \psi_0 (s)|\mathcal{C}\rangle = \frac{\langle \psi_0 (s)|\mathcal{C}\rangle}{\langle\psi_0 (0)|\psi_0(s)\rangle}\cdot \langle\psi_0 (0)|\psi_0(s)\rangle \equiv Z(s)\cdot \exp\left[\frac{1}{2}W(s)\right]\,,
\end{align}
where $|\psi_0(0)\rangle$ is the unperturbed ground state. Both the universal scaling functions $Z(s)$ and $W(s)$ can be formulated the formal perturbation series.

For $Z(s)$, we have
\begin{align}
    Z(s) = \lim_{\beta\to \infty} \frac{\langle \psi_0 (0)|e^{-\beta H}|\mathcal{C}\rangle}{\langle \psi_0 (0)|e^{-\beta H}|\psi_0(0)\rangle} = \lim_{\beta\to \infty}\frac{\langle \psi_0(0)|\mathcal{T} e^{-\int_0^\beta \mathrm{d}\tau \, H_1(\tau)}|\mathcal{C}\rangle}{\langle \psi_0(0)|\mathcal{T} e^{-\int_0^\beta \mathrm{d}\tau  \, H_1(\tau)} |\psi_0(0)\rangle} \,,
    \label{eq:Z-formal-series-SM}
\end{align}
where the unique perturbed ground state $|\psi_0(s)\rangle$ for finite $L$ is projected out in the limit $\beta \to\infty$. In the second equality, we work in the interaction picture via 
\begin{align}
    e^{-\beta H}=e^{-\beta H_0}\mathcal{T}e^{-\int_0^\beta \mathrm{d}\tau H_1(\tau)}\,,
\end{align}
where $H_1 (\tau)= e^{\tau H_0}H_1e^{-\tau H_0}$ is the perturbation term in the interaction picture.

For $W(s) = \ln |\langle\psi_0 (0)|\psi_0(s)\rangle|^2$, we can consider
\begin{align}
    \langle\psi_0(0)|e^{-\beta H}|\psi_0 (0)\rangle = e^{-\beta E_0 (s,L)}\left[ \exp ( W(s) ) + \mathcal{O}(e^{-\beta \Delta E (s,L)})\right]\,,\quad \beta \to \infty\,,
    \label{eq:W-formal-series-SM}
\end{align}
where we expect that the perturbed ground-state energy scales as $E_0 = -\frac{\pi}{6L}c(s)$ and the (finite-size) energy gap scales as $\Delta E (s,L) \sim \frac{1}{L}$, with $s$ fixed. Here, $c(s)$, with $c(0)=c$, is the ``running'' central charge, denoting the deformed central charge under the perturbation. As $\beta \to \infty$, the contribution of the excited states decays exponentially, leaving only the ground-state contribution. Therefore, in the interaction picture, we can extract $W(s)$ as follows:
\begin{align}
    W(s) =\lim_{\beta\to\infty} \left[\ln \langle \psi_0(0)|e^{\beta H_0}e^{-\beta H}|\psi_0(0)\rangle -\frac{\pi\beta}{6L} \delta c(s) \right] =\lim_{\beta\to\infty}\left[\left\langle \mathcal{T} e^{-\int_0^\beta \mathrm{d}\tau \, H_1(\tau)}\right\rangle_\mathrm{c} -\frac{\pi\beta}{6L} \delta c(s) \right]\,,
\end{align}
where $\langle \mathcal{T}e^{-\int_0^\beta \mathrm{d}\tau \, H_1(\tau)}\rangle_\mathrm{c}$ denotes the \emph{connected} contribution of $\langle \psi_0(0)|\mathcal{T}e^{-\int_0^\beta \mathrm{d}\tau \, H_1(\tau)}|\psi_0(0)\rangle$, via the linked cluster theorem: $\ln\langle \psi_0(0)|\mathcal{T}e^{-\int_0^\beta \mathrm{d}\tau \, H_1(\tau)}|\psi_0(0)\rangle = \langle \mathcal{T}e^{-\int_0^\beta \mathrm{d}\tau \, H_1(\tau)}\rangle_\mathrm{c}$, and 
\begin{align}
    \delta c(s)= c(s) - c = \lim_{\beta\to \infty} \frac{6L}{\pi\beta}\langle \mathcal{T}e^{-\int_0^\beta \mathrm{d}\tau \, H_1(\tau)}\rangle_{\mathrm{c}}
\end{align}
is the change of ``running'' central charge.

The formal perturbation series can be calculated by expanding the time-ordered exponential $\mathcal{T}e^{-\int_0^\beta \mathrm{d}\tau \, H_1(\tau)}$ in powers of the coupling $g$. Specifically, for $W(s)$, we expand $\langle \psi_0(0)|\mathcal{T}e^{-\int_0^\beta \mathrm{d}\tau \, H_1(\tau)}|\psi_0(0)\rangle$ as
\begin{align}
    \lim_{\beta\to\infty} \langle \psi_0(0)|\mathcal{T}e^{-\int_0^\beta \mathrm{d}\tau \, H_1(\tau)}|\psi_0(0)\rangle_{\mathrm{c}} &= \sum_{n=1}^\infty \frac{1}{(2n)!} \int_0^\infty \mathrm{d}\tau_1 \cdots \mathrm{d}\tau_{2n} \, \langle H_1(\tau_1)\cdots H_1(\tau_{2n})\rangle_{\mathrm{c}}\nonumber\\
    &\equiv \sum_{n=1}^\infty \left[ \frac{2\beta}{6 L} C_{2n} + W_{2n}\right]\cdot s^{2n}\,,
\end{align}
where each order of the perturbation series can be split into a divergent term $\frac{2\beta}{6 L} C_{2n}$, scaled as $\frac{\beta}{L}$ when $\beta\to \infty$, and a subleading finite term $W_{2n}$. The series expansion of $ W(s)$ is obtained as $W(s) = \sum_{n=1}^\infty W_{2n} s^{2n}$ and the ``running'' central charge is obtained as $c(s) = c + \sum_{n=1}^\infty C_{2n} s^{2n}$.

\subsection{B. Perturbation correction up to second order}

We perform the perturbation expansion up to the second order to illustrate how the conformal perturbation theory works.

A crosscap state $|\mathcal{C}\rangle$ can be characterized by the one-point conformal crosscap correlator of primary fields $\varphi_a$, 
\begin{align}
    \langle \psi_0 (0)|\varphi_a (w,\bar{w})|\mathcal{C}\rangle = \frac{\mathcal{A}_a}{(1+|w|^2)^{2h_a}}\,,
    \label{eq:cross-correlator-1-SM}
\end{align}
where $\mathcal{A}_a$ is the crosscap coefficient. The overall phase of the crosscap state can always be fixed by requiring the coefficient associated with the identity operator to be positive, $\mathcal{A}_0 > 0$.  Furthermore, the two-point conformal crosscap correlator can be determined up to an unknown prefactor $G_a(\eta)$ as a function of the crosscap cross ratio~\cite{Fioravanti1994,Nakayama2016a,Nakayama2016b}:
\begin{align}
    \langle \psi_0 (0)|\varphi_a (w_1,\bar{w}_1)\varphi_a (w_2,\bar{w}_2)|\mathcal{C}\rangle = \frac{\mathcal{A}_0\cdot G_a(\eta)}{|w_1-w_2|^{4h_a}}\,,
    \label{eq:cross-correlator-2-SM}
\end{align}
where the crosscap correlators of the $\varepsilon$ and $\sigma$ fields given in Eqs.~\eqref{eq:cross-correlator-epsilon-1-SM}, ~\eqref{eq:cross-correlator-epsilon-2-SM} and ~\eqref{eq:cross-correlator-sigma-SM} are the specific cases.

We consider the CFT perturbed by the primary field $\varphi_a$, with $h_\alpha = \bar{h}_\alpha <1$, 
\begin{align}
    H = H_0 + H_1 \,,\quad H_1 = -g\int_0^L \mathrm{d}x \, \varphi_a(x)\,,
\end{align}
in which the first-order perturbation of the crosscap overlap is non-vanished if $\mathcal{A}_a \neq 0$. Up to the second order, we denote the perturbative expansion as
\begin{align}
    Z(s) &= Z_0 + Z_{1} \cdot s + Z_{2} \cdot s^2 + \mathcal{O} (s^3)\,,\nonumber\\
    W(s) &= W_{2}\cdot s^2 + \mathcal{O} (s^4)
\end{align}
with $Z_0 = \mathcal{A}_0$ by definition. Then, the perturbative expansion of the crosscap overlap up to second order is given by
\begin{align}
    \langle\psi_0 (s)|\mathcal{C}\rangle = \mathcal{A}_0 + Z_{1}\cdot s + \left(Z_{2} + \frac{1}{2}\mathcal{A}_0\, W_{2}\right) \cdot s^2 +\mathcal{O}(s^3)\,.
\end{align}

For $Z(s)$, expanding Eq.~\eqref{eq:Z-formal-series-SM} up to the second order, we have
\begin{align}
    Z_{1} \cdot s = -\int_0^\infty \mathrm{d}\tau \langle\psi_0(0)|H_1(\tau)|\mathcal{C}\rangle = g\int_0^\infty \mathrm{d}\tau \int_0^L \mathrm{d}x \, \langle\psi_0(0)|\varphi_a (z,\bar{z})|\mathcal{C}\rangle \, ,
\label{eq:Z1-SM}
\end{align}
and
\begin{align}
    Z_{2} \cdot s^2 &= \frac{1}{2}\int_0^\infty \mathrm{d}\tau_1 \int_0^\infty \mathrm{d}\tau_2 \left[\langle\psi_0(0)|H_1(\tau_1)H_1(\tau_2)|\mathcal{C}\rangle - \mathcal{A}_0 \langle\psi_0(0)|H_1(\tau_1)H_1(\tau_2)|\psi_0 (0)\rangle\right]\nonumber\\
    &= \frac{g^2}{2}\int_0^\infty \mathrm{d}\tau_1 \int_0^L \mathrm{d}x_1\int_0^\infty \mathrm{d}\tau_2 \int_0^L \mathrm{d}x_2 \left[\langle \psi_0 (0)|\varphi_a (z_1,\bar{z}_1)\varphi_a (z_2,\bar{z}_2)|\mathcal{C}\rangle - \mathcal{A}_0 \langle \psi_0 (0)|\varphi_a (z_1,\bar{z}_1)\varphi_a (z_2,\bar{z}_2)|\psi_0(0)\rangle\right] \, ,
\label{eq:Z2-SM}
\end{align}
where $z = \tau + ix$ is the complex coordinate on the cylinder.

Considering the conformal transformation $w=e^{\frac{2\pi}{L}z}$, the primary field $\varphi_a$ transforms as
\begin{align}
    \varphi_a (z,\bar{z})= \left(\frac{2\pi |w|}{L}\right)^{2h_a} \varphi_a (w,\bar{w})\,,
\end{align}
and the space-time integral in Eqs.~\eqref{eq:Z1-SM} and \eqref{eq:Z2-SM} can be transformed into integrals on the $\mathbb{RP}^2$ manifold. 

For $Z_1$, we have
\begin{align}
    Z_1 \cdot s &= g \left(\frac{2\pi}{L}\right)^{2h_a} \int_0^\infty \mathrm{d}\tau \int_0^L \mathrm{d}x \, |w|^{2h_a}\langle\psi_0(0)|\varphi_a (w,\bar{w})|\mathcal{C}\rangle \nonumber\\
    &= g \left(\frac{2\pi}{L}\right)^{2h_a} \mathcal{A}_a\int_0^\infty \mathrm{d}\tau \int_0^L \mathrm{d}x \, \frac{|w|^{2h_a}}{(1+|w|^2)^{2h_a}} \, ,
\end{align}
and further changing the integral variables $(\rho, \theta) = (e^{-\frac{2\pi}{L}\tau},\frac{2\pi x}{L})$, we obtain
\begin{align}
    Z_1 \cdot s &= g \left(\frac{L}{2\pi}\right)^{2-2h_a}\mathcal{A}_a \cdot 2\pi \int_0^1 \mathrm{d}\rho \frac{\rho^{2h_a -1}}{(1+\rho^2)^{2h_a}}\nonumber\\
    &= \left[(2\pi)^{2h_a -1} \mathcal{A}_a\cdot \frac{_2F_1(h_a,2h_a,h_a +1;-1)}{2h_a} \right]\cdot s\,,
\end{align}
where $s= gL^{2-2h_a}$ and $_2F_1(a,b,c;x)$ is the hypergeomertic function.

For $Z_2$, the calculation is similar:
\begin{align}
    Z_2 \cdot s^2 &= \frac{g^2}{2} \left(\frac{2\pi}{L}\right)^{4h_a} \mathcal{A}_0 \int_0^\infty \mathrm{d}\tau_1 \int_0^L \mathrm{d}x_1\int_0^\infty \mathrm{d}\tau_2 \int_0^L \mathrm{d}x_2 \frac{|w_1w_2|^{2h_a}}{|w_1-w_2|^{4h_a}} \left(G_a(\eta) -1\right) \nonumber\\
    &= \left[\frac{\mathcal{A}_0}{2  (2\pi)^{3-4h_a}} \int_{0}^1 \mathrm{d}\rho_1 \int_{0}^1 \mathrm{d}\rho_2\int_0^{2\pi}\mathrm{d}\theta \frac{(\rho_1\rho_2)^{2h_a-1}}{|\rho_1 -\rho_2 e^{i\theta}|^{4h_a}} \left(G_a(\eta)-1\right)\right]\cdot s^2\,,
\end{align}
where the new integral variables read $(\rho_1,\rho_2,\theta) = (e^{-\frac{2\pi}{L}\tau_1},e^{-\frac{2\pi}{L}\tau_2},\frac{2\pi (x_1-x_2)}{L} )$. In the above derivation, the one-point and two-point crosscap correlators [Eqs.~\eqref{eq:cross-correlator-1-SM} and \eqref{eq:cross-correlator-2-SM}] and the two-point plane correlator have been used. The integral for $Z_2$ is convergent, and we expect that every order of the perturbation expansion of $Z(s)$ is convergent, since the divergent \emph{bulk} contributions from the numerator $\langle \psi_0(0)|\mathcal{T} e^{-\int_0^\beta \mathrm{d}\tau  H_1(\tau)}|\mathcal{C}\rangle$ and the denominator $\langle \psi_0(0)|\mathcal{T} e^{-\int_0^\beta \mathrm{d}\tau  H_1(\tau)}|\psi_0(0)\rangle$ cancel at each order. In fact, according to the fusion rule $\varphi_a\varphi_a \sim \mathbbm{1} + \cdots$, we should have $G_a(\eta \to 0) \to 1$ and thus $G_a(\eta \to 0) -1 \to 0$ as $|w_1-w_2|\to 0$, which cancels the divergent \emph{bulk} contribution in the integral.

Next, we consider the perturbation of $W(s)$. We have
\begin{align}
    \lim_{\beta\to \infty}\langle \psi_0(0)|\mathcal{T}e^{-\int_0^\beta \mathrm{d}\tau \, H_1(\tau)}|\psi_0(0)\rangle_{\mathrm{c}} = \lim_{\beta\to \infty} \int_0^\beta \mathrm{d}\tau_1 \int_0^{\tau_1} \mathrm{d}\tau_2 \;\langle \psi_0(0)|H_1(\tau_1)H_1(\tau_2)|\psi_0(0)\rangle +\mathcal{O}(g^4) \, ,
\end{align}
where the leading correction appears at the second order:
\begin{align}
     &\quad\lim_{\beta\to \infty} \int_0^\beta \mathrm{d}\tau_1 \int_0^{\tau_1} \mathrm{d}\tau_2 \;\langle\psi_0(0)|H_1(\tau_1)H_1(\tau_2)|\psi_0(0)\rangle \nonumber\\
     &= g^2 \left(\frac{2\pi}{L}\right)^{4h_a} \lim_{\beta\to \infty} \int_0^\beta \mathrm{d}\tau_1 \int_0^L \mathrm{d}x_1\int_0^{\tau_1} \mathrm{d}\tau_2 \int_0^L \mathrm{d}x_2 \frac{|w_1w_2|^{2h_a}}{|w_1-w_2|^{4h_a}}\nonumber\\
    &= g^2\left(\frac{L}{2\pi}\right)^{4-4h_a } \lim_{\rho\to 0} 2\pi\int_{\rho}^1 \mathrm{d}\rho_1 \int_{\rho_1}^1 \mathrm{d}\rho_2\int_0^{2\pi}\mathrm{d}\theta \frac{(\rho_1\rho_2)^{2h_a-1}}{|\rho_1 -\rho_2 e^{i\theta}|^{4h_a}} \nonumber\\
    &=\left[(2\pi)^{4h_a-3}\lim_{\rho\to 0} \int_\rho^1 \frac{\mathrm{d}\rho_1}{\rho_1^{1-2h_a}} \int_{\rho_1}^1 \frac{\mathrm{d}\rho_2}{\rho_2^{1+2h_a}} \int_0^{2\pi} \frac{\mathrm{d}\theta}{\left|1-\frac{\rho_1}{\rho_2}e^{-i\theta}\right|^{4h_a}}\right]\cdot s^2
\end{align}
with $\rho = e^{-\frac{2\pi\beta}{L}} \to 0$ as $\beta \to \infty$. To separate out the contribution of the ``running'' central charge $c(s)$ and $W(s)$ from the above integral, we expand the angular part of the integral as a power series~\cite{Saleur1987}:
\begin{align}
    \int_0^{2\pi} \frac{\mathrm{d}\theta}{\left|1-\frac{\rho_1}{\rho_2}e^{-i\theta}\right|^{4h_a}} &= \int_0^{2\pi} \mathrm{d}\theta \left(1-\frac{\rho_1}{\rho_2}e^{-i\theta}\right)^{-2h_a}\left(1-\frac{\rho_1}{\rho_2}e^{i\theta}\right)^{-2h_a}\nonumber\\
    &=\sum_{n,m=0}^\infty \left(\frac{\rho_1}{\rho_2}\right)^{n+m} \frac{\Gamma (n+2h_a)}{n!\Gamma (2h_a)} \frac{\Gamma (m+2h_a)}{m!\Gamma (2h_a)}\int_0^{2\pi} \mathrm{d}\theta \, e^{-in\theta} e^{im\theta}\nonumber\\
    &= 2\pi\sum_{n=0}^\infty \left[\frac{\Gamma (n+2h_a)}{n!\Gamma (2h_a)}\right]^2 \cdot \left(\frac{\rho_1}{\rho_2}\right)^{2n}\,,
\end{align}
where $\Gamma (x)$ is the gamma function. Then, the radial integrals of $\rho_1$ and $\rho_2$ can be integrated term by term,
\begin{align}
    &\phantom{=} \; 2\pi \sum_{n=0}^\infty \left[\frac{\Gamma (n+2h_a)}{n!\Gamma (2h_a)}\right]^2 \lim_{\rho\to 0}\int_\rho^1 \frac{\mathrm{d}\rho_1}{\rho_1^{1-2h_a}} \int_{\rho_1}^1 \frac{\mathrm{d}\rho_2}{\rho_2^{1+2h_a}}  \left(\frac{\rho_1}{\rho_2}\right)^{2n}   \nonumber \\
    &= 2\pi \sum_{n=0}^\infty \left[\frac{\Gamma (n+2h_a)}{n!\Gamma (2h_a)}\right]^2 \frac{1}{2n+2h_a}\int_\rho^1 \frac{\mathrm{d}x}{x}\left(1-x^{2n+2h_a}\right) \nonumber \\
    &= 2\pi \sum_{n=0}^\infty \frac{1}{2n+2h_a}\left[\frac{\Gamma (n+2h_a)}{n!\Gamma (2h_a)}\right]^2 \, \lim_{\rho\to 0}\left[-\ln\rho - \frac{1-\rho^{2n+2h_a}}{2n+2h_a}\right]\,,
\end{align}
from which we separated out the second-order contribution of the ``running'' central charge, which is proportional to $-\ln\rho = \frac{2\pi\beta}{L}$, and the second-order contribution of $W(s)$:
\begin{align}
    W_2 &= (2\pi)^{4h_a-3} \cdot 2\pi \sum_{n=0}^\infty \frac{1}{2n+2h_a}\left[\frac{\Gamma (n+2h_a)}{n!\Gamma (2h_a)}\right]^2 \lim_{\rho\to 0}\left( - \frac{1-\rho^{2n+2h_a}}{2n+2h_a}\right)\nonumber\\
    &= -(2\pi)^{4h_a-2}\sum_{n=0}^\infty \frac{1}{(2n+2h_a)^2}\left[\frac{\Gamma (n+2h_a)}{n!\Gamma (2h_a)}\right]^2\,.
\end{align}

In summary, up to the second order, the perturbation expansion of the crosscap overlap is given by
\begin{align}
    \langle\psi_0(s)|\mathcal{C}\rangle 
    &= \mathcal{A}_0 + \mathcal{A}_a\cdot \frac{_2F_1(h_a,2h_a,h_a +1;-1)}{2(2\pi)^{1-2h_a} h_a} \cdot s \nonumber\\
    &+\frac{\mathcal{A}_0}{2(2\pi)^{2-4h_a}}\left[\frac{1}{2\pi} \int_{0}^1 \mathrm{d}\rho_1 \int_{0}^1 \mathrm{d}\rho_2\int_0^{2\pi}\mathrm{d}\theta \frac{(\rho_1\rho_2)^{2h_a-1}}{|\rho_1 -\rho_2 e^{i\theta}|^{4h_a}} \left(G_a(\eta)-1\right) -\sum_{n=0}^\infty \frac{1}{(2n+2h_a)^2}\left(\frac{\Gamma (n+2h_a)}{n!\Gamma (2h_a)}\right)^2\right]\cdot s^2 \nonumber\\
    &+\mathcal{O}(s^3) \, ,
\label{eq:cross-ovlp-2nd-perturb-SM}
\end{align}
where $\mathcal{A}_a$ is the crosscap coefficient associated with the primary field $\varphi_a$, as defined in Eq.~\eqref{eq:cross-correlator-1-SM}.

In rational CFTs, a crosscap state $|\mathcal{C}\rangle$ can be expressed as a linear combination of crosscap Ishibashi states:
\begin{align}
|\mathcal{C}\rangle = \sum_{a} \mathcal{A}_a \, |a\rangle\rangle_{\mathcal{C}}\, ,
\end{align}
where the expansion coefficients are precisely the crosscap coefficients. The absolute values of these coefficients can be determined from the loop channel-tree channel correspondence between the Klein bottle partition function and the cylinder partition function with crosscap boundaries. In particular, for rational CFTs with charge conjugation modular invariants, one can always construct the PSS crosscap state $|\mathcal{C}_0\rangle$ associated with the identity field, whose crosscap coefficients are 
\begin{align}
    \mathcal{A}_a = \frac{P_{0a}}{\sqrt{S_{0a}}}\, ,
    \label{eq:cross-coefficient-PSS-SM}
\end{align}
as shown in Eq.~\eqref{eq:gPSS-RCFT-SM}.

The conformal perturbation theory provides a perturbative approach to investigate the conjectured monotonicity of the Klein bottle entropy (norm-square of the crosscap overlap) under the renormalization flow~\cite{ZhangYS2023}. Specifically, the first-order correction of the crosscap overlap in Eq.~\eqref{eq:cross-ovlp-2nd-perturb-SM} takes a remarkably simple form. 
From the series expansion of the hypergeometric function,
\begin{align}
    _2F_1(h_a,2h_a,h_a+1;-1) = \sum_{n=0}^\infty \frac{(2h_a)_{2n}}{(2n)!}\frac{h_a}{2n+h_a}\left(1-\frac{2n+2h_a}{2n+1+h_a}\frac{2n+h_a}{2n+1}\right)\, ,
\end{align}
where $(q)_n = q(q+1)\cdots (q+n-1)$ is the Pochhammer symbol, one sees that $_2F_1(h_a,2h_a,h_a +1;-1)$ is strictly positive for $h_a <1$.  Consequently, whenever the first-order correction is non-vanishing, its sign is determined solely by the crosscap coefficient $\mathcal{A}_a$. 

\subsection{C. Perturbed $A$-series unitary minimal models}

We apply the conformal perturbation theory to the $A$-series unitary minimal models, with central charge $c= 1-\frac{6}{m(m+1)}$, $m\geq 3$. The primary fields are labeled by integer pairs $(r,s)$ within the conformal grid $1 \leq r \leq m-1$, $1 \leq s \leq m$, subject to the condition that $r+s$ is even to avoid double counting. 

The PSS crosscap state associated with the identity field $(1,1)$ can be expressed in terms of the Ishibashi states as
\begin{align}
    |\mathcal{C}_{(1,1)}\rangle = \sum_{1\leq r\leq m-1\,, 1\leq s\leq m}^{r+s = \mathrm{even}} \mathcal{A}_{(r,s)} |(r,s)\rangle\rangle_{\mathcal{C}} \, ,
\end{align}
where the crosscap coefficients $\mathcal{A}_{(r,s)}$ can be computed via Eq.~\eqref{eq:cross-coefficient-PSS-SM}.

The modular $\mathcal{S}$-matrix is
\begin{align}
    S_{(r,s),(r',s')} &= \delta^{(2)}_{r+s,0}\delta^{(2)}_{r'+s',0}\sqrt{\frac{8}{m(m+1)}} \sin \left(\frac{\pi r r'}{m}\right) \sin\left(\frac{\pi s s'}{m+1}\right)\, ,
\end{align}
and the modular $\mathcal{T}$-matrix is determined via the conformal weights of the primary fields
\begin{align}
    h_{(r,s)} = \frac{[r(m+1)-sm]^2-1}{4m(m+1)}\, .
\end{align}
Using Eq.~\eqref{eq:P-matrix-SM}, one can calculate the associated $P$-matrix as~\cite{Bianchi1991,Pradisi1995},
\begin{align}
    &\quad P_{(r,s),(r',s')} \nonumber\\
    &= \frac{4}{\sqrt{m(m+1)}} \left[\delta^{(2)}_{r,0}\delta^{(2)}_{s,0}\delta^{(2)}_{r',0}\delta^{(2)}_{s',0}
    \sin \left(\frac{\pi r r'}{2m}\right) \sin \left(\frac{\pi s s'}{2(m+1)}\right)
    + \delta^{(2)}_{r,1}\delta^{(2)}_{s,1}\delta^{(2)}_{r',1}\delta^{(2)}_{s',1}
    \cos \left(\frac{\pi r r'}{2m}\right) \cos \left(\frac{\pi s s'}{2(m+1)}\right)
    \right]\, ,
\end{align}
where $\delta^{(2)}_{n,m} \equiv \frac{1+ (-1)^{n+m}}{2}$ is the Kronecker-$\delta$ function with periodic $2$. 

Therefore, the crosscap coefficients of the standard PSS state $|\mathcal{C}_{(1,1)}\rangle$ read
\begin{align}
    \mathcal{A}_{(r,s)} 
    = \frac{P_{(1,1),(r,s)}}{\sqrt{S_{(1,1),(r,s)}}}
    = \delta^{(2)}_{r,1} \delta^{(2)}_{s,1} \sqrt{\frac{\sqrt{2}\cot \left(\frac{\pi r}{2m}\right) \cot \left(\frac{\pi s}{2(m+1)}\right)}{\sqrt{m(m+1)}}}\, ,
\end{align}
which are non-negative for all $(r,s)$. This provides a perturbative proof of the monotonicity conjecture~\cite{ZhangYS2023} for the crosscap overlap (and hence Klein bottle entropy) under the renormalization group flow in $A$-series unitary minimal models whenever the crosscap coefficient of the perturbing field is non-vanishing.

\subsection{D. Ising CFT with thermal and magnetic perturbations}

For the Ising field theory, we have already obtained the crosscap correlators via the bosonziation technique. It is straightforward to apply these results in Eq.~\eqref{eq:cross-ovlp-2nd-perturb-SM}, where $|\mathcal{C}_\pm\rangle = \mathcal{A}_\mathbbm{1} |\mathbbm{1}\rangle\rangle_\mathcal{C} \pm \mathcal{A}_\varepsilon |\varepsilon\rangle\rangle_\mathcal{C}$ with $\mathcal{A}_{\mathbbm{1}} = \sqrt{\frac{2+\sqrt{2}}{2}}$ and $\mathcal{A}_\varepsilon = \sqrt{\frac{2-\sqrt{2}}{2}}$.

For the thermal perturbation ($h_\varepsilon = \bar{h}_\varepsilon = 1/2$), we have $G_\varepsilon (\eta) = 1+ \eta^2/(1-\eta )$ for the two-point crosscap correlator of the $\varepsilon$ field [Eq.~\eqref{eq:cross-correlator-epsilon-2-SM}], therefore, substituting it into Eq.~\eqref{eq:cross-ovlp-2nd-perturb-SM}, we obtain 
\begin{align}
    \langle\psi_0(s_1)|\mathcal{C}_\pm\rangle = \sqrt{\frac{2+\sqrt{2}}{2}}  \pm\sqrt{\frac{2-\sqrt{2}}{2}}\, \frac{\pi}{4}\cdot s_1-\sqrt{\frac{2+\sqrt{2}}{2}} \, \frac{\pi^2}{32}\cdot s_1^2 + \mathcal{O}(s_1^3)
\end{align}
with $s_1 = g_1 L$. This result is consistent with the exact solution in Eq.~\eqref{eq:Ising-crossovlp-thermal-SM}, as first reported in Ref.~\cite{ZhangYS2023}.

For the magnetic perturbation ($h_\sigma =\bar{h}_\sigma = 1/16$), since $\mathcal{A}_\sigma =0$, the leading contribution appears at the second order. The function $G_\pm^\sigma (\eta)$ for the two-point crosscap correlator of the $\sigma$ field [Eq.~\eqref{eq:cross-correlator-sigma-SM}] can be expanded in powers of $\eta$ as
\begin{align}
    G_\pm^\sigma (\eta) &= (1-\eta)^{-1/8}\left[\frac{\sqrt{2}}{2}\sqrt{1+\sqrt{1-\eta}} \pm \frac{2-\sqrt{2}}{2} \sqrt{1-\sqrt{1-\eta}}\right] \nonumber\\
    &= (1-\eta)^{-1/8}\left[1-\sum_{n=1}^\infty \frac{(4n-2)!(n-1)!}{[(2n-1)!]^2 n!}\left(\frac{\eta}{16}\right)^n \pm \frac{2-\sqrt{2}}{2}\sqrt{\frac{\eta}{2}}\sum_{n=0}^\infty \frac{(4n)!}{(2n+1)!(2n)!}\left(\frac{\eta}{16}\right)^n \right]\nonumber\\
    &=1 + \sum_{n=1}^\infty \frac{\Gamma (n+1/8)}{n!\Gamma (1/8)}\eta^n - \frac{1}{(1-\eta)^{1/8}}\left[\sum_{n=1}^\infty \frac{(4n-2)!(n-1)!}{[(2n-1)!]^2 n!}\left(\frac{\eta}{16}\right)^n \mp \frac{\sqrt{2}-1}{2}\sqrt{\eta}\sum_{n=0}^\infty \frac{(4n)!}{(2n+1)!(2n)!}\left(\frac{\eta}{16}\right)^n \right]\,,
\end{align}
which converges quickly as a power series in $\eta$. The expansion $G_\pm^\sigma (\eta) = (1-\eta)^{-2h_\sigma}(\sum_{n=0}^\infty a_n\eta^n \pm \eta^{h_\varepsilon} \sum_{n=0}^\infty b_n\eta^n)$ is precisely the conformal partial wave decomposition~\cite{Nakayama2016a,Nakayama2016b}, consistent with the fusion rule $\sigma\times\sigma \sim \mathbbm{1}+\varepsilon$. By choosing a suitable cutoff ($n_{\mathrm{max}} \approx 40$) to truncate the contributions from the descendant fields with large conformal weights, we can perform the numerical integration in Eq.~\eqref{eq:cross-ovlp-2nd-perturb-SM} with very high precision. The final results are given by
\begin{align}
    \langle\psi_0(s_2)|\mathcal{C}_+\rangle &=\sqrt{\frac{2+\sqrt{2}}{2}} -1.63528 \cdot s_2^2 +\mathcal{O}(s_2^4)\,,\nonumber\\
    \langle\psi_0(s_2)|\mathcal{C}_-\rangle &=\sqrt{\frac{2+\sqrt{2}}{2}} -1.71711 \cdot s_2^2 +\mathcal{O}(s_2^4)
\label{eq:Ising-2nd-SM}
\end{align}
with $s_2 = g_2 L^{15/8}$. 

Below we demonstrate this result numerically using the transverse-field Ising chain in the presence of a longitudinal field:
\begin{align}
H = -\sum_{j=1}^N \sigma^x_j\sigma^x_{j+1} - \sum_{j=1}^N\sigma^z_j - h_2 \sum_{j=1}^N\sigma^x_j \, ,
\label{eq:quantum-Ising-perturbed-SM}
\end{align}
where periodic boundary condition is adopted. For $h_2 = 0$, Ising CFT is realized with velocity $v=2$. The longitudinal field term (with coupling $h_2$) plays the role of the magnetic perturbation, and its normalization is given by $\mathcal{N}_\sigma=\lim_{r\rightarrow \infty}\lim_{N\rightarrow \infty} r^{1/4} \langle \sigma^x_j \sigma^x_{j+r} \rangle_{\mathrm{c}} \approx 0.645$~\cite{Pfeuty1970}. The dimensionless coupling for the lattice model~\eqref{eq:quantum-Ising-perturbed-SM} is $s_2 =\frac{\sqrt{\mathcal{N}_\sigma}}{v}h_2(v N)^{15/8}$~\cite{ZhangYS2023}. The DMRG results for the crosscap overlap between the ground state of Eq.~\eqref{eq:quantum-Ising-perturbed-SM} and the lattice crosscap state $|\mathcal{C}_{\mathrm{latt}}^+\rangle = \prod_{j=1}^{N/2} \left( |\uparrow\rangle_j |\uparrow\rangle_{j+N/2} + |\downarrow\rangle_j |\downarrow\rangle_{j+N/2}\right)$ are displayed in Fig.~\ref{fig:Figure1-SM}, which again show good agreement with the conformal perturbation theory result [Eq.~\eqref{eq:Ising-2nd-SM}].

\begin{figure}[ht]
\centering
\includegraphics[width=0.6\textwidth]{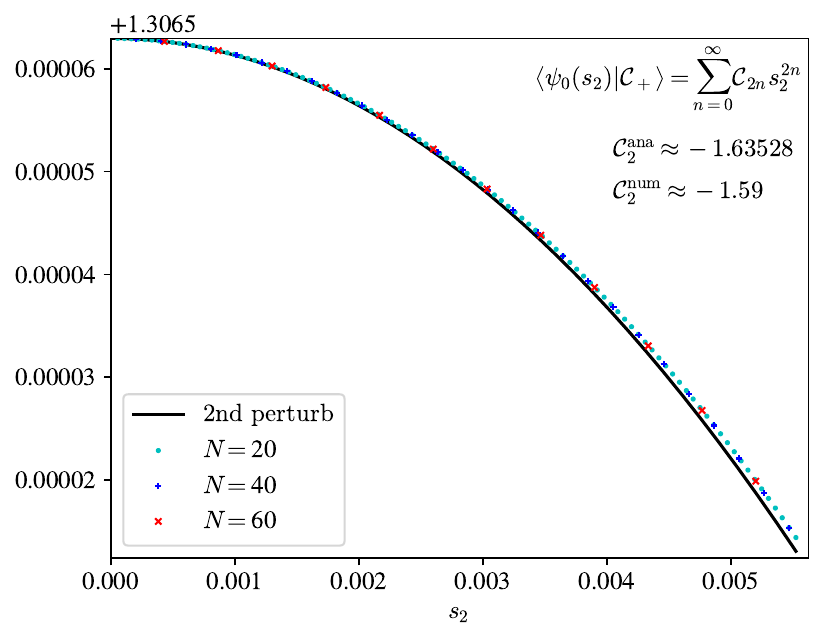}
\caption{Crosscap overlap of the transverse field Ising chain with a longitudinal field compared with the leading-order (2nd order) conformal perturbation prediction. The numerical data with three different chain lengths ($N=20,40,60$), shown with different colored symbols, clearly collapse on the same line, corresponding to the universal scaling function. For three data sets obtained with different chain lengths, parabola fit gives (almost) the same numerical value $\mathcal{C}^{\mathrm{num}}_2 \approx -1.59$.}
\label{fig:Figure1-SM}
\end{figure}

\subsection{E. $\mathbb{Z}_3$ parafermion CFT with thermal perturbation}

The $\mathbb{Z}_3$ parafermion CFT possesses a non-diagonal torus partition function with respect to the $\mathcal{M}(6,5)$ Virasoro symmetry. Its corresponding Klein bottle partition function is a linear combination of $\mathcal{M}(6,5)$ Virasoro characters~\cite{TangW2017},
\begin{align}
    Z^{\mathcal{K}}= \chi_{(1,1)} + \chi_{(1,5)} + \chi_{(3,5)} + \chi_{(3,1)} + 2\chi_{(3,3)} + 2\chi_{(1,3)}\, ,
\end{align}
where the identity and thermal primary fields are identified as $\mathbbm{1} = (1,1)$ and $\varepsilon = (3,5)$, respectively. The modular $\mathcal{S}$ transformation of this Klein bottle partition function determines the associated crosscap coefficients to be $\mathcal{A}_{\mathbbm{1}} = (3+6/\sqrt{5})^{1/4}$ and $\mathcal{A}_\varepsilon = (3-6/\sqrt{5})^{1/4}$. For the $\mathbb{Z}_3$ parafermion CFT with thermal perturbation (perturbation operator has conformal weight $h_\varepsilon = \bar{h}_\varepsilon = 2/5$), the first-order perturbation correction to the crosscap overlap is given by
\begin{align}
\langle\psi_0(s)|\mathcal{C}_\pm^{\mathbb{Z}_3}\rangle = \left(3+\frac{6}{\sqrt{5}}\right)^{1/4} \pm \frac{5}{8(2\pi)^{1/5}}\left(3-\frac{6}{\sqrt{5}}\right)^{1/4}\,\frac{\Gamma (2/5)\Gamma (7/5)}{\Gamma (4/5)} \cdot s + \mathcal{O}(s^2) = 1.54401 \pm 0.54881\cdot s + \mathcal{O}(s^2)\,,
\label{eq:Z3-crosscap-correction-SM}
\end{align}
where $\pm$ denotes the crosscap state and its Kramers-Wannier dual for the $\mathbb{Z}_3$ parafermion CFT.

Leading-order expansion coefficients of the crosscap overlap are expected to be useful in identifying critical theories in lattice models simulations. As an example, we consider the three-state quantum clock chain
\begin{align}
H = -\sum_{j=1}^N (\sigma^{\dag}_j\sigma_{j+1} + \sigma^{\dag}_{j+1}\sigma_{j}) - h_3 \sum_{j=1}^N (\tau_j + \tau^{\dag}_j) 
\label{eq:quantum-clock-SM}
\end{align}
with
\begin{align}
\tau = \begin{pmatrix}
0 & 1 & 0  \\
0 & 0 & 1 \\
1 & 0 & 0
\end{pmatrix},  \quad
\sigma = \begin{pmatrix}
1 & 0 & 0  \\
0 & e^{2\pi \mathrm{i}/3} & 0 \\
0 & 0 & e^{4\pi \mathrm{i}/3}
\end{pmatrix}
\end{align}
being $\mathbb{Z}_3$ spin matrices, where periodic boundary condition is imposed. At the critical point $h_3 = 1$, the model~\eqref{eq:quantum-clock-SM} is described by the $\mathbb{Z}_3$ parafermion CFT, with velocity $v=\frac{3\sqrt{3}}{2}$ determined from the exact solution~\cite{Albertini1992}. Away from (but near) $h_3 = 1$, the model is ordered (with spontaneous $\mathbb{Z}_3$ symmetry broken) for $h_3<1$ and disordered for $h_3>1$, whose low-energy effective theory is just the $\mathbb{Z}_3$ parafermion CFT with thermal perturbation. The dimensionless coupling $s$ for the lattice model~\eqref{eq:quantum-clock-SM} is given by $s = \frac{\sqrt{\mathcal{N}_{\varepsilon}}}{v} (1-h_3) (v N)^{6/5}$~\cite{ZhangYS2023}, where $\mathcal{N}_{\varepsilon}=\lim_{r\rightarrow \infty}\lim_{N\rightarrow \infty} r^{8/5} \langle (\tau^{\dag}_j + \tau_j) (\tau^{\dag}_{j+r} + \tau_{j+r}) \rangle_{\mathrm{c}} = 0.315$ ($\langle \cdots \rangle_{\mathrm{c}}$: connected correlator evaluated at the critical point $h_3 = 1$) is the normalization of the perturbation operator on the lattice. The crosscap overlap between the ground state of Eq.~\eqref{eq:quantum-clock-SM} and the lattice crosscap state $|\mathcal{C}_{\mathrm{latt}}\rangle = \prod_{j=1}^{N/2} \left( \sum_{\alpha=1}^{3}|\alpha\rangle_j |\alpha\rangle_{j+N/2} \right)$ has been calculated using DMRG. The numerical results shown in Fig.~\ref{fig:Figure2-SM} are in good agreement with the conformal perturbation theory result in Eq.~\eqref{eq:Z3-crosscap-correction-SM}.

\begin{figure}[ht]
\centering
\includegraphics[width=0.6\textwidth]{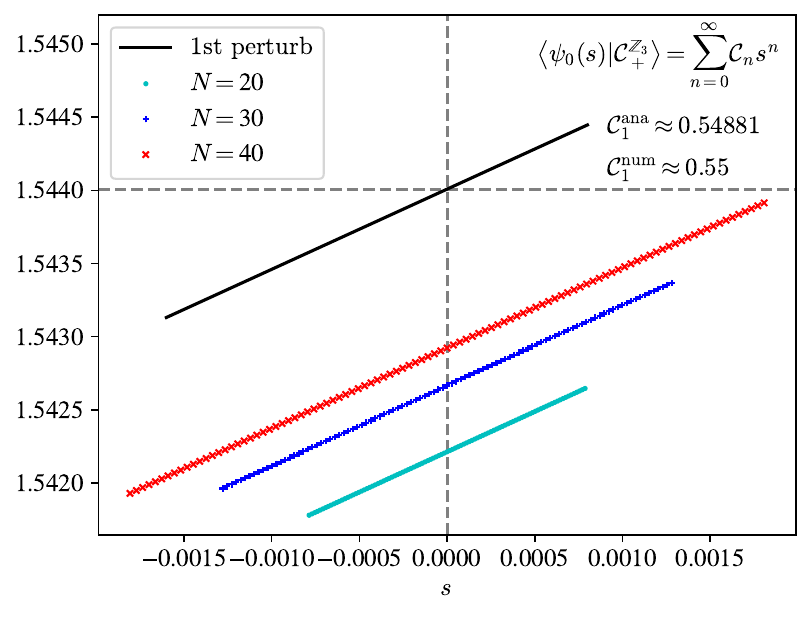}
\caption{Crosscap overlap of the three-state clock model compared with the leading-order (1st order) conformal perturbation prediction. Numerical results with three different chain lengths ($N=20,30,40$) are shown with different colored symbols. For all three data sets, fitting their slopes gives (almost) the same numerical value $\mathcal{C}^{\mathrm{num}}_1 \approx 0.55$.}
\label{fig:Figure2-SM}
\end{figure}

\end{widetext}

\end{document}